\documentclass[sigconf,9pt]{acmart}

\usepackage[english]{babel}

\copyrightyear{2026}
\acmYear{2026}
\setcopyright{cc}
\setcctype{by}
\acmConference[SIGCOMM '26]{ACM SIGCOMM 2026 Conference}{August 17--21, 2026}{Denver, CO, USA}
\acmBooktitle{ACM SIGCOMM 2026 Conference (SIGCOMM '26), August 17--21, 2026, Denver, CO, USA}
\acmDOI{10.1145/3789240.3829196}
\acmISBN{979-8-4007-2467-1/2026/08}

\usepackage{tikz}
\usepackage{amsmath}
\usepackage{xcolor}
\usepackage{xspace}
\usepackage{graphicx}
\usepackage{subcaption}
\usepackage{url}
\usepackage{enumitem}
\usepackage{multirow}
\usepackage[noabbrev, capitalise]{cleveref}
\usepackage[linesnumbered, ruled, vlined]{algorithm2e}
\usepackage{pgfplots}
\pgfplotsset{compat=1.18}
\usetikzlibrary{patterns}
\usepackage{soul}
\usepackage{siunitx}
\usepackage{makecell} 
\usepackage{diagbox} 
\usepackage{hyperref}
\usepackage{pifont}
\usepackage{siunitx} 
\usepackage{multirow}
\usepackage{fontawesome5}
\let\oldfaStar\faStar
\let\oldfaStarHalf\faStarHalf

\renewcommand{\faStar}{\scriptsize\oldfaStar}
\renewcommand{\faStarHalf}{\scriptsize\oldfaStarHalf}

\usepgfplotslibrary{polar}
\crefformat{section}{\S#2#1#3}

\usetikzlibrary{external}
\tikzset{external/system call={}}

\sethlcolor{yellow} 

\definecolor{darkblue}{RGB}{0, 0, 150}

\definecolor{darkred}{RGB}{220, 0, 0}

\usepackage{enumitem}
\setlist[itemize]{leftmargin=0cm,itemindent=.5cm,labelwidth=0.6\itemindent,labelsep=0cm,align=left, itemsep=.0cm, topsep=0cm}
\setlist[enumerate]{leftmargin=0cm,itemindent=.3cm,labelwidth=0.6\itemindent,labelsep=0.1cm,align=left, itemsep=.0cm, topsep=0cm}

\crefformat{section}{\S#2#1#3} 
\crefformat{subsection}{\S#2#1#3}
\crefformat{subsubsection}{\S#2#1#3}

\newcommand{\parab}[1]{\noindent\textbf{#1}\xspace}

\newcommand{\sysname}{EPIC\xspace} 


\usepackage[justification=RaggedRight,labelfont=bf,font=small,textfont=bf,tableposition=below]{caption}

\begin{document}
\title{\sysname: Abstraction and Polymorphism of In-Network Collectives on Ethernet}


\author{
    Yitao Yuan\textsuperscript{1},
    Jianglong Nie\textsuperscript{1},
    Tianyu Bai\textsuperscript{1},
    Ruizhe Zhou\textsuperscript{1},
    Siyuan Cao\textsuperscript{1},
    Xujie Fan\textsuperscript{1}, 
    Yuchen Xu\textsuperscript{1}, \\
    Junkai Chen\textsuperscript{1},
    Chenqi Zhao\textsuperscript{1},
    Nengyuan Zhang\textsuperscript{1},
    Shaoke Fang\textsuperscript{1}, 
    Jiangyuan Chen\textsuperscript{2},
    Yuanfeng Chen\textsuperscript{3}, \\
    Jiaqi Sun\textsuperscript{1},
    Zhan Wang\textsuperscript{4},
    Xiaohua Xu\textsuperscript{2}, 
    Yuchao Zhang\textsuperscript{5},
    Yang Liu\textsuperscript{5},
    Xiangrui Yang\textsuperscript{3}, 
    Jing Lin\textsuperscript{6}, \\
    Xiaohe Hu\textsuperscript{6},
    Yang Li\textsuperscript{7}, 
    Chao Jiang\textsuperscript{7},
    Limin Xiao\textsuperscript{7},
    Weifeng Zhang\textsuperscript{7},
    Junjie Wang\textsuperscript{8},
    Wei Cheng\textsuperscript{8}, \\
    Yazhu Lan\textsuperscript{9}, 
    Jianbo Dong\textsuperscript{9},
    Binzhang Fu\textsuperscript{9},
    Wenfei Wu\textsuperscript{1}
}
\authornote{Wenfei Wu is the corresponding author; contact him at wenfeiwu@pku.edu.cn. Wenfei Wu is also affiliated with Beijing Key Laboratory of Software and Hardware Cooperative Artificial Intelligence Systems.
    Authors are members of ETH+ Consortium.}
\affiliation{
  \institution{
    1 PKU,
    2 USTC,
    3 NUDT,
    4 ICT, CAS, 
    5 BUPT,
    6 Infrawaves,
    7 Lenovo Research,  
    8 Centec,
    9 Alibaba Cloud \\
    ETH+ Consortium
    }
    \city{}
    \state{}
    \country{}
}


\renewcommand{\authors}{Yitao Yuan, Jianglong Nie, Tianyu Bai, Ruizhe Zhou, Siyuan Cao, Xujie Fan, Yuchen Xu, Junkai Chen, Chenqi Zhao, Nengyuan Zhang, Shaoke Fang, Jiangyuan Chen, Yuanfeng Chen, Jiaqi Sun, Zhan Wang, Xiaohua Xu, Yuchao Zhang, Yang Liu, Xiangrui Yang, Jing Lin, Xiaohe Hu, Yang Li, Chao Jiang, Limin Xiao, Weifeng Zhang, Junjie Wang, Wei Cheng, Yazhu Lan, Jianbo Dong, Binzhang Fu, Wenfei Wu}

\renewcommand{\shortauthors}{Yitao Yuan, Jianglong Nie, Tianyu Bai, Ruizhe Zhou, Siyuan Cao, Xujie Fan, et al.}
\begin{abstract}
In-Network Collective (INC) acceleration holds immense potential for optimizing AI training and inference; however, its cross-layer nature has historically hindered investment and adoption within the open Ethernet ecosystem. To bridge this gap, we propose \sysname (Ethernet Polymorphic In-network Collectives), an INC protocol specification and reference system built on the principle of ``Unified Abstraction, Polymorphic Realization.'' \sysname introduces an abstraction compatible with standard Ethernet that aligns functional boundaries with participant roles, while offering polymorphic realizations tailored to varying hardware capabilities.

We address three fundamental challenges: first, we employ a modular design that enables an evolutionary path from simple to complex implementations, allowing vendors to iterate their hardware incrementally; second, we apply formal verification methodologies to prove the correctness of all proposed polymorphic modes; and third, we develop a unified resource management model versatile enough for diverse INC scenarios. 
We conduct extensive experiments—model checking, packet/flow simulations, VM emulation, Tofino/NP Testbed, and FPGA/RTL verification—to validate \sysname's correctness, performance gain, and feasibility. 

\end{abstract}


\maketitle



\section{Introduction}
\label{sec:introduction}


As processing power for large-scale ML model training surges, communication overhead has become a critical performance factor~\cite{dong2024boosting,sapio2021scaling,li2024understanding,hoefler2024hammingmesh,wang2023topoopt,cao2024crux}. \textit{In-Network Collective (INC)} communication, a new network communication paradigm, addresses this by offloading collective operations like AllReduce to switches, thereby compressing traffic volume, reducing transmission time, and lowering overhead~\cite{hoefler2014energy,hoefler2007implementation,hoefler2025network}. This efficacy has been proven by products such as NVIDIA SHARP~\cite{graham2016scalable,graham2020scalable} and various prototypes~\cite{lao2021atp,sapio2021scaling,liu2023network,broadcom_tomahawk5}. Meanwhile, the open \textit{Ethernet} ecosystem, one of the most mature and widely deployed networks~\cite{qi2025sglb, qian2024alibaba,meng2025astral,gangidi2024rdma,bai2023empowering,gao2021cloud}, is evolving into networking AI clusters and integrating INC capabilities.


Enabling INC within the open Ethernet ecosystem faces a core challenge: \textit{single participants struggle to implement cross-layer INC systems independently}. Building INC requires the joint design of communication libraries, network stacks, NICs, switches, and controllers, which are developed by different vendors and operators in Ethernet. The system's integrity relies heavily on the interoperability of these components. This fragmentation poses a major risk for single participants, particularly switch vendors: 
the complexity of INC necessitates excessive upfront investment and prolonged coordination, but isolated efforts within each scope fail to yield system-level benefits without cross-vendor support, which could lead to shelved features and wasted investment.




As a protocol standardization working group, we propose \sysname (\underline Ethernet \underline Polymorphic \underline In-network \underline Collectives), an INC protocol leveraging ``\textbf{unified abstraction, polymorphic realization}'' as a viable path forward on Ethernet.
\textbf{(i)} It establishes a unified abstraction defining INC behaviors, participant functional scopes, and rigorous interoperability interfaces, while retaining implementation flexibility within each scope (\cref{sec:overview}).
\textbf{(ii)} \sysname's data plane aligns components with ecosystem roles via strict semantics with three-mode polymorphic realizations (\cref{sec:data-plane}). 
\textbf{(iii)} \sysname's control plane operates via software-defined networking (SDN), where a central controller manages global resources, policies, and rule dissemination (\cref{sec:resource-management}). 
\sysname addresses three critical challenges: the prohibitive development overhead for vendors to support various polymorphic realizations, the complexity of verifying protocol correctness across these modes, and the lack of a unified resource model to support various management policies (\cref{subsec:challenges}).



\begin{table}[t]
\caption{Implementations and Key Results}
\label{tab:properties}
\footnotesize
\setlength{\tabcolsep}{3pt}
\begin{tabular}{|c|l|l|}
\hline
\textbf{Implementation} & \multicolumn{1}{c|}{\textbf{Target Property}} & \multicolumn{1}{c|}{\textbf{Key Results}} \\ \hline
Model Checking          & Correctness                                    & All 3 modes are correct.                             \\ \hline
Tofino Testbed          & Performance                                    & Up to $1.59\times$ for collectives and \\
                        & Acceleration                                   & $1.38\times$ for training.                                                     \\ \hline
Emulation               & Interoperability,                              & All 3 modes work with RoCE.                          \\
                        & Evolvability                                   & Mode-III reuses 61\% of Mode-II.                       \\ \hline
Flow-level              & Gain in Resource                               & Reduce GPT-3 JCT by up to 45.8\%,                      \\
Simulation              & Management                                     & multi-job P99 JCT by 30.9\%                                                    \\ \hline
Packet-level            & Loss Tolerance                                 & Mode-III outperforms Mode-II.                        \\
Simulation              &                                                &                                                      \\ \hline
FPGA                    & Latency,                                       & \SI{53}{ns} processing latency                              \\
                        & Resource Cost                                  &  O(1) MB SRAM                                      \\ \hline
RTL                     & Chip Feasibility                               & \SI{3.2}{Tbps}; \SI{4.89}{\milli\meter\squared}@ \SI{28}{nm}                                \\ \hline
\end{tabular}
\end{table}

The \sysname specification was distributed to multiple organizations for comprehensive validation. The working group members produced various implementations and validated \sysname's properties as in \cref{tab:properties}.
In this paper, we make the following contributions.
\begin{itemize}
    \item A unified abstraction of six-primitive collectives to enable INC systems in Ethernet ecosystem,
    \item A polymorphic data plane enabling switch vendors with diverse hardware capabilities to realize switches,
    \item A modularized data plane enabling switch vendors to evolve INC switches between polymorphic modes,
    \item Analysis of properties of polymorphic modes, including model-checking verified correctness, transmission efficiency, space complexity, logic complexity, and fault tolerance,
    \item Resource model derived from the abstraction that enables various resource management policies.
\end{itemize}

{\textit{This paper does not raise any ethical issues.}}

\section{Background}
\label{sec:background}

\subsection{\sysname's Application Scope and Benefits}
\label{subsec:scope-benefits}


\parab{\sysname focuses on scale-out networks.} AI clusters use dual-interconnects: scale-out (NIC/switch-based) and scale-up (e.g., NVLink~\cite{nvidia_nvlink_2014}, UALink~\cite{ualink_consortium}). \sysname targets the Ethernet/RoCE ecosystem~\cite{li2025revisiting,ibta_rocev2_2014} for scale-out expansion, embedding INC semantics in RoCE header fields, which is the de facto standard on Ethernet~\cite{bai2023empowering,gao2021cloud}. 
In contrast, scale-up protocols have not yet converged with multi-vendor interoperability; embedding INC semantics into them involves a prohibitive hardware barrier of redesigning accelerator I/O dies.




\parab{\sysname focuses on regular collectives.} We distinguish primitives by data patterns: 
\textbf{(1)} Regular: AllReduce, Reduce, and Broadcast. These maintain strict alignment between element indices, packet sequences, and memory addresses. 
\textbf{(2)} Irregular: MoE token dispatch/combine (AlltoAllv)~\cite{liu2024deepseek}. These involve asymmetric traffic and lack positional correspondence.
Designing for irregular patterns requires complex index-to-memory mapping and congestion management. 
Its numerous dispatch/combine traffic patterns also strain the forwarding-table capacity if implemented using IP multicast and \sysname's lookup table.
MoE AlltoAllv's abstraction and realization would be considered separately as future work.



\parab{\sysname benefits communication and end-to-end model training/inference.} INC enhances communication by reducing latency (fewer hops), minimizing traffic congestion, and lowering end-host overhead to better saturate bandwidth. 
For end-to-end tasks, INC provides direct acceleration in communication-bound clusters. In high-bandwidth (and expensive) environments, it still yields substantial gains by freeing GPU resources for computation, raising the performance floor and simplifying system tuning.

\subsection{Requirements of INC on Ethernet}
\label{subsec:requirements}

\sysname should satisfy the requirements from ecosystem organizations, hardware-offloading trends, and AI collective patterns.



\parab{(1) \sysname must strictly adhere to Ethernet functional scopes and interoperability.} It partitions components according to the industry division of labor, where functionality results from collaboration between CCL developers, NIC/switch vendors, and operators via standard protocols like IP and RoCE.
Standardizing these interaction primitives is essential to avoid the risks of closed, vertically integrated solutions. By defining clear interoperability specifications, \sysname enables cross-layer optimization without sacrificing vendor neutrality. This alignment is a prerequisite for breaking ecosystem deadlock and enabling deployable INC.


\parab{(2) \sysname should align with RoCE}, one of the de facto standards for AI interconnects~\cite{gangidi2024rdma}. As bandwidth scales toward \SI{1.6}{Tbps}, hardware transport offload via RoCE is essential to bypass the CPU ``performance wall'' and maintain zero-copy, low-latency transfers.
Unlike previous software-based INC (e.g., SwitchML~\footnote{SwitchML is developed on DPDK or RDMA UC.} or ATP), \sysname leverages mature hardware engines to sustain line-rate speeds. This approach treats INC as an enhancement of existing infrastructure rather than a replacement. By inheriting legacy RoCE, \sysname enables vendors to scale bandwidth without the cost of proprietary protocols, ensuring a production-ready and evolvable ecosystem.


\parab{(3) \sysname must support a comprehensive suite of collective primitives beyond AllReduce.} To keep pace with evolving AI workloads—such as FSDP (ZeRO)~\cite{zhao2023pytorch,rajbhandari2020zero}, which relies on ReduceScatter and AllGather—a modern INC framework must offer versatile support for diverse patterns.
Exhaustive primitive support is critical, as it is both future-proof and production-ready for AI training and inference.
Broad functional coverage eliminates the need for hybrid deployments, where the lack of hardware-offloaded primitives forces a mix of INC and conventional CCLs and causes associated orchestration overhead.


\subsection{Challenges}
\label{subsec:challenges}


We overcome the challenges of ensuring evolvability and correctness in a polymorphic data plane while maintaining control plane generality.


\parab{(1) Diverse performance-complexity tradeoffs (\cref{sec:analysis}) and high development overhead hinder universal polymorphic INC support for switch vendors.} Ethernet switch capabilities vary: fixed-function ASICs prioritize throughput over flexibility, while programmable ones (e.g., Tofino~\cite{tofino}) are limited by SRAM and ALU resources. 
Choosing a restrictive single-mode solution limits the vendor's opportunity to address a broader range of market scenarios, whereas pursuing all modes incurs the risk of prohibitive investment.


To resolve this, \sysname \textit{modularizes switch INC functionality}, enabling a low-cost transition from simple to complex modes. By decoupling functionalities into reusable components, vendors achieve scaling through module composition. This adaptive approach allows implementations to align with specific hardware constraints—from basic packet replication to full RoCE endpoints—facilitating an incremental and sustainable deployment path for the vendors (\cref{sec:data-plane}).


\parab{(2) Polymorphic realizations complicate the verification of computational integrity.} Ensuring protocol termination and server-equivalent results is challenging, as each mode introduces unique intermediate states and cross-layer interactions. Analyzing these diverse designs individually incurs prohibitive analytical overhead.

To address this, we performed a \textit{formal analysis} of the \sysname protocol suite. Using automated model checking, we systematically explored the state space of the polymorphic data plane and network conditions (e.g., packet loss). This rigorous approach automatically verifies protocol termination and computational accuracy across all modes, providing provable guarantees of system integrity while eliminating manual, error-prone proofs (\cref{sec:analysis}).



\parab{(3) Distributed switch resources lack a unified model to support diverse, scenario-specific management policies.} The increasing heterogeneity of AI workloads—ranging from multi-tenant environments to varied cluster scales— demands unique allocation policies. Manually coordinating customized INC management for every infrastructure (switch modes) combination is infeasible due to extreme complexity.

To bridge this gap, we propose a \textit{unified INC resource model} based on the SDN paradigm. By virtualizing switch resources and decoupling the control from the data plane, this model provides generic interfaces for centralized orchestration. This architecture supports diverse policies—including isolation and contention-based sharing—transforming fragmented resources into a flexible, programmable fabric. This approach minimizes administrative overhead while maximizing resource utilization across dynamic AI scenarios (\cref{sec:resource-management}).

\begin{figure}[tb]
    \centering
    \includegraphics[width=0.9\linewidth]{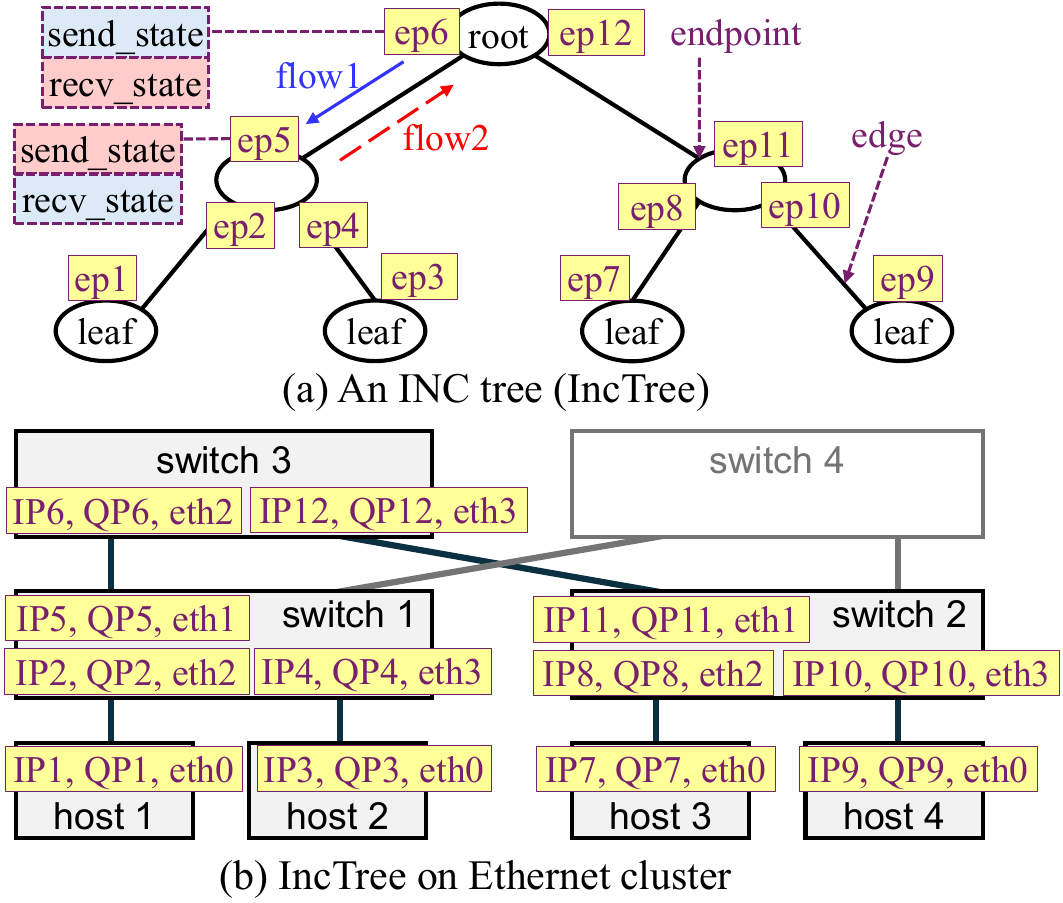}
    \caption{\sysname's abstraction}
    \label{fig:abstraction}
\end{figure}

\section{\sysname Overview}
\label{sec:overview}

\subsection{Abstraction}
\label{subsec:abstraction}

We define \sysname's abstraction and map it to Ethernet clusters. \sysname's realization can support six collective primitives.

\parab{INC Tree.}
Collective communication is performed on a group of parallel application processes called \textit{ranks}. \sysname employs an \textit{INC Tree (IncTree)} to describe the logical topology of a communication group (\cref{fig:abstraction}). Each rank maps to a leaf node, while non-leaf nodes act as intermediate aggregation or replication points within the network fabric. The tree is assigned a \textit{root}. An \textit{edge} connects two nodes on the tree, and is undirected. Data units (packets) on an edge with the same direction form a \textit{flow}. 

We define an \textit{endpoint} to describe flow routing and transmission. An edge has two \textit{endpoints}, each on one of the edge's two nodes. Note that a node can connect to multiple edges, but those edges' endpoints on that node are distinct. On an edge, a flow's \textit{transmission states} consist of sending states on its source endpoint and receiving states on its sink endpoint (\cref{fig:abstraction}a). In a node, a flow's \textit{routing} from one edge to another can be described as from one endpoint to another.

\begin{figure}[tb]
    \centering
    \includegraphics[width=\linewidth]{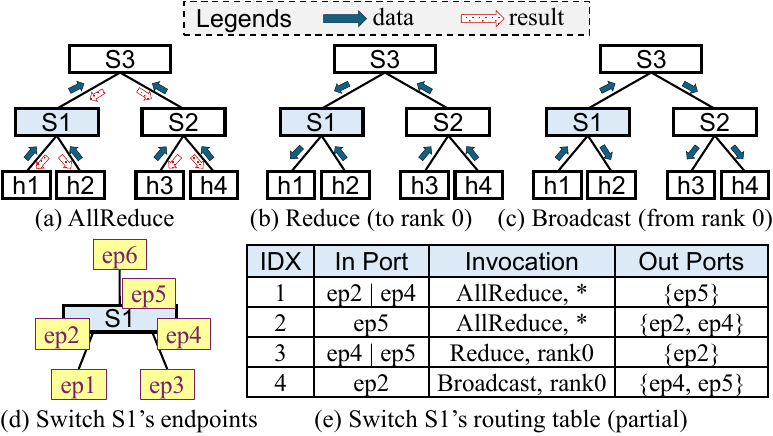}
    \caption{Data Flow and Routing in IncTree}
    \label{fig:aggtree-routing}
\end{figure}

\parab{Collectives on IncTree.}
\sysname performs various collective primitives on the IncTree. It supports six primitives: AllReduce, Reduce, Broadcast, Barrier, ReduceScatter, and AllGather. 
The latter three can be derived from the former three (\cref{fig:aggtree-routing}a to \cref{fig:aggtree-routing}c): Barrier is equivalent to an AllReduce with an empty data payload, ReduceScatter is the sequential execution of multiple Reduces, and AllGather is, similarly, multiple Broadcasts (Appendix~\cref{sec:app:collectives}). Thus, in the following text, we focus on the design of AllReduce, Reduce, and Broadcast.

\parab{IncTree on Ethernet Cluster.} In Ethernet-based Clos topologies, hosts reside at the edge while switches form the core. An IncTree maps leaf nodes to hosts and intermediate nodes to switches. Each endpoint is defined as an <IP, QP> tuple associated with a specific host NIC or switch port (\cref{fig:abstraction}b). Consequently, an IncTree edge may span multiple physical links defined by the path between its two endpoints. 
The IncTree abstraction also applies to accelerator-centric topologies~\cite{jouppi2023tpu}, with non-leaf nodes on accelerators.

During collectives, switches execute local actions via match-action tables. These tables recognize and forward packets to ports that align with the IncTree's data flow. For instance, in an AllReduce operation (\cref{fig:aggtree-routing}e), switch $S1$ aggregates packets from child endpoints ($ep2$, $ep4$) toward the parent ($ep5$) and distributes results back down those same paths.

\subsection{Architecture}
\label{subsec:architecture}

\begin{figure}[tb]
    \centering
    \includegraphics[width=\columnwidth]{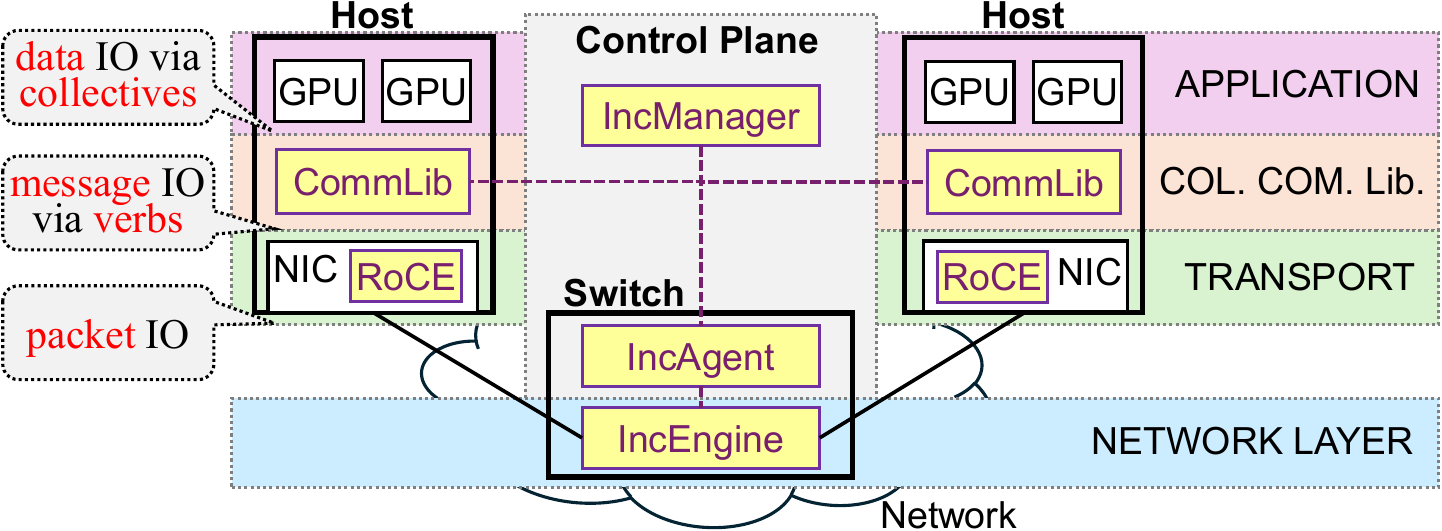}
    \caption{\sysname Architecture}
    \label{fig:architecture}
\end{figure}

\parab{Architecture.} \sysname decouples control and data planes, aligning with industry functional boundaries. The data plane comprises host-side \textit{CommLib}, standard NIC \textit{RoCE}, and switch-based \textit{IncEngine}. The control plane features switch-resident \textit{IncAgents} and a centralized \textit{IncManager} for resource discovery, policy formulation, and rule dissemination (\cref{fig:architecture}).

\parab{Interfaces.} Components interact through standardized interfaces across two paths: \textbf{(1) Control Path}: The IncManager coordinates with CommLib and IncAgent via RPC. The IncAgent then configures the IncEngine using internal register or table updates; \textbf{(2) Data Path}: A layered collaboration ensures interoperability.

The data path consists of three components. \textbf{(1)} CommLib: Provides collective primitives (e.g., {\tt AllReduce}) to applications and exchanges data with the NIC via standard RDMA Verbs. \textbf{(2)} RoCE: Executes standard RDMA operations without hardware changes, interfacing with the IncEngine via RoCE packets over physical links. \textbf{(3)} IncEngine: Performs packet-level parsing, processing (aggregation/replication), and forwarding between NICs or peer engines.

\subsection{Workflow}
\label{subsec:workflow}


\subsubsection{System and Group Lifecycle}

\parab{Bootup.}
The lifecycle begins with component initialization: CommLib daemons launch on hosts, and IncEngine is enabled on switches. All instances report local states to the IncManager, which constructs the global topology and manages resources (\cref{subsec:resource-model}).

When an application calls {\tt InitGroup()}, the IncManager computes and maps a logical IncTree onto the physical fabric. It then disseminates configurations: host CommLibs receive flow information, and switch IncAgents install local routing states into their respective IncEngines.

\parab{Connection and Routing.} Following the setup, CommLib establishes RoCE connections to neighboring switch nodes. Switches support polymorphic connection handling, ranging from full stack functionality to state-less translation (\cref{subsec:mode-i}, \cref{subsec:mode-ii}, \cref{subsec:mode-iii}). During initialization, IncManager pre-computes rules for all $2N+1$ traffic patterns (AllReduce, Reduce, Broadcast) and configures forwarding for both RoCE data and ACK packets.

\parab{Teardown.} Upon completion, {\tt DestroyGroup()} triggers the IncManager to instruct IncAgents to delete local states and forwarding rules. The IncManager then releases group-specific resource reservations for future allocation.

\subsubsection{Runtime Collective Invocation} 

\parab{Control Signaling.} \sysname supports multiple runtime primitives using in-band signaling, avoiding the custom INC headers like SHARP~\cite{graham2016scalable} or the CPU-intensive header prepending like NetReduce~\cite{liu2023network}. Before data transmission, the CommLib sends a standalone RDMA {\tt Send with Immediate} message\footnote{Switches recognize it from other verbs by its distinct RoCE OP Code.} to notify IncTree nodes of the collective type, root, and data size. This signaling is pipelined with subsequent data to hide latency. If lost, the switch refuses data processing until retransmission, ensuring protocol safety through a validated PSN range (derived from the data size).

\parab{Data Processing.} The execution follows a five-step pipeline:
\textbf{(1)} Chunking \& Flow Control: CommLib partitions tensors into messages and applies application-level flow control to prevent switch buffer overflow.
\textbf{(2)} Standard Transport: The commodity NIC encapsulates messages into RoCE packets for delivery to the IncEngine.
\textbf{(3)} Switch Processing: The IncEngine performs aggregation, replication, or forwarding. It identifies the collective setting via the preceding control message and uses pre-configured rules to route packets through the IncTree.
\textbf{(4)} Reliable Reception: Receiver NICs reassemble packets and signal CommLib via Work Completions (WC). RoCE ACK/NAK packets are processed by the IncEngine on the reverse path to manage state release (\cref{sec:data-plane}).
\textbf{(5)} Completion: CommLib assembles the results and returns control to the application.

\subsubsection{System Fault Tolerance}
Running distributed systems with \sysname must handle runtime switch, link, and rank failures. \sysname can run as a network slice (separate CommLib and IncEngine instances from traditional CCL and switch forwarding), and set up traditional CCL (e.g., NCCL) as failover; the failover process can be enabled via services like MCCS~\cite{wu2024mccs}.

\sysname itself handles failures by reinitializing groups~\cite{uec2024specupdate}. System-level failover requires sandboxing ranks and runtime group member change; \sysname can apply Continuum~\cite{lao2026continuum}.

\section{Polymorphic Data Plane}
\label{sec:data-plane}

\sysname modularizes IncEngine; composing modules in different ways creates different modes, and reusing modules evolves modes from simple to complex.\footnote{\sysname enables a vendor to evolve modes, but not interoperation between modes due to their different workflows.}

\begin{figure*}[tb]
\begin{minipage}[b]{0.73\textwidth}
    \centering
    \includegraphics[width=\textwidth]{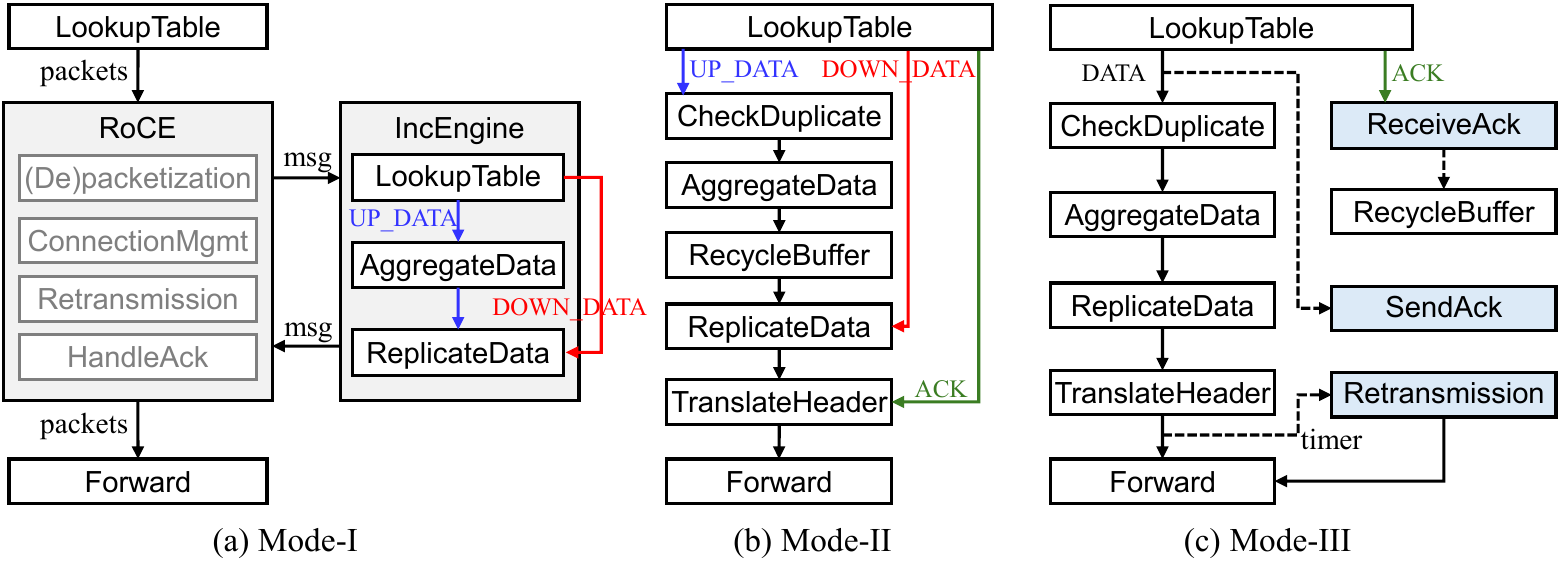}
    \caption{Modules and data flow in three modes}
    \label{fig:modes-modules}
\end{minipage}
\hfill
\begin{minipage}[b]{0.24\textwidth}
    \centering
    \includegraphics[width=\textwidth]{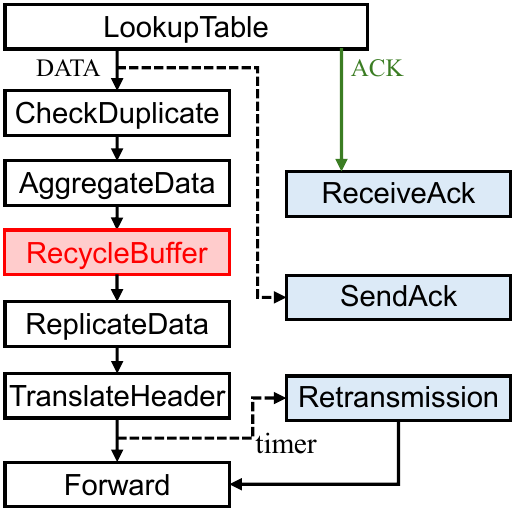}
    \caption{A pitfall in Mode-III design}
    \label{fig:pitfall}
\end{minipage}
\end{figure*}

\subsection{Modularize IncEngine}
\label{subsec:modulize}

We decouple the runtime communication group states from the processing logic, and decompose all functionalities in IncEngine into independent modules. At runtime, these modules retrieve the corresponding states using the header of the current packet, and execute specific logic.

\parab{States and Context.} The collection of all states involved in an INC group at runtime is referred to as the group's \textit{Context}. While an INC group logically possesses a global context across the entire cluster, physically, each switch is only required to install the \textit{Local Context} relevant to its operation; where there is no ambiguity, we use ``context'' to refer to ``switch local context''.

On an IncTree switch node, the states include: (1) routing states, in the format of rules to look up packet headers and decide the output ports, (2) transmission states, where each endpoint involves sending states for its outgoing flow and receiving states for its incoming flow, and (3) computation states, which are buffers to temporarily store intermediate states. At runtime, modules utilize the header of the current packet to retrieve its context. 
These three kinds of states are described and retrieved by the endpoint in the IncTree abstraction.

\parab{Function Modules.}
We prepare the following modules to construct IncEngine. There are modules for state retrieval and routing based on the packet header: {\tt LookupTable},  
{\tt TranslateHeader}, and {\tt Forward}. There are modules for flow transmission: {\tt ReceiveAck}, {\tt SendAck}, and {\tt Retransmission}. There are modules for data (packet payload) operation: 
{\tt CheckDuplicate}, {\tt AggregateData}, {\tt RecycleBuffer}, and 
{\tt ReplicateData}. Module internal logic is elaborated in Algorithm~\ref{alg:mode-II} and \ref{alg:mode-III} in Appendix~\cref{sec:app:algorithms}.

\subsection{Mode-I: Connection Terminated}
\label{subsec:mode-i}

Mode-I provides a ``heavyweight'' realization of the \sysname abstraction for high-end switches capable of supporting a full RoCE stack. In this mode, the switch functions as a Parameter Server (PS), terminating host connections and maintaining complete transport states to ensure reliable transmission. The architecture utilizes a layered protocol stack where the switch handles deduplication, ACK processing, and message reassembly.


\begin{figure}[t]
    \centering
    \includegraphics[width=0.8\linewidth]{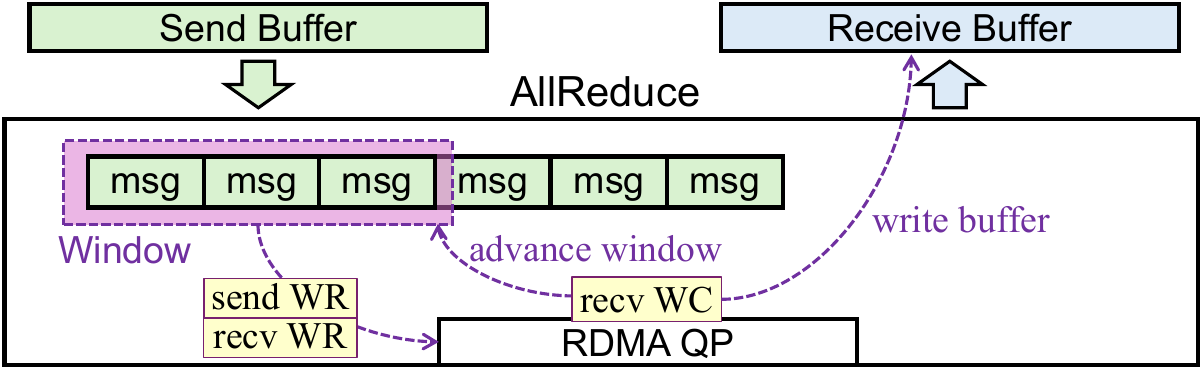}
    \caption{Flow Control in CommLib}
    \label{fig:flow-control}
\end{figure}


\parab{AllReduce workflow.} The host-side CommLib initiates communication with the control signal.
To manage traffic, the system employs message-granularity flow control using paired Send-Recv verbs. This choice is critical, as one-sided Write verbs do not provide work completion (WC) signals that confirm ``result'' readiness. Flow control is achieved by setting the RoCE outstanding WR to be $W$ messages, where $W$ is configured as a system parameter window size (\cref{fig:flow-control}).

Intermediate switches identify \sysname packets via lookup tables, directing data to internal transport layers for hop-by-hop reliable delivery, and to the internal application layer for data aggregation (performed by the {\tt AggregateData} module). Once the IncTree root completes data reduction, the {\tt ReplicateData} module broadcasts results back to child nodes via the transport layer again; the child nodes relay the results towards host receivers (\cref{fig:modes-modules}a).

\parab{Reduce and Broadcast workflows.} They serve as simplified variations of the AllReduce process without bidirectional transmission. In a Reduce operation, data flows unidirectionally from multiple senders to a single receiver through switch-based aggregation, with host CommLibs executing only Send or Recv verbs. Conversely, Broadcast transmits data from one sender to multiple receivers using the {\tt ReplicateData} module. Both primitives leverage the same pre-configured routing rules and RoCE functions.


\begin{table}[t]
\centering
\caption{Symbols and meaning}
\footnotesize
\label{tab:symbols}
\begin{tabular}{|l|l|}
\hline
\multicolumn{2}{|l|}{$U$ \textbf{:} Maximum Transmission Unit (MTU); unit is byte}       \\ \hline
\multicolumn{2}{|l|}{$M$ \textbf{:} message size; unit is MTU; a message is $UM$ bytes}    \\ \hline
\multicolumn{2}{|l|}{$W$ \textbf{:} window size; unit is message;  a window is $UMW$ bytes}  \\ \hline
\multicolumn{2}{|l|}{$N$ \textbf{:} array size in IncTree state} \\ \hline
$H$ \textbf{:} aggregation tree depth   &  $D$ \textbf{:} node degree           \\ \hline
$B$ \textbf{:} link bandwidth           &  $L$ \textbf{:} link latency          \\ \hline
\end{tabular}
\end{table}

\parab{Interaction with RoCE Flow Control and Congestion Control.}
\sysname's flow control means keeping the inflight data volume from overwhelming the switch buffers. The three mechanisms --- \sysname flow control, RoCE congestion control, and RoCE flow control (e.g., PFC or CBFC) --- are layered in order, and work independently.
The runtime actual traffic rate, progress, and volume are constrained by the minimum/slowest of the three mechanisms.
Such independence of \sysname flow control from RoCE congestion control and flow control holds for all the three modes.

\subsection{Mode-II: Connection Translated}
\label{subsec:mode-ii}

Mode-II offers a minimalist design for switch vendors by substituting full transport termination with a ``connection translation'' approach. Instead of maintaining complete RoCE states, the switch modifies and forwards packets while relying on end-hosts for reliability, ensuring transport transparency. This mode requires all NIC endpoints to be initialized with identical Packet Sequence Numbers (PSNs). \footnote{PSN can be set through {\tt ibv\_modify\_qp()}.}

\parab{AllReduce Workflow.} \textbf{(1)} CommLib works in the same way as Mode-I.
The switch maintains a {\tt payload} buffer (array of MTU) and a {\tt degree} buffer (array of counters), both of size {\tt N}.

\begin{figure}[tb]
    \centering
    \includegraphics[width=\columnwidth]{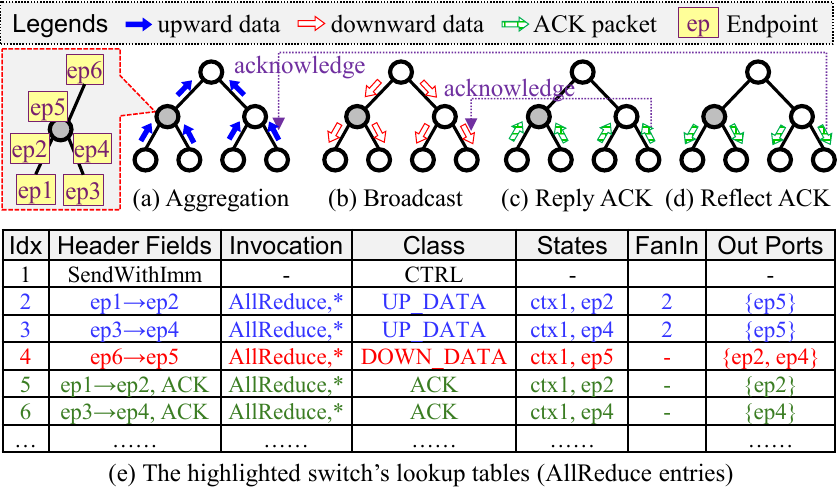}
    \caption{Lookup Table in Mode-II}
    \label{fig:workflow-mode-II}
\end{figure}

\textbf{(2)} The switch identifies upward data packets by header fields (\cref{fig:workflow-mode-II}e), and processes them by modules.
{\tt CheckDuplicate} identifies retransmissions.
For the first arrival of a packet, {\tt AggregateData} module sums up its payload into {\tt payload} buffer (at {\tt idx=pkt.psn\%N})\footnote{If reproducibility is needed (adding numbers in a deterministic order)~\cite{wang2023roar,de2021flare}, IncEngine needs an extra buffer to store data first, and adds data up when the degree is full.} and increments the {\tt degree} at {\tt idx}; for retransmission, aggregation is skipped. If aggregation is incomplete ({\tt degree[idx]<FAN-IN}), the packet is dropped; otherwise, the result ({\tt payload[idx]}) is copied back to the packet, forming a \textit{result packet}. {\tt ReplicateData} further clones packets for the next hops, and {\tt TranslateHeader} modifies header fields (e.g., Dest IP, Dest QP). Finally, {\tt Forward} sends the packet to the corresponding switch port.\footnote{The endpoint states also include RDMA {\tt memory\_region (MR) address} and {\tt rkey} to enable the write admission to the RDMA registered memory region. {\tt address} and {\tt rkey} are distributed (from host CommLib to switch IncEngine via IncManager) during group initialization.}

\underline{\texttt{RecycleBuffer}:} The \textit{payload/degree} size is set to twice the window ($2MW$ MTUs). To support streaming with minimum switch SRAM space, \textit{payload} is used circularly: upon completion of aggregation ({\tt degree[idx]==\text{FAN-IN}}), the slot at $(psn + MW) \% (2MW)$ is cleared for subsequent packets; this is a safe operation as that slot is out of all senders' windows \cite{liu2023network,zhao2023netrpc}.

\textbf{(3)} The IncTree root sends result packets downward. Switches identify downstream packets (\cref{fig:workflow-mode-II}b, \cref{fig:workflow-mode-II}e) and process them via {\tt ReplicateData}, {\tt TranslateHeader}, and {\tt Forward} modules (\cref{fig:modes-modules}).

\textbf{(4)} The result packet finally reaches the RoCE NIC, and triggers an ACK towards switches (\cref{fig:workflow-mode-II}c). This ACK is reflected by the first-hop switch back to the RoCE NIC (with header modification) to acknowledge the upstream data packet (\cref{fig:workflow-mode-II}d). RoCE ACKs are cumulative, and this mechanism ensures correctness even if ACKs are coalesced. Negative ACKs (NAKs) generated due to packet loss are handled in the same way as ACKs: reflected to trigger standard RoCE retransmission at the sender. In the case of packet retransmission, each step of the workflow is idempotent (repeatable without affecting the result).

\parab{Reduce and Broadcast Workflows.}
\textbf{(1)} Reduce: Control and data packets are handled the same as AllReduce; the receiver's RoCE NIC replies ACKs, which are broadcast along the tree to the senders. NAKs are handled similarly.
\textbf{(2)} Broadcast: Control and data packets are sent to receivers along the IncTree, replicated at each node. Receivers reply with ACKs to the sender, which are aggregated by IncTree nodes. ACKs are cumulative; intermediate nodes maintain an $ackPsn$ for each endpoint to record the progress of the cumulative ACK. A node also records a $nodeAckPsn$ as the minimum of all child $ackPsn$ values. An ACK is forwarded upstream only if it updates the $nodeAckPsn$, effectively preventing ACK amplification. NAKs caused by packet loss are not aggregated but are forwarded to the sender to trigger retransmission~\cite{li2024cepheus,huang2023mc}.

\parab{Group PSN Synchronization.}
A primary challenge in multi-collective support is PSN divergence: asymmetric primitives like Reduce or Broadcast cause sequence numbers to drift between ranks, breaking PSN (and switch addressing) alignment for subsequent (possible) AllReduce. 
A ``work-around'' method for this issue is to mandate a single primitive in each group and use multiple subgroups for complex operations like ReduceScatter. 
This method increases the lookup table entries: an $N$-member group needs to configure $2N+1$ \sysname groups, $N$ for reduce, $N$ for broadcast, and $1$ for AllReduce.
Considering a switch only holds a few local groups and $N$'s scale, the table entry cost is acceptable. The switch memory is time-division multiplexed among collective invocations.

We also suggest a ``RoCE-Refined'' approach: allowing a northbound interface to manually synchronize sequence numbers on NICs after asymmetric operations. This is a minimal hardware modification, ensuring Mode-II correctness across diverse traffic patterns.


\subsection{Mode-III: Connection Augmented}
\label{subsec:mode-iii}



\begin{figure}[tb]
    \centering
    \includegraphics[width=\columnwidth]{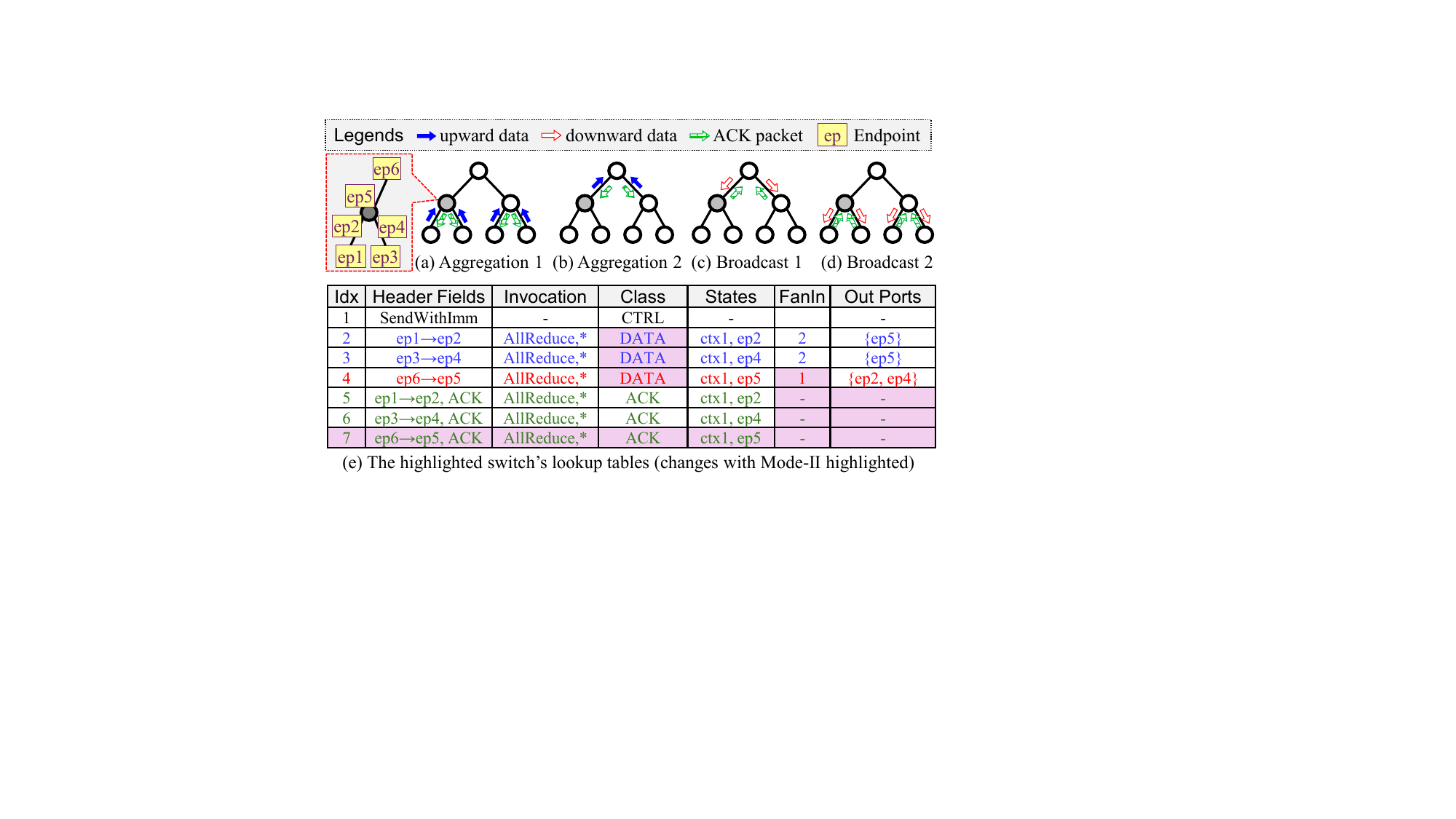}
    \caption{Lookup Table in Mode-III}
    \label{fig:mode-III-tables}
\end{figure}

Mode-III serves as a middle ground between the full-stack complexity of Mode-I and the minimalist nature of Mode-II, providing hop-by-hop retransmission 
without the overhead of complete message packetization (\cref{fig:mode-III-tables}).

\begin{figure}[tb]
    \centering
    \includegraphics[width=\linewidth]{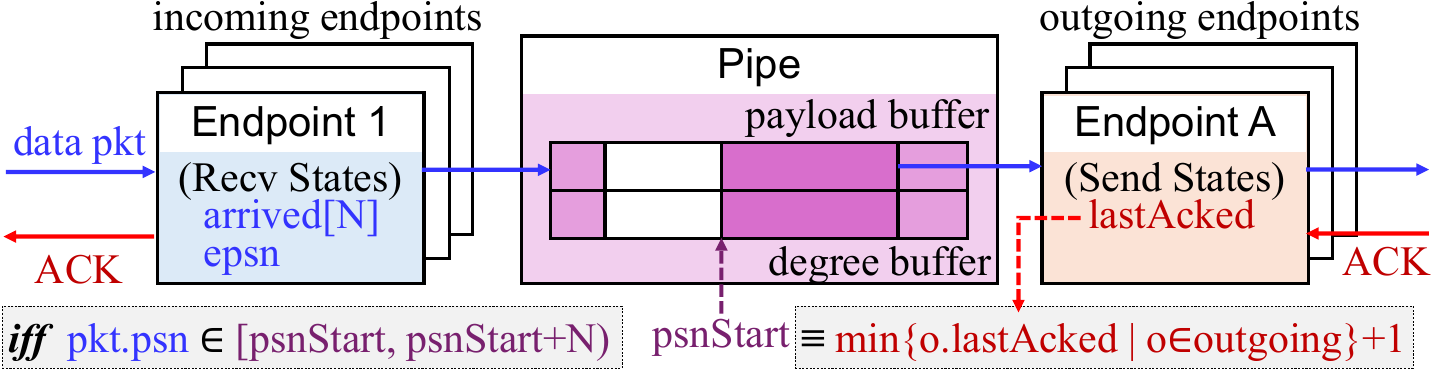}
    \caption{The Pipe Abstraction}
    \label{fig:pipe}
\end{figure}

\parab{A Pipe Abstraction.} This mode introduces a \textit{pipe} abstraction to manage buffer readiness and transport reliability at each switch hop (\cref{fig:pipe}). Each pipe consists of a {\tt payload} buffer and a {\tt degree} buffer—both sized {\tt N}—and maintains a {\tt psnStart} value representing the valid PSN range {\tt [psnStart, psnStart+N)}. 
A pipe is associated with one or several incoming endpoints (whose count is the {\tt FAN-IN} degree) and one or several outgoing endpoints. 

When an incoming endpoint receives a data packet, the switch checks buffer readiness, transmission states, and aggregation states. If the PSN exceeds the valid range, the switch drops the packet and replies a NAK; otherwise, the switch replies an ACK and proceeds. If the packet is within the valid range and is checked as the first arrival, the switch aggregates the packet and proceeds; otherwise, the packet is dropped. If the PSN's aggregation degree is full, the switch sends the packet further to downstream endpoints.


When an outgoing endpoint receives an ACK, the endpoint maintains and updates a {\tt lastAcked} state, which is the maximum ACK PSN it has observed. This value is used to dynamically update the pipe's {\tt psnStart} (defined as the minimum of all outgoing endpoints' {\tt lastAcked + 1}), thereby advancing the sliding window and freeing buffer space for subsequent packets (Algorithm~\ref{alg:mode-III} in Appendix~\cref{sec:app:algorithms}).

\parab{AllReduce Workflow.} \textbf{(1)} CommLib and RoCE NIC remain consistent with Mode-I/II. \textbf{(2)} Upward data packets are handled by an aggregation pipe, with multiple incoming endpoints and one outgoing endpoint. 
\textbf{(3)} Downward packets are handled similarly by a broadcast pipe, with one incoming endpoint and multiple outgoing endpoints.
\textbf{(4)} RoCE NIC sends NAKs when observing packet loss, and the switch handles NAK by retransmitting packets starting from $lastAcked+1$. Whether a switch actively sends NAKs when it observes packet loss (e.g., out-of-order) is optional; both choices work correctly, but sending NAKs is recommended for performance issues (Appendix~\cref{subsec:app:pitfalls}).

Reduce and Broadcast work with one pipe on each switch, enabling unidirectional data flow.

\parab{Group PSN Synchronization.}
IncTree edges allow bidirectional transmission; unidirectional operations (Reduce/Broadcast) desynchronize the sequence numbers of the reverse flow on the same edge. To resolve this in Mode-III, when an ACK for a flow in one direction arrives at an endpoint, the system must simultaneously update the PSN progress of the reverse flow (i.e., {\tt ePsn}) on that endpoint to match the group progress.
We leave the discussion of other possible modes and switch microarchitectures in Appendix~\cref{sec:app:other-modes} and \cref{sec:app:microarchitecture}.

\parab{Rate Synchronization by Congestion Control.}
In Mode-III, rank windows progress independently. If ranks experience different congestion conditions, the congested ranks progress slowly, and the uncongested ones progress fast. It is a better choice to have the switch slow down the faster ranks to avoid excessive sending and dropping (due to being out of pipe PSN range). On the first hop switch, if a packet is dropped due to exceeding the PSN range, the switch had better send the sender a congestion signal (e.g., CNP for DCQCN) to slow down that rank.

\section{Analysis of Modes}
\label{sec:analysis}


\begin{figure}[tb]
    \centering
\begin{tikzpicture}
\begin{polaraxis}[
    width=0.45\columnwidth,
    height=0.45\columnwidth,
    xtick={0, 72, 144, 216, 288},
    xticklabels={
        Correctness,
        Transmission Efficiency,
        Logic Simplicity,
        Space Simplicity,
        Loss Tolerance
    },
    xticklabel style={
        xshift={
            \ticknum == 0 ? 2pt : (
            \ticknum == 1 ? 40pt : (
            \ticknum == 2 ? -8pt : (
            \ticknum == 3 ? -8pt : 25pt)))
        },
        yshift={
            \ticknum == 0 ? 0pt : (
            \ticknum == 1 ? -20pt : (
            \ticknum == 2 ? -15pt : (
            \ticknum == 3 ? 15pt : 15pt)))
        }
    },
    ymin=0, ymax=30,
    ytick={0, 10, 20, 30},
    yticklabels={, , ,},
    grid=both,
    major grid style={gray!30},
    axis line style={gray!50},
    legend pos=outer north east,
    legend style={
        font=\footnotesize,
        at={(1.8,0.5)},
        anchor=west,
    },
    legend cell align={left},
    font=\footnotesize
]

\addplot[color=blue, fill=blue, fill opacity=0.1, mark=square*,mark options={fill=blue},] coordinates {
    (0,30) (72,25) (144,10) (216,25) (288,30)
} -- cycle;
\addlegendentry{Mode-I}

\addplot[color=red, fill=red, fill opacity=0.1, mark=*,mark options={fill=red},] coordinates {
    (0,30) (72,30) (144,30) (216,10) (288,10)
} -- cycle;
\addlegendentry{Mode-II}

\addplot[color=green!60!black, fill=green, fill opacity=0.1, mark=triangle*, mark size=2.5pt, mark options={fill=green!60!black}] coordinates {
    (0,30) (72,30) (144,20) (216,30) (288,30)
} -- cycle;
\addlegendentry{Mode-III}

\end{polaraxis}
\end{tikzpicture}

    \caption{Comparison of Modes (outer is better)}
    \label{fig:radar-comparison}
\end{figure}
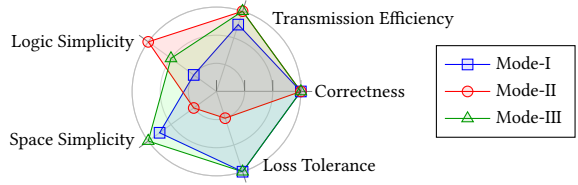

\cref{fig:radar-comparison} compares and shows the advantages of the three modes.



\subsection{Correctness}
\label{subsec:correctness}




\parab{Analysis.} 
The correctness of the \sysname protocol is established through two primary pillars: computational equivalence to single-node processing and guaranteed protocol termination.

In Mode-I, correctness is inherited from the underlying RoCE transport layer, which ensures that IncEngine only processes ordered, unique, and valid packets. This reliability allows the engine to index data elements in the same order as PSN, providing accurate aggregation.

For Mode-II, end-host retransmission ensures data completeness, while the {\tt CheckDuplicate} module prevents redundant computation. Addressing conflicts are resolved by the ``aggregate-then-forward'' mechanism: it constrains the skew between the fastest and slowest ranks, and ensures the PSN range of inflight packets never exceeds twice the window size ($2W$). By sizing switch buffers to $2W$, we prevent packet collisions and incorrect buffer mapping.

Mode-III similarly utilizes hop-by-hop retransmissions and redundancy detection. Here, each computation buffer maintains a strict permissible PSN range exactly matching its capacity ($N$). This prevents addressing conflicts where packets with a PSN difference greater than $N$ might otherwise alias to the same memory location. 

Across all modes, a robust timeout-retransmission failsafe guarantees eventual protocol termination under any network condition.


\parab{Model Checking.}
The correctness of \sysname is verified via formal model checking~\cite{bai2026verifying}, where each network node is modeled as a Non-deterministic Finite Automaton (NFA). State transitions are triggered by packet processing, and non-determinism arises from the concurrent nature of multi-port reception and transmission.

The global protocol state is represented by the composition of these individual NFAs into a unified, complex system. To verify the protocol, the model checker exhaustively simulates all execution branches from every state, accounting for network uncertainties such as packet loss and out-of-order delivery. For each transmitted packet, the checker explores every possible outcome—success, loss, or reordering (with existing packets)—creating separate execution paths for each scenario.

The protocol is deemed correct only if every explored branch satisfies two invariant properties: computational accuracy (the final result matches server-side reduction) and liveness (the protocol eventually reaches a terminal state). This rigorous exploration of the state space ensures that the polymorphic data plane remains robust under any possible network condition.

The \sysname model checker provides a specification language to describe an INC system, including topology and node logic, and a compiler to compile a user program to a TLA+ program; it runs TLA+ model checker to verify the system correctness.

The \sysname model checker overcomes two challenges. First, it enables describing network nondeterminisms including reliable, unreliable, and out-of-order packet delivery. Second, it handles state explosion by input space reduction, nondeterminism constraints, and symmetric state elimination (Appendix~\cref{sec:app:model-checking}).

\parab{Fixing a Pitfall.} 
Model checking revealed a critical pitfall when evolving from Mode-II to Mode-III regarding buffer management. In Mode-II, buffer slots out of the current window are recycled when an aggregation event occurs inside the current window, as the ``aggregate-then-forward'' mechanism naturally synchronizes window advancement across all ranks ({\tt RecycleBuffer} in \cref{subsec:mode-ii}).

However, direct application of this logic to Mode-III is incorrect (\cref{fig:pitfall}). In Mode-III, window progression is independently governed by ACKs, meaning the completion of an aggregation at a specific $PSN$ does not guarantee that the $PSN + MW$ position is outside the active windows of ``all'' participants. Clearing the buffer at $PSN + MW$ prematurely risks erasing data from faster ranks, resulting in computational corruption. To resolve this, we introduced the pipe abstraction ({\tt pipe} in \cref{subsec:mode-iii}), which mandates a unified writable range to forcibly synchronize sender progress.

In addition, the \sysname model checker also finds design risks in ATP, SwitchML, and NetReduce; we fix them in the technical report~\cite{bai2026verifying}. We put other insights in Appendix~\cref{sec:app:model-checking}.




%

\subsection{Transmission Efficiency}
\label{subsec:app:transmission-efficiency}

\begin{figure}[t]
    \centering
    \includegraphics[width=\linewidth]{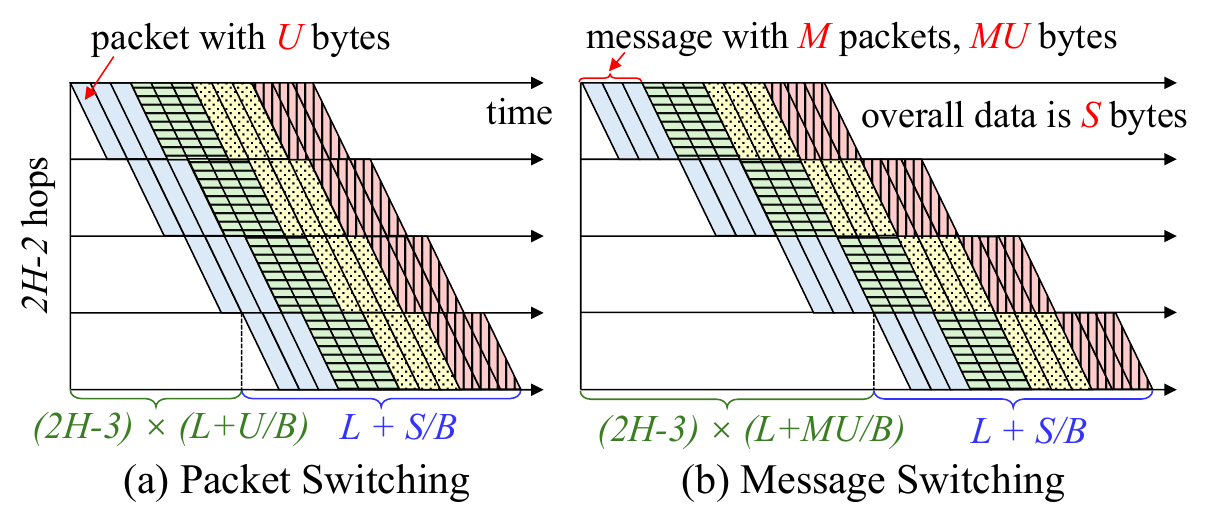}
    \caption{Transmission Efficiency}
    \label{fig:transmission-efficiency}
\end{figure}

Switches operate in a Store-and-Forward model but at different granularities across modes.
Following the definitions in \cref{tab:symbols}, Mode-I switches operate with a message granularity ($M$ MTUs), requiring the reception of a full message before processing. In contrast, Mode-II/III operate with a packet (MTU) granularity (\cref{fig:transmission-efficiency}).

Assuming an AllReduce operation on a depth-$H$ IncTree, the path connecting two leaf nodes contains $2H-2$ edges.
The total time to transmit application data of size $S$ is composed of propagation delay and transmission delay.
In Mode-I, the total time includes propagation delay over $2H-2$ edges, store-and-forward delay over $2H-3$ intermediate switches with message granularity $M \times U$, and the final transmission time:
\begin{equation}
    T_{Mode-I} = (2H-2)L + (2H-3){MU}/{B} + {S}/{B}
\end{equation}
In Mode-II/III, due to packet-level pipelining, the delay is dominated by the single packet serialization time:
\begin{equation}
    T_{Mode-II/III} = (2H-2)L + (2H-3){U}/{B} + {S}/{B}
\end{equation}

Mode-II/III reduces the transmission time relative to Mode-I by $(2H-3)(M-1)U/B$.
In practice, if $S\gg MU$ (very large messages) or $L \gg S/B$ (very small messages), this advantage is not significant.

\subsection{Logic Complexity}
\label{subsec:app:logic-complexity}

Comparing the number of modules and lines of code in implementation, the order of modes from simple to complex is Mode-II, Mode-III, and Mode-I.

\subsection{Space Complexity}
\label{subsec:app:space-complexity}

Mode-I uses hop-by-hop flow control, requiring buffers for one-hop bandwidth-delay product (BDP) ($2BL$). For aggregation with degree $D$, it needs $D-1$ receive buffers, 1 aggregation buffer, and 1 broadcast buffer. Total $\approx (D+1) \times 2BL$.
    
Mode-II relies on end-to-end flow control, requiring buffers to cover the path BDP ($2 \times (2H-2)BL$). Total usage is $4(H-1)BL$. Supporting reproducibility needs a separate buffer for data packets, increasing the space to $4(H-1)(D+1)BL$.

Mode-III introduces hop-by-hop reliability, buffering for single-hop BDP ($2BL$). Total usage is $4BL$ for aggregate and broadcast. With reproducibility, it becomes $(D+1) \times 2BL$.

The order of space overhead from small to large is Mode-III = Mode-I < Mode-II. Mode-I forces reproducibility, but Mode-III provides non-reproducibility options which achieve a smaller space.

\subsection{Loss Tolerance}
\label{subsec:app:loss-tolerance}

Let $1-r$ denote the packet loss rate, i.e., $r$ is the packet success rate. Endpoint throughput achieves bandwidth saturation with $r=1$, and the throughput decreases by a factor of $F(r)$ with $r<1$. We consider the AllReduce on a depth-$H$ IncTree, there are $2H-2$ hops.

We first model the impact of packet loss on throughput in AllReduce. Let $r$ be the transmission success rate ($1-r$ is the loss rate) and $F(r)$ be the throughput decay factor. With an aggregation of depth $H$, AllReduce data packets traverse $2H-2$ hops; hop $i$ to $i+1$ is upward when $1\le i\le H-1$, and downward when $H\le i \le 2H-2$.

In Mode-I/III, in aggregation and broadcast, transmissions between sibling nodes are independent due to independent loss retransmission. The parent's throughput is synchronized by the slowest child. Let $T_i$ be the receive throughput at hop $i$. Thus,

\[
T_i =
\begin{cases}
B,      & \text{if } i = 1 \\
T_{i-1}\cdot \min_{j\in i.\text{children}} F(r_j),            & \text{if } 2\le i < H \\
T_{i-1}\cdot \min_{j\in i.\text{siblings}} F(r_j),   & \text{if } i \ge H.
\end{cases}
\]

In Mode-II, during aggregation, the success of aggregating all packets with the same PSN is forced to synchronize with the slowest one; in broadcast, the ranks affected by packet loss would spare extra bandwidth to retransmit, affecting the sending of new packets and also slowing down other lossless ranks (due to synchronization). We model the overall throughput loss by a multiplicative model: 

\[
T_i =
\begin{cases}
B,      & \text{if } i = 1 \\
T_{i-1}\cdot \prod_{j\in i.\text{children}} F(r_j),            & \text{if } 2\le i < H \\
T_{i-1}\cdot \prod_{j\in i.\text{siblings}} F(r_j),   & \text{if } i \ge H.
\end{cases}
\]

We simplify the model by assuming $r_j = r$ and $F(r)=r$ for further analysis\footnote{In practical systems, $F(r)$ could be worse than linear because packet loss triggers congestion control, which decreases the sending rate or congestion window.}. Mode-I \& Mode-III achieve: $T \approx B \cdot r^{2(H-1)}$; Mode-II achieves: $T \approx B \cdot (r^{D-1})^{2(H-1)} = B \cdot r^{2(H-1)(D-1)}$.

Mode-I and Mode-III outperform Mode-II in loss tolerance. But if lossless layer-2 mechanisms (PFC/CBFC) are applied in practice where $r \approx 1$, the three modes perform similarly.

\section{INC Resource Management}
\label{sec:resource-management}

\subsection{Resource Model}
\label{subsec:resource-model}

\begin{figure}[tb]
    \centering
    \includegraphics[width=0.9\linewidth]{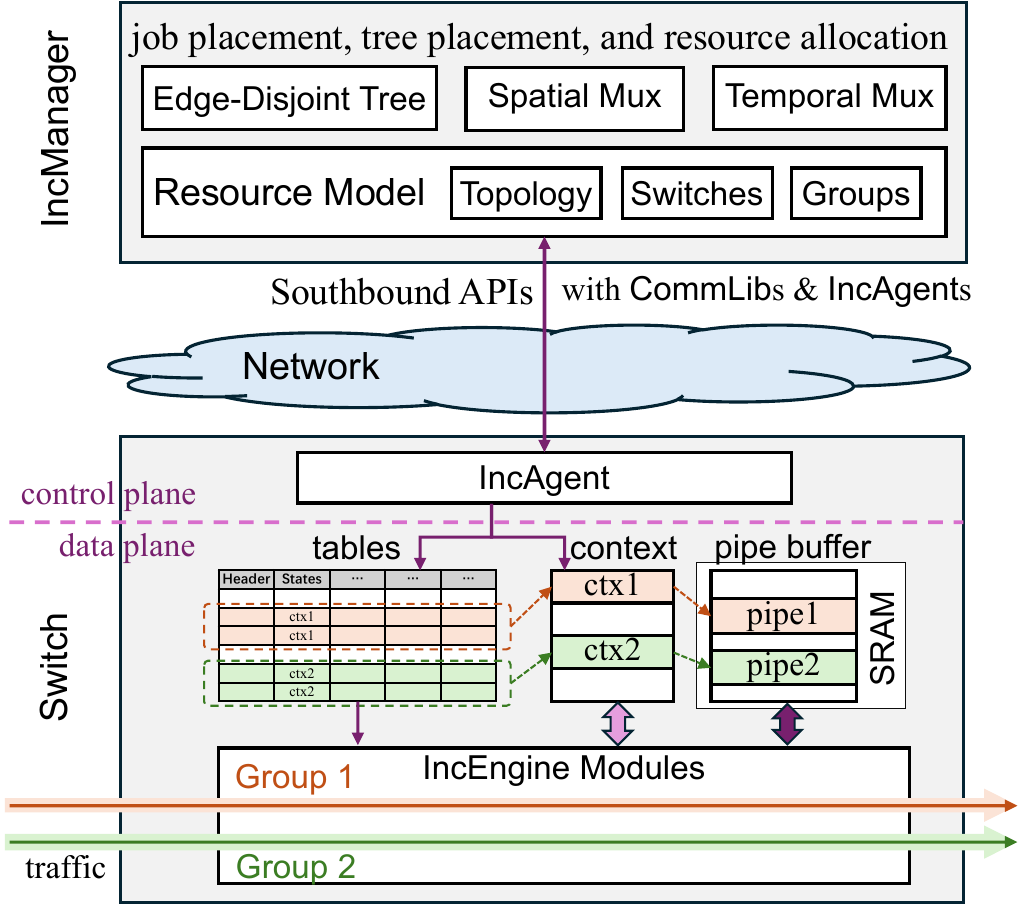}
    \caption{Resource Model and Policies}
    \label{fig:resource-model}
\end{figure}

\parab{Architecture.}
The \sysname control plane enables intelligent resource management through a centralized IncManager and distributed IncAgents, as shown in \cref{fig:resource-model}. During initialization, the IncManager constructs a global view by aggregating reported device capabilities, port statuses, and available on-chip SRAM from each IncAgent, while simultaneously performing global topology discovery via link-state reporting.

At runtime, the IncManager serves as the decision-making hub, instantiating communication groups by issuing instructions to IncAgents. These instructions configure the switch-local context, including lookup tables and computation states such as aggregation buffer offsets. By decoupling policy formulation from hardware execution, the framework supports dynamic resource reallocation and real-time state updates based on network conditions. This architecture ensures fine-grained, centralized control over distributed INC resources, maximizing efficiency across the cluster.

\parab{Indirection Layer for Resource Allocation.}
In the \sysname architecture, switches manage group-specific rules in match-action tables and associated contexts in high-speed SRAM. Direct context reallocation via table modification is computationally expensive due to the numerous rules. Analysis (\cref{subsec:app:space-complexity}) reveals that most SRAM is consumed by transient states—such as payload and degree buffers ($O(\text{BDP})$)—which remain idle between collective invocations, while persistent endpoint states ($O(D)$) remain minimal.

To optimize utilization, \sysname introduces an indirection layer that decouples these large transient states from the fixed context. By utilizing pointers within the context to reference dynamic SRAM regions, the IncManager can calculate and assign memory offsets in real time. This pointer-based approach facilitates agile, fine-grained resource reallocation and significantly enhances SRAM utilization without the overhead of reconfiguring global forwarding tables.

\subsection{Supporting Various Policies}
\label{subsec:policies}

Cluster-wide INC resource management essentially needs to place an IncTree on the topology and decide the resource allocation on each of its switch nodes. The \sysname resource model can support various existing and new policies, where we list three classes below.

\parab{Edge-Disjoint Tree.} Legacy INC solutions often utilize an Edge-Disjoint Tree (EDT) approach~\cite{rottenstreich2024edge, dong2025mina} necessitated by fixed-function hardware. These systems reuse static ingress buffers that cannot be virtualized, meaning a physical port can typically serve only one communication group. This creates a strict topological constraint: multiple aggregation trees must not share edges to avoid resource contention.

The \sysname resource model maintains compatibility with EDT by decoupling IncTree placement from resource allocation. \sysname first determines an EDT for a group and then independently allocates SRAM resources to bind the tree. We provide an algorithm to calculate candidate EDTs within Clos networks. The process removes edges occupied by active EDTs, then iteratively scans switches from lower to upper tiers. It identifies the lowest level containing switches reachable by all group members without traversing higher levels. The algorithm returns these switches as potential IncTree roots or indicates if no such candidate exists.

\parab{Spatial Multiplexing.} Legacy solutions like SwitchML~\cite{sapio2021scaling} and NetReduce~\cite{liu2023network} achieve isolation by statically partitioning switch SRAM, typically for single-tier scenarios. \sysname extends this to cluster-wide, multi-layer topologies, enabling non-interfering concurrent groups and robust multi-tenancy.

To manage this, \sysname jointly optimizes IncTree placement and resource allocation by evaluating both logical connectivity and ``path width''—a metric combining available bandwidth and SRAM capacity. Notably, the system may prioritize longer paths if their larger aggregation buffers support higher overall throughput.
For Clos networks, we employ a greedy iterative algorithm to identify the Pareto frontier of candidate IncTrees. By scanning from lower to upper levels, the algorithm identifies potential roots that balance tree depth against resource capacity. This provides administrators with a selection of optimal solutions, ranging from low-latency trees for synchronization to high-bandwidth trees for data-intensive training.

\parab{Temporal Multiplexing.} To maximize efficiency, \sysname supports temporal resource sharing like ATP~\cite{lao2021atp,li2023a2tp,li2023pa}. It models switch capacity as a combination of unallocated space and oversubscribed blocks weighted by the duty cycle. This allows the IncManager to perform joint placement and allocation even in congested environments, effectively multiplexing physical SRAM across multiple communication groups.

Mutual exclusivity can be maintained primarily through scheduled coordination, which invokes collectives in a non-overlapping manner. This is particularly effective for 3D parallel training, where Tensor Parallel (TP) and Data Parallel (DP) groups communicate in interleaved phases, enabling them to share physical hardware without contention.

In decentralized or multi-tenant scenarios where scheduling is infeasible, \sysname employs a contention-and-fallback mechanism. Groups attempt to ``lock'' resources via a first-come-first-served (FCFS) policy at switch-resident recorders. If a group fails to secure the complete IncTree, it performs an all-or-nothing release and falls back to host-based communication (e.g., NCCL), preventing partial occupancy and ensuring system-wide progress.

\section{Evaluation}
\label{sec:evaluation}

\subsection{Experiment Settings}
\label{subsec:settings}

\parab{Implementation.} The \sysname standard specification is provided to the organizations in the working group, and the group contributes several different implementations, each of which can validate the specific attributes of \sysname as described in \cref{tab:properties}. The implementations are open-sourced~\footnote{\url{https://github.com/In-Net/EPIC}}. Model checking is built in TLA+~\cite{tlaplus}, testbed prototype on Tofino and NP switches~\cite{tofino, sun2026rearchitecting}, emulation on OVS~\cite{ovs} connecting VMs, flow simulation in OMNeT++~\cite{omnetpp}, and packet simulation in NS3~\cite{ns3,sridharan2023chakra,wang2025simai}. 

\parab{Environments.} \textbf{(1)} Our testbed has four servers and one switch in a star topology. Each server has two NVIDIA GeForce RTX 3090 GPUs, two Xeon(R) Silver 4316 CPUs with
40 physical cores, \SI{192}{GB} memory, \SI{1}{TB} SSD, and two \SI{100}{GbE} CX-5 NICs (each NIC bonded with a GPU); the switch has \SI{100}{Gbps} ports connecting NICs. 
Tofino and NP switches have different programmabilities; we implement Mode-II on Tofino and Mode-II/III on NP.
We integrate \sysname with PyTorch, where the CommLib needs to pipeline the memory copy between the GPU and the RDMA-registered buffer and \sysname's packet I/O.
\textbf{(2)} Emulation runs on VMs connected by OVS~\cite{ovs}, where host VMs run SoftRoCE, and Mode-I and Mode-II/III switch VMs have IncEngine built atop {\tt libibverbs} and {\tt libpcap}. 
\textbf{(3)} Simulations are based on NS3 or OMNeT++, which run on a PC.



\parab{Topology.} We use the following notation to describe the topology ``Tree-$x$-$y$'', where $x$ means tree depth and $y$ means tree branches at non-leaf nodes. For example, a star topology with four servers is denoted as Tree-2-4 (servers counted as one tier and the switch as another).

\parab{Workloads.} We run three kinds of workloads: (1) Collective communication of the six primitives to study communication acceleration, (2) a single training job to study \sysname's acceleration, and (3) multi-tenant training to study resource management. In multi-tenant training, we use job traces from production clusters~\cite{cao2024crux}.

\parab{Baselines.} In testbed experiments, we compare \sysname with NCCL (Ring and Tree); for AllReduce specifically, we compare \sysname with ATP and SwitchML\footnote{We compare with the SwitchML DPDK version. SwitchML provides an RDMA UC version, where the RDMA is configured without reliability guarantee, and the host still needs to spare CPU to handle transmission failures.}. In packet/flow-level simulation, we compare \sysname with a ring-based algorithm. In emulation, we compare \sysname with MPI (which uses CPU memory instead of GPU; more suitable for VMs). In the experiment description, we use \sysname-I/II/III to denote \sysname's three modes.

\parab{Metrics.} We measure the following metrics to compare solutions: (1) the collective's algorithm throughput, which is the application data size of the ranks divided by the overall collective completion time, and (2) job completion time (JCT).

\subsection{Feasibility}
\label{subsec:feasibility}


\parab{Correctness.} We run model checking with the Tree-3-2 topology, and verify Mode-II/III's AllReduce, Reduce, and Broadcast. The network environment is set to reliable, lossy, and out-of-order. In all environments, \sysname is correct: the computation result is equivalent to a single server's result, and the protocol eventually terminates.



\begin{table}[t]
\caption{[Emulation, Tree-3-2] AllReduce Algorithm Throughput (Mbps)}
\label{tab:emulation:allreduce}
\centering
\footnotesize
\setlength{\tabcolsep}{1pt}
\begin{tabular}{@{}c *{9}{|c} @{}}
\hline
\textbf{Msg. Size (B)} & \textbf{4K} & \textbf{16K} & \textbf{64K} & \textbf{256K} & \textbf{1M} & \textbf{4M} & \textbf{16M} & \textbf{64M} & \textbf{256M} \\
\hline
 \sysname-I   & 111 & 316 & 596 & 744 & 786 & 876 & 915 & 1015 & 1176 \\
 \sysname-II  & 109 & 201 & 361 & 355 & 447 & 484 & 483 & 529  & 558  \\
 \sysname-III & 158 & 256 & 367 & 420 & 463 & 494 & 512 & 615  & 660 \\
 MPI          & 74.9  & 185 & 330 & 346 & 496 & 519 & 514 & 508  & 517 \\
\hline
\end{tabular}
\end{table}

\parab{Interoperability.} \cref{tab:emulation:allreduce} shows AllReduce throughput in emulation. All three modes in \sysname interoperate with host-side SoftRoCE correctly. They all provide correct results and finish normally. 
\sysname shows reasonably high throughput (details in Appendix \cref{sec:app:emulation}); as all emulated components share the server CPU cores, the throughput does not represent the actual acceleration in a hardware environment (which is not the purpose of emulation).


\parab{Evolvability.} The development of \sysname-II and III follows the modular design. \sysname-II and III consist of 2,803 and 2,741 lines of code, respectively, and reuse 1,557 lines of code. Switch vendors can iterate the INC feature from \sysname-II to \sysname-III, with correctness guarantee while continuing to improve performance (Appendix \cref{sec:app:emulation}).

\parab{Resource Affordability.}
Across various hardware, \sysname requires only 1–\SI{16}{MB} of on-chip SRAM, with platforms such as Intel Tofino, Agilex 7 FPGA, and Xilinx VU13P supporting \SI{8}{MB}, \SI{16}{MB}, and \SI{1}{MB}, respectively (Appendix \cref{sec:app:fpga}, \cref{sec:app:testbed}). For Mode-II, the most memory-intensive realization, a \SI{100}{Gbps} network with \SI{10}{\us} RTT necessitates \SI{250}{KB} per job (twice the path BDP). Simulations confirm that even a \SI{1}{MB} allocation—supporting four concurrent groups—significantly boosts performance in 3D training and multi-tenant environments, thereby reducing JCT (\cref{subsec:eval-policies}). Thus, the resource cost of \sysname is affordable in production scenarios.

\parab{Chip Area.} Evaluated by a chip vendor, when synthesized using a \SI{28}{nm} process technology, \sysname's IncEngine instance, configured with 512 FP32 ALUs, 512 UINT ALUs, and a \SI{1}{MB} payload buffer, occupies an area of \SI{4.89}{\milli\meter\squared}. The total area overhead for the eight integrated engines (\SI{25.6}{Tbps}) amounts to \SI{39.12}{\milli\meter\squared}, demonstrating an affordable footprint for high-bandwidth INC.

\subsection{Performance Acceleration from INC}
\label{subsec:performance-acceleration}

\begin{figure}[t]
    \centering
    \small
    \begin{tikzpicture}[scale=0.85]
        \begin{axis}[
            name=axis0,
            ymode=log,
            xlabel={Message size (B)}, 
            ylabel={Tput (Gbps)}, 
            xtick={1,2,3,4,5,6,7,8,9,10},
            xticklabels={4K,16K,64K,256K,1M,4M,16M,64M,256M,1G}, 
            xmax=5,
            xtick=data, 
            xtick align=inside, 
            width=0.6\linewidth,  
            height=0.45\linewidth,
            grid=major,                
            grid style={dashed, gray!30},
            legend style={
                font=\footnotesize,
                at={(1.02,1.05)},   
                anchor=south,
                legend columns=4,
            },
        ]
            
        \addplot[color=black, solid, mark=o, thick] coordinates { 
        (1,5.42) (2,16.84) (3,41.72) (4,59.46) (5,60.79)
        (6,62.65) (7,64.35) (8,61.7) (9,66.64) (10,65.29)
        };

        \addplot[color=blue, solid, mark=square, thick] coordinates { 
        (1,3.85) (2,13.10) (3,32.73) (4,52.34) (5,61.56) (6,64.39) (7,65.14) (8,65.37) (9,65.37) (10,65.38)
        };

        \addplot[color=black, dotted, mark=*, thick] coordinates { 
        (1,0.015) (2,0.058) (3,0.243) (4,0.91) (5,3.47) (6,13.34) (7,29.06) (8,40.99) (9,46.33) (10,47.97)
        };

        \addplot[color=blue, dotted, mark=square*, thick] coordinates  {  
        (1,0.015) (2,0.06) (3,0.251) (4,0.82) (5,3.71) (6,12.10) (7,26.55) (8,36.01) (9,39.73) (10,41.07)
        };

        \addplot[color=red, solid, mark=triangle, thick] coordinates  { 
        (1,1.98) (2,7.22) (3,21.32) (4,41.63) (5,54.64)
        (6,59.28) (7,60.56) (8,60.89) (9,60.97) (10,60.99)
        };

        \addplot[color=cyan, solid, mark=diamond, thick] coordinates  { 
        (1,1.92) (2,6.47) (3,15.93) (4,25.07) (5,29.27) (6,30.54) (7,30.88) (8,30.96) (9,30.96) (10,30.94)
        };
        
        \addplot[color=purple, dashdotted, mark=triangle*, thick] coordinates { 
        (1,0.015) (2,0.06) (3,0.244) (4,0.89) (5,3.17) (6,11.05) (7,24.85) (8,31.89) (9,33.51) (10,34.10)
        };

        \addplot[color=orange, dashdotted, mark=pentagon*, thick] coordinates  {  
        (1,0.015) (2,0.06) (3,0.223) (4,0.86) (5,3.12) (6,10.12) (7,21.39) (8,27.37) (9,29.01) (10,29.29)
        };

        \legend{\sysname-II, SwitchML,  NCCL-Ring-1KB, NCCL-Ring-256B, ATP 7-to-1, ATP 8-to-1, NCCL-Tree-1KB, NCCL-Tree-256B} 
        \end{axis}        
        \begin{axis}[
            name=axis1,
            xshift=0.5\linewidth,
            xlabel={Message size (B)}, 
            ylabel={}, 
            xtick={1,2,3,4,5,6,7,8,9,10},
            xticklabels={4K,16K,64K,256K,1M,4M,16M,64M,256M,1G}, 
            xmin=5,
            ymin=0,
            xtick=data, 
            xtick align=inside, 
            width=0.68\linewidth,  
            height=0.45\linewidth,
            grid=major,                
            grid style={dashed, gray!30},
            legend style={
                font=\fontsize{6pt}{7.2pt}\selectfont,
                at={(0.45,1.05)},   
                anchor=south,
                legend columns=4,
            },
        ]
            
        \addplot[color=black, solid, mark=o, thick] coordinates { 
        (1,5.42) (2,16.84) (3,41.72) (4,59.46) (5,60.79)
        (6,62.65) (7,64.35) (8,61.7) (9,66.64) (10,65.29)
        };

        \addplot[color=blue, solid, mark=square, thick] coordinates { 
        (1,3.85) (2,13.10) (3,32.73) (4,52.34) (5,61.56) (6,64.39) (7,65.14) (8,65.37) (9,65.37) (10,65.38)
        };

        \addplot[color=black, dotted, mark=*, thick] coordinates { 
        (1,0.015) (2,0.058) (3,0.243) (4,0.91) (5,3.47) (6,13.34) (7,29.06) (8,40.99) (9,46.33) (10,47.97)
        };

        \addplot[color=blue, dotted, mark=square*, thick]  coordinates  {  
        (1,0.015) (2,0.06) (3,0.251) (4,0.82) (5,3.71) (6,12.10) (7,26.55) (8,36.01) (9,39.73) (10,41.07)
        };

        \addplot[color=red, solid, mark=triangle, thick] coordinates  { 
        (1,1.98) (2,7.22) (3,21.32) (4,41.63) (5,54.64)
        (6,59.28) (7,60.56) (8,60.89) (9,60.97) (10,60.99)
        };

        \addplot[color=cyan, solid, mark=diamond, thick] coordinates  { 
        (1,1.92) (2,6.47) (3,15.93) (4,25.07) (5,29.27) (6,30.54) (7,30.88) (8,30.96) (9,30.96) (10,30.94)
        };
        
        \addplot[color=purple, dashdotted, mark=triangle*, thick]  coordinates { 
        (1,0.015) (2,0.06) (3,0.244) (4,0.89) (5,3.17) (6,11.05) (7,24.85) (8,31.89) (9,33.51) (10,34.10)
        };

        \addplot[color=orange, dashdotted, mark=pentagon*, thick] coordinates  {  
        (1,0.015) (2,0.06) (3,0.223) (4,0.86) (5,3.12) (6,10.12) (7,21.39) (8,27.37) (9,29.01) (10,29.29)
        };

        \end{axis}
        
    \end{tikzpicture}
    \caption{[Testbed, Tree-2-8] AllReduce Algorithm Throughput}
    \label{fig:testbed-allreduce}
\end{figure}
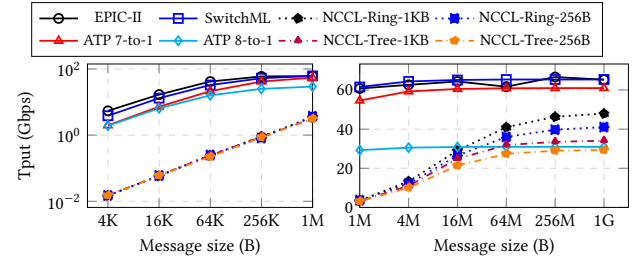

\parab{Collective Communication.} \cref{fig:testbed-allreduce} shows AllReduce performance on the testbed. All INC solutions consistently outperform the baseline NCCL. For small messages, INC provides a significantly lower end-to-end latency due to reduced hop counts and a thinner software stack. For large messages, the performance gain stems from the substantial reduction in network traffic volume. 

When the link capacity is saturated for all solutions, INC solutions---\sysname, ATP, and SwitchML---outperform NCCL. But \sysname spends less CPU resources to saturate links due to its RoCE-compatibility and hardware transmission offloading.
SwitchML achieves high throughput using a DPDK-based end-host implementation; however, because DPDK lacks transport-layer reliability, SwitchML must handle packetization and retransmissions, leading to extra CPU overhead compared to \sysname (6 cores vs. 1 core). While ATP performs well with a dedicated Parameter Server (PS) on additional links, its small-message latency remains high, and throughput drops when the PS competes for bandwidth with workers (ATP 8-to-1).

\begin{figure*}[t]
    \small
\begin{minipage}[t]{0.34\textwidth}
    \centering
    \begin{tikzpicture}
        \begin{axis}[
            name=axis0,
            ybar=0pt,
            ymode=log,
            xlabel={4KB Message}, 
            ylabel={Throughput (Gbps)}, 
            xtick={1,2,3,4},
            xticklabels={Reduce, RS, Bcast, AG}, 
            bar width=0.4, 
            xmin=0.4, xmax=4.6,
            ymin=-2,
            xtick align=inside, 
            width=0.575\linewidth,  
            height=0.5\linewidth,
            grid=major,                
            grid style={dashed, gray!30},
            legend style={
                at={(1,1.05)},   
                anchor=south,
                legend columns=4,
                column sep=5pt,
            },
            xticklabel style={rotate=30, anchor=north east},
            ylabel style={yshift=-3pt},
        ]
        \addplot[fill=blue!30, postaction={pattern=grid}] coordinates { 
        (1,4.19) (2,3.97) (3,4.45) (4,4.37) 
        };

        \addplot[fill=red!30, postaction={pattern=crosshatch}] coordinates { 
        (1,0.016) (2,0.014) (3,0.015) (4,0.015) 
        };
        \legend{\sysname-II, NCCL-Ring} 
        \end{axis}        
        \begin{axis}[
            name=axis1,
            xshift=0.4\linewidth,
            ybar=0pt,
            xlabel={1GB Message}, 
            ylabel={}, 
            xtick={1,2,3,4},
            xticklabels={Reduce, RS, Bcast, AG}, 
            bar width=0.4, 
            xmin=0.4, xmax=4.6,
            ymin=0,
            xtick align=inside, 
            width=0.575\linewidth,  
            height=0.5\linewidth,
            grid=major,                
            grid style={dashed, gray!30},
            legend style={
                font=\fontsize{6pt}{7.2pt}\selectfont,
                at={(0.45,1.05)},   
                anchor=south,
                legend columns=4,
            },
            xticklabel style={rotate=30, anchor=north east},
        ]
            
        \addplot[fill=blue!30, postaction={pattern=grid}] coordinates { 
        (1,66.68) (2,66.3) (3,88.96) (4,90.55) 
        };

        \addplot[fill=red!30, postaction={pattern=crosshatch}] coordinates { 
        (1,67.26) (2,95.58) (3,66.60) (4,95.58) 
        };
        
        \end{axis}
    \end{tikzpicture} 
    \caption{[Testbed, Tree-2-8] Collective \\ Algorithm Throughput}
    \label{fig:testbed-collective}
\end{minipage}
\hfill
\begin{minipage}[t]{0.3\textwidth}
    \centering
	\begin{tikzpicture}
\begin{axis}[
    ybar=0pt,
    width=1\linewidth,
    height=0.63\linewidth,
    enlarge x limits=0.1,
    ylabel={Throughput (Gbps)},
    xtick={6,7,8,9,10},
    xticklabels={16MB,32MB,64MB,128MB,256MB},
    xmin=5.8, xmax=10.2,
    ymin=0,    ymax=150,
    legend style={at={(0.45,1.05)}, anchor=south, legend columns=3, font=\scriptsize},
    ymajorgrids=true,
    grid style=dashed,
    bar width=0.12, 
    xtick align=inside, 
    xticklabel style={rotate=30, anchor=north east},
]

\addplot [fill=blue!10, postaction={pattern=horizontal lines}] coordinates  {
(6,89.267) (7,76.006) (8,79.986) (9,80.624) (10,81.953)
};
\addlegendentry{NCCL-RS}

\addplot [fill=blue!20, postaction={pattern=vertical lines}]coordinates {
(6,90.457) (7,76.144) (8,78.639) (9,79.6) (10,80.098)
};
\addlegendentry{NCCL-AG}

\addplot [fill=blue!30, postaction={pattern=grid}]coordinates {
(6,89.862) (7,76.075) (8,79.312) (9,80.112) (10,81.025)
};
\addlegendentry{NCCL-RS+AG}

\addplot [fill=red!10, postaction={pattern=north east lines}]coordinates {
(6,55.881) (7,60.415) (8,62.49) (9,63.895) (10,64.382)
};
\addlegendentry{\sysname-RS}

\addplot [fill=red!20, postaction={pattern=north west lines}] coordinates {
(6,58.714) (7,60.002) (8,64.773) (9,65.567) (10,66.072)
};
\addlegendentry{\sysname-AG}



\addplot [fill=red!30, postaction={pattern=crosshatch}] coordinates {
(6,112.531) (7,118.249) (8,124.971) (9,127.131) (10,128.105)
};
\addlegendentry{\sysname-RS+AG}

\end{axis}
\end{tikzpicture}
	\caption{[Testbed, Tree-2-8] \\ RS \& AG Complements}
	\label{fig:app:testbed:RS+AG}
\end{minipage}
\hfill
\begin{minipage}[t]{0.34\textwidth}
    \centering
    \begin{tikzpicture}
        \begin{axis}[
            name=axis0,
            xlabel={Loss Rate (\%) @ 1 link}, 
            xlabel shift=-5pt,
            ylabel={Throughput (Gbps)}, 
            xtick={1,2,3,4,5,6,7,8},
            xticklabels={0,0.001,0.003,0.01,0.03,0.1,0.3,1}, 
            ymin=40,
            xtick=data, 
            xtick align=inside, 
            xticklabel style={rotate=60, anchor=north east},
            width=0.65\linewidth,  
            height=0.53\linewidth,
            grid=major,                
            grid style={dashed, gray!30},
            legend style={
                at={(0.95,1.05)},   
                anchor=south,
                legend columns=4,
                column sep=5pt,
            },
            ylabel style={yshift=-3pt},
        ]
            
        \addplot[color=blue, solid, mark=*, thick] coordinates { 
        (1, 88.25) (2, 88.25)(3, 88.19) (4, 88.05)(5, 87.55)(6, 86.66)(7, 68.08)(8, 52.66)

        };

        
        \addplot[color=red, solid, mark=triangle*, thick] coordinates  { 
        (1, 89.88)(2, 89.88)(3, 89.85) (4, 89.74)(5, 89.40)(6, 88.29)(7, 87.92)(8, 87.61)
        };



        


        \legend{\sysname-II,  \sysname-III} 
        \end{axis}        
        \begin{axis}[
            name=axis1,
            xshift=0.48\linewidth,
            xlabel={Links with 0.1\% loss}, 
            ylabel={}, 
            xtick={1,2,3,4,5,6,7,8,9},
            xticklabels={0, 1, 2, 3, 4, 5, 6, 7, 8}, 
            ymin=40,
            xtick=data, 
            xtick align=inside, 
            width=0.6\linewidth,  
            height=0.53\linewidth,
            grid=major,                
            grid style={dashed, gray!30},
            legend style={
                at={(0.45,1.05)},   
                anchor=south,
                legend columns=4,
            },
        ]
            
        \addplot[color=blue, solid, mark=*, thick] coordinates { 
       
            (0,88.25) (1,86.66) (2,82.31) (3,68.08) (4,65.54)
            (5,62.93) (6,60.56) (7,58.39) (8,56.38)
        };

        
        \addplot[color=red, solid, mark=triangle*, thick] coordinates  { 
        
            (0,89.88) (1,88.29) (2,88.39) (3,87.64) (4,88.00)
            (5,87.40) (6,87.77) (7,87.28) (8,87.63)
        };



        


        \end{axis}
    \end{tikzpicture}
    \caption{[Packet Simulation, Tree-2-8] AllReduce Algorithm Throughput}
    \label{fig:pkt-sim-allreduce}
\end{minipage}
\end{figure*}

\begin{table*}[t]
\centering
\footnotesize 
\setlength{\tabcolsep}{2pt}
    \begin{minipage}[t]{0.38\textwidth}
\centering
    \caption{[Testbed, Tree-2-8] Model Training Iteration Time (s), context\_length=2K, batch\_size=256.}
    \label{tab:testbed-model-training}



\begin{tabular}{cc|c|c|c|c}
\hline
\multicolumn{2}{c|}{\textbf{Model}}                                & \textbf{\sysname-II }  & \textbf{SwitchML} & \textbf{ATP}    & \textbf{NCCL}   \\ 
\hline
\multicolumn{1}{c|}{\multirow{3}{*}{DP=8}} & GPT-3 Large  & 16.7 & 16.7 & 16.8 & 16.9 \\ \cline{2-6} 
\multicolumn{1}{c|}{}                      & OPT-350m & 12.2 & 12.3 & 12.3 & 12.4 \\ \cline{2-6} 
\multicolumn{1}{c|}{}                      & Llama-3.2-1B & 24.5 & 24.6 & 24.7 & 24.8 \\ \hline
\multicolumn{1}{c|}{\multirow{3}{*}{TP=8}} & GPT-3 Large  & 28.3 & 31.6 & 34.5 & 38.1 \\ \cline{2-6} 
\multicolumn{1}{c|}{}                      & OPT-350m & 20.0 & 22.5 & 24.8 & 27.5 \\ \cline{2-6} 
\multicolumn{1}{c|}{}                      & Llama-3.2-1B & 35.1 & 38.8 & 42.1 & 46.1 \\ 
\hline
\end{tabular}

    \end{minipage}
\hfill
    \begin{minipage}[t]{0.26\textwidth}
\centering
    \caption{[Testbed, Tree-2-4] Inference Performance}
    \label{tab:testbed:inference}



\begin{tabular}{c|c|c|c}
\hline
\multicolumn{2}{c|}{\textbf{\makecell{Performance\\(ms)}}} & \textbf{\sysname-II} & \textbf{NCCL} \\ \hline
\multirow{2}{*}{\textbf{GPT-2 Small}} & \textbf{TTFT} & 12.0 & 13.0 \\ \cline{2-4}
                                      & \textbf{TPOT} & 8.81 & 10.1 \\ \hline
\multirow{2}{*}{\textbf{GPT-2 Large}} & \textbf{TTFT} & 26.8 & 37.9 \\ \cline{2-4}
                                      & \textbf{TPOT} & 25.1 & 36.3 \\ \hline
\end{tabular}
    \end{minipage}
\hfill
    \begin{minipage}[t]{0.3\textwidth}
\centering
    \caption{[Flow Simulation] JCT (3 iterations) of GPT-3-175B on 128-GPU Fat-tree}
    \label{tab:flow-sim:GPT3}
\begin{tabular}{c|cc|cc}
\hline
\multirow{2}{*}{\textbf{SRAM (Unit)}} & \multicolumn{2}{c|}{\textbf{w/o scale-up}}    & \multicolumn{2}{c}{\textbf{w/ scale-up}}     \\ \cline{2-5} 
                               & \multicolumn{1}{c|}{\textbf{4}} & \textbf{8} & \multicolumn{1}{c|}{\textbf{4}} & \textbf{8} \\ \hline
\textbf{Ring}                  & \multicolumn{1}{c|}{253}        & 253        & \multicolumn{1}{c|}{106}        & 106        \\ \hline
\textbf{EDT}                   & \multicolumn{1}{c|}{179}        & 179        & \multicolumn{1}{c|}{84.6}       & 84.6       \\ \hline
\textbf{Spatial Mux}           & \multicolumn{1}{c|}{179}        & 158        & \multicolumn{1}{c|}{84.6}       & 63.6       \\ \hline
\textbf{Temporal Mux}          & \multicolumn{1}{c|}{158}        & 137        & \multicolumn{1}{c|}{84.6}       & 63.6       \\ \hline
\end{tabular}
    \end{minipage}
\end{table*}

For collectives requiring payload aggregation (AllReduce, Reduce, and ReduceScatter), Tofino constrains \sysname's packet payload size to \SI{256}{B} due to limited switch aggregation hardware resources. Broadcast and AllGather do not require payload aggregation and thus do not have this limitation, so we configure their packet payload size as \SI{1}{KB}, matching NCCL's default payload size.
Compared with NCCL at \SI{256}{B} payloads, \sysname achieves a $1.59\times$ speedup. NCCL with \SI{1}{KB} payloads improves efficiency, suggesting that future \sysname support for larger payloads in aggregation collectives would yield greater gains. Finally, the NCCL Tree algorithm suffers from the overhead of multiple binomial rounds, proving less effective than the pipelined Ring approach.
When performing Barrier, \sysname takes \textasciitilde\SI{5}{\us} for one operation, while NCCL takes \textasciitilde\SI{2}{\ms}. \sysname outperforms NCCL by \textasciitilde400$\times$.



\cref{fig:testbed-collective} shows results across four primitives. \sysname provides a significant advantage for small messages due to reduced hop counts and minimal software overhead.
For large-scale data transfers, NCCL's Reduce/Broadcast show a similar throughput due to their ring-based algorithm; \sysname outperforms NCCL in Broadcast because it reduces switch traffic and avoids rank-to-rank forwarding, but it is equal to NCCL in Reduce because the \SI{256}{B} aggregation payload lowers goodput and offsets the benefit of INC. In terms of ReduceScatter and AllGather, NCCL performs concurrent rings, so its throughput surges; \sysname keeps RS/AG similar to Reduce/Broadcast because of its sequential execution, but \sysname causes less traffic to the network.

Appendix \cref{ssec:np-performance} shows the performance where the switch is Huawei NetEngine 8000~\cite{sun2026rearchitecting}. \sysname-II and \sysname-III are implemented on this testbed, validating the IncEngine's interoperability with RoCE. The throughput is not high (about \SI{11}{Gbps} for \SI{1}{GB} message for both modes); but if we comment out the payload aggregation instruction in the program, the throughput reaches about \SI{72}{Gbps}. The experiment indicates that INC requires high-speed hardware to perform vector summation, and SRAM access throughput is the key factor influencing the speed.


\parab{Traffic Volume Reduction.}
We evaluate overlapping ReduceScatter (RS) and AllGather (AG) operations across two communication groups in NCCL and \sysname. The result is shown in \cref{fig:app:testbed:RS+AG}. In this experiment, we also configure AllGather's packet payload size as \SI{256}{B} instead of the \SI{1}{KB} setting used above to keep the RS and AG traffic symmetric. In NCCL, aggregate throughput is capped by line rate because nodes must simultaneously transmit and receive for both ring-based collectives, causing bandwidth contention. Conversely, \sysname breaks this limit by exploiting directional link independence. By orchestrating nodes to transmit for RS while receiving for AG (and vice versa), \sysname ensures non-overlapping directional flows. Consequently, \sysname achieves a theoretical throughput of $2\times$ the link bandwidth, as evidenced by the additive throughput observed in \cref{fig:app:testbed:RS+AG}.


\parab{Training Applications.} We train three models (GPT-3 Large, OPT-350m, Llama-3.2-1B) and report their iteration times in \cref{tab:testbed-model-training}. \sysname accelerates DP and TP by 1.2\% and 31.3\% (Llama-3.2). Meanwhile, packet-level simulation based on SimAI~\cite{wang2025simai} and NS3 reports a 12\%--22\% speedup for Llama-7B to GPT-3-175B (Appendix \cref{sec:app:packet-simulation}).

An INC-capable network ideally serves as a transparent accelerator that does not degrade performance.
\sysname's gain matches Amdahl's law, scaling with the DP/TP communication fraction in the overall execution time; the speedup is more pronounced in TP due to its higher communication intensity, whereas DP yields more modest gains from its relatively lower volume.


In several scenarios, \sysname improves overall job efficiency obviously: (1) The TP group is large and spans servers via the scale-out network. (2) In a multi-tenancy network, GPU fragmentation makes some jobs run on the scale-out (\cref{subsec:eval-policies}); (3) The cluster itself is constructed with consumer-grade or mid-range GPUs~\cite{lim2024accelerating,xu2026multipath,liu2025train,athlur2022varuna}, and thus has collective traffic on the scale-out (e.g., Appendix \cref{sec:app:testbed}); (4) It takes tremendous effort to tune a large-scale training program/cluster to fully exploit parallelism and hide communication cost, enabling \sysname makes the system run at a higher communication/overall efficiency as the program/cluster is being tuned, exploiting the GPUs in this fine-tuning cycle.





\parab{Inference Applications.} We run inference on GPT-2 small/large with TP=4 and sequence length=1024 (\cref{tab:testbed:inference}). \sysname reduces GPT-2 large's TTFT/TPOT by 29.3\%/30.9\%, respectively. For inference's small-message patterns, \sysname's reduced hops and latency translate directly into end-to-end performance gains.

\subsection{Loss Tolerance}
\label{subsec:eval-loss-tolerance}

To study loss tolerance, we conduct simulations using an idealized model where congestion control is disabled, forcing the sender to saturate link bandwidth while prioritizing retransmissions for lost packets. 





\parab{Throughput vs Loss Rate.} 
This setup allows for a rigorous comparison of \sysname-II and \sysname-III by tuning the packet loss rate on a single link, as shown in the left panel of \cref{fig:pkt-sim-allreduce}. While both modes see throughput declines as loss rates rise, \sysname-II experiences a sharp collapse as the loss rate increases beyond 0.1\%. This occurs because a single lost packet in \sysname-II triggers redundant retransmissions across all ranks for that sequence number, whereas \sysname-III allows lossless ranks to receive ACKs and continue, preventing excessive retransmission.

\parab{Throughput vs Lossy Links.}
As shown in the right panel of \cref{fig:pkt-sim-allreduce}, further experiments maintaining a constant 0.1\% loss rate while increasing the number of lossy links reveal that \sysname-II suffers more severe degradation than \sysname-III. In \sysname-II, the throughput decline is the cumulative result of losses across all affected links, creating a stacked decrease effect. Conversely, in \sysname-III, the overall performance is determined only by the link with the highest loss rate, as the retransmission times on other links are effectively masked by the most bottlenecked one.

\parab{Rate Synchronization for EPIC-III.} We run AllReduce on Tree-2-8 with \SI{200}{Gbps} link capacity. One link is congested by \SI{100}{Gbps} background traffic. We apply DCQCN on ranks. Full results are in \cref{tab:app:simulation:switch-cnp} of Appendix~\cref{sec:app:packet-simulation}.
Without rate synchronization (i.e., without switch replying CNP to faster ranks), the algorithm throughput is \SI{43.3}{Gbps} (\SI{16}{MB} message); with rate synchronization, the throughput becomes \SI{95.9}{Gbps}. Applying congestion control-based rate synchronization can avoid excessive sending and dropping, improving the overall collective throughput.

\subsection{Performance Gain from Policy}
\label{subsec:eval-policies}


\parab{Single 3D Parallel Training Job.} 
\cref{tab:flow-sim:GPT3} shows a single training job. We tune switch SRAM sizes based on BDP-relative units (\cref{sec:analysis}). Groups select candidate IncTrees greedily, following our algorithms (\cref{subsec:policies}). Ring-based NCCL serves as the baseline, yielding the longest JCT due to the lack of INC acceleration. The EDT policy suffers from placement conflicts, leaving several communication groups without INC resources.

In contrast, Spatial Mux grants more groups INC access when resources permit. Temporal Mux further maximizes benefits by sharing limited resources across groups, benefiting the entire workload (up to 1.85$\times$). Integrating scale-up networks yields similar conclusions, though overall INC acceleration gains diminish as high-bandwidth scale-up links absorb significant traffic, reducing relative network transmission time (1.67$\times$).

\begin{figure}[tb]
    \centering
\begin{tikzpicture}[scale=0.9]
\begin{axis}[
    ymin=0.85, ymax=1,
    xlabel={JCT (s)},
    ylabel={CDF},
    legend style={
        at={(1.05,0.5)},
        anchor=west,
        legend columns=1,        
        column sep=0.4em,        
        row sep=0.1em,           
        font=\footnotesize         
    },
    height=0.4\linewidth,
    width=0.7\linewidth,
    xmajorgrids=true,
    ymajorgrids=true,
    grid style=dashed,
]

    \addplot[solid, thick, color=blue,  mark=o, mark repeat=6, mark phase=0]
    table[
      x expr={\thisrowno{2}==1 ? \thisrowno{0} : nan},
      y expr={\thisrowno{2}==1 ? \thisrowno{1} : nan}
    ]{plots/flow_simulation/data_fattree_tenant_trace2_tail.dat};
    
    \addplot[solid, thick, color=red,   mark=diamond, mark size=2.5pt,  mark repeat=6, mark phase=2]
    table[
      x expr={\thisrowno{2}==2 ? \thisrowno{0} : nan},
      y expr={\thisrowno{2}==2 ? \thisrowno{1} : nan}
    ]{plots/flow_simulation/data_fattree_tenant_trace2_tail.dat};
    
    \addplot[solid, thick, color=black, mark=square,   mark repeat=6, mark phase=4]
    table[
      x expr={\thisrowno{2}==3 ? \thisrowno{0} : nan},
      y expr={\thisrowno{2}==3 ? \thisrowno{1} : nan}
    ]{plots/flow_simulation/data_fattree_tenant_trace2_tail.dat};
    
    \addplot[solid, thick, color=orange,mark=triangle, mark size=2.5pt, mark repeat=6, mark phase=1]
    table[
      x expr={\thisrowno{2}==4 ? \thisrowno{0} : nan},
      y expr={\thisrowno{2}==4 ? \thisrowno{1} : nan}
    ]{plots/flow_simulation/data_fattree_tenant_trace2_tail.dat};

\legend{Ring, EDT, Spatial Mux, Temporal Mux}
\end{axis}
\end{tikzpicture}
\caption{[Flow Simulation] Tail 15\% JCT of Alibaba Trace on 2048-GPU Fat-tree}
    \label{fig:flow-simulation:multi-tenancy}
\end{figure}
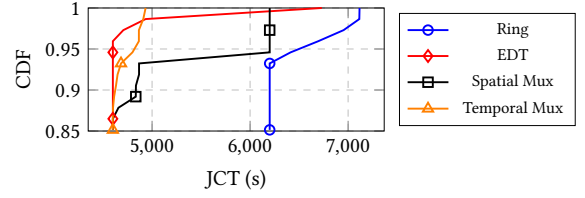

\parab{Multi-tenant Training Jobs.}  \cref{fig:flow-simulation:multi-tenancy} shows the simulation results of multi-tenant training on a 2,048-GPU cluster with production trace~\cite{cao2024crux}, varying INC management policies. Compared to the baseline, all INC-enabled policies significantly reduce the average JCT. Among these policies, Temporal Mux achieves the highest overall acceleration. 

JCT distribution analysis reveals that while Temporal Mux is less efficient than EDT for jobs in the 88th to 98th percentiles, it drastically reduces tail latency. At the 99th percentile, JCTs for Temporal Mux, EDT, and Ring are \SI{4917}{s}, \SI{5420}{s}, and \SI{7115}{s}, respectively. This improvement stems from Temporal Mux allowing all jobs to contend for INC resources; previously blocked jobs gain acceleration, while formerly exclusive jobs face minor interference.

\section{Related Work}

\parab{INC Primitive Design.} Many existing programmable switch-based INC solutions focus on certain primitives (e.g., AllReduce~\cite{lao2021atp, sapio2021scaling, liu2023network, li2019accelerating, gebara2021network, klenk2020network, di2021risc, yang2022using, li2023a2tp, li2023pa}, Broadcast~\cite{li2024cepheus, huang2023mc, khalilov2024network}), which belongs to \sysname Mode-II; \sysname supports six primitives and hybrid invocations through specialized control signals and group PSN synchronization, and fixes their bugs by model checking. There are product switches with six collectives supported~\cite{graham2016scalable, graham2020scalable, broadcom_tomahawk5}, which belong to \sysname Mode-I; \sysname proposes Mode-III and a modular design that supports evolvability. The Ultra Ethernet Consortium is finalizing the INC standard~\cite{uec2024specupdate}, which is a clean-slate, Mode-I-like design with INC header and UET transport.
There are in-network computing solutions for irregular traffic patterns~\cite{he2023generic, huang2025traci} and generic computing~\cite{xu2023clickinc, zhao2023netrpc, wan2025generic}, which we will consider in the future. 

\parab{INC Resource Management.} INC management in a cluster should consider function placement~\cite{blocher2021switches, segal2022constrained}, routing~\cite{zhu2025dina, xia2024accelerating, fang2023grid, rottenstreich2024edge, segal2021soar,dong2025mina}, job placement~\cite{zhao2024training}, and resource allocation~\cite{zhao2023enabling}; \sysname provides a unified resource model to enable these policies.

\parab{INC Switch Microarchitecture.} \sysname focuses on protocol design and is compatible with existing switch microarchitectures~\cite{wang2023roar, de2021flare}. 

\parab{Modular Network Design.} There were modular designs of software switches/routers~\cite{kohler2000click, dobrescu2009routebricks, panda2016netbricks}, network functions~\cite{meng2019micronf,gallo2018clicknf}, and network layers~\cite{ee2006modular}. 
\sysname differs from them in the application scenario: it implements the same functionality with different algorithms and applies modularity to enable algorithm evolution. ClickINC~\cite{xu2023clickinc} implements the same algorithms on different platforms. Click~\cite{kohler2000click} uses modules to flexibly compose different functionalities. P4~\cite{bosshart2014p4} provides a platform-independent language to program different functionalities.


\parab{Collective Communication Libraries.} There are various collective communication libraries and their optimizations~\cite{kim2024tccl, nccl, rccl, oneccl, dong2021accl, msccl, shah2023taccl, hidayetoglu2024hiccl}; \sysname's CommLib works at the same layer and should be integrated within mainstream CCLs.

\section{Conclusion}

To address Ethernet fragmentation and encourage vendors with diverse technology stacks in INC, we propose an \sysname protocol featuring unified abstraction and polymorphic realization. Its modular design allows seamless, incremental hardware evolution while maintaining protocol integrity. Validated via model checking, \sysname ensures robust performance across diverse conditions. \sysname unifies resource management and enhances control-plane agility. \sysname significantly accelerates collective communication and reduces single- and multi-tenant training JCT.


\section*{Acknowledgment}
We thank the anonymous reviewers and shepherd. Wenfei Wu is supported by the National Natural Science Foundation of China (General Program No. 62472009),  ``the Fundamental Research Funds for the Central Universities, Peking University'', the Beijing Natural Science Foundation (Haidian Original Innovation Joint Fund, Key Program L252013), and the Beijing Municipal Science and Technology Project (No. Z241100004224024).

\bibliographystyle{ACM-Reference-Format}
\bibliography{reference}

@inproceedings {zheng2025p4,
author = {Hao Zheng and Xin Yan and Wenbo Li and Jiaqi Zheng and Xiaoliang Wang and Qingqing Zhao and Luyou He and Xiaofei Lai and Feng Gao and Fuguang Huang and Wanchun Dou and Guihai Chen and Chen Tian},
title = {When P4 Meets Run-to-completion Architecture},
booktitle = {22nd USENIX Symposium on Networked Systems Design and Implementation (NSDI 25)},
year = {2025},
isbn = {978-1-939133-46-5},
address = {Philadelphia, PA},
pages = {1487--1505},
url = {https://www.usenix.org/conference/nsdi25/presentation/zheng-hao},
publisher = {USENIX Association},
month = apr
}

@misc{sridharan2023chakra,
      title={Chakra: Advancing Performance Benchmarking and Co-design using Standardized Execution Traces},
      author={Srinivas Sridharan and Taekyung Heo and Louis Feng and Zhaodong Wang and Matt Bergeron and Wenyin Fu and Shengbao Zheng and Brian Coutinho and Saeed Rashidi and Changhai Man and Tushar Krishna},
      year={2023},
      eprint={2305.14516},
      archivePrefix={arXiv},
      primaryClass={cs.LG},
      url={https://arxiv.org/abs/2305.14516},
}

@misc{bai2026verifying,
      title={Verifying In-Network Computing Systems for Design Risks},
      author={Tianyu Bai and Ying Zhang and Xiaoxi Zhang and Wenfei Wu},
      year={2026},
      eprint={2604.10186},
      archivePrefix={arXiv},
      primaryClass={cs.DC},
      url={https://arxiv.org/abs/2604.10186},
}

@inproceedings{xu2026multipath,
author = {Xu, Yuchen and Nie, Jianglong and Li, Baojia and Chen, Mingzhuo and Lu, Hao and Qu, Guanyu and Liu, Zhenchuan and Yin, Shuangshuang and Huang, Xiaojie and He, Chunzhi and Xia, Yinben and Wen, Quan and Li, Xiang and He, Zekun and Wang, Yachen and Zou, Xianneng and Miao, Congcong and Wu, Wenfei},
title = {Multipath Collective Communication Beyond Scale-up Networks in GPU Clouds},
year = {2026},
isbn = {9798400722127},
publisher = {Association for Computing Machinery},
address = {New York, NY, USA},
url = {https://doi.org/10.1145/3767295.3769330},
doi = {10.1145/3767295.3769330},
booktitle = {Proceedings of the 21st European Conference on Computer Systems},
pages = {564–585},
numpages = {22},
location = {McEwan Hall/The University of Edinburgh, Edinburgh, Scotland UK},
series = {EUROSYS '26}
}

@article{liu2025train,
  title={How to Train a Model on a Cheap Cluster with Low Cost using Block Coordinate Descent},
  author={Liu, Zeyu and Zhang, Yunquan and Zhang, Boyang and Jiang, Guoyong and Cheng, Daning},
  journal={arXiv e-prints},
  pages={arXiv--2506},
  year={2025}
}

@inproceedings{athlur2022varuna,
  title={Varuna: scalable, low-cost training of massive deep learning models},
  author={Athlur, Sanjith and Saran, Nitika and Sivathanu, Muthian and Ramjee, Ramachandran and Kwatra, Nipun},
  booktitle={Proceedings of the Seventeenth European Conference on Computer Systems},
  pages={472--487},
  year={2022}
}

@inproceedings{lim2024accelerating,
  title={Accelerating model training in multi-cluster environments with consumer-grade gpus},
  author={Lim, Hwijoon and Ye, Juncheol and Abdu Jyothi, Sangeetha and Han, Dongsu},
  booktitle={Proceedings of the ACM SIGCOMM 2024 Conference},
  pages={707--720},
  year={2024}
}

@inproceedings{sun2026rearchitecting,
    author = {Sun, Haifeng and Liu, Bing and Tian, Taixu and Sun, Jinbo and He, Jintao and Huang, Qun and He, Luyou and Wang, Xuan and Gao, Feng and Wang, Liguo and Xu, Xiangcan and Guo, Junyi and Zhu, Xiaoping and Yang, Yongqiang},
    title = {Rearchitecting Programmable Networks For In-Network Computing: From Hardware To Language},
    year = {2026},
    isbn = {9798400722127},
    publisher = {Association for Computing Machinery},
    address = {New York, NY, USA},
    url = {https://doi.org/10.1145/3767295.3769344},
    doi = {10.1145/3767295.3769344},
    booktitle = {Proceedings of the 21st European Conference on Computer Systems},
    pages = {1094–1110},
    numpages = {17},
    location = {McEwan Hall/The University of Edinburgh, Edinburgh, Scotland UK},
    series = {EUROSYS '26}
}

@article{bosshart2014p4,
author = {Bosshart, Pat and Daly, Dan and Gibb, Glen and Izzard, Martin and McKeown, Nick and Rexford, Jennifer and Schlesinger, Cole and Talayco, Dan and Vahdat, Amin and Varghese, George and Walker, David},
title = {P4: programming protocol-independent packet processors},
year = {2014},
issue_date = {July 2014},
publisher = {Association for Computing Machinery},
address = {New York, NY, USA},
volume = {44},
number = {3},
issn = {0146-4833},
url = {https://doi.org/10.1145/2656877.2656890},
doi = {10.1145/2656877.2656890},
journal = {SIGCOMM Comput. Commun. Rev.},
month = jul,
pages = {87–95},
numpages = {9}
}

@inproceedings{jouppi2023tpu,
author = {Jouppi, Norm and Kurian, George and Li, Sheng and Ma, Peter and Nagarajan, Rahul and Nai, Lifeng and Patil, Nishant and Subramanian, Suvinay and Swing, Andy and Towles, Brian and Young, Clifford and Zhou, Xiang and Zhou, Zongwei and Patterson, David A},
title = {TPU v4: An Optically Reconfigurable Supercomputer for Machine Learning with Hardware Support for Embeddings},
year = {2023},
isbn = {9798400700958},
publisher = {Association for Computing Machinery},
address = {New York, NY, USA},
url = {https://doi.org/10.1145/3579371.3589350},
doi = {10.1145/3579371.3589350},
booktitle = {Proceedings of the 50th Annual International Symposium on Computer Architecture},
articleno = {82},
numpages = {14},
location = {Orlando, FL, USA},
series = {ISCA '23}
}

@inproceedings{lao2026continuum,
  title={Continuum: An Interruption-Resilient Runtime for ML Training},
  author={ChonLam Lao and Jiaqi Gao and Jiamin Cao and Zhipeng Zhang and Pengcheng Zhang and Jiangfei Duan and Minlan Yu and Aditya Akella and Zhilong Zheng and Yu Guan and Yichi Xu and Yong Li and Ennan Zhai and Dennis Cai and Zhengping Qian and Jingren Zhou},
  booktitle={OSDI},
  year={2026}
}

@inproceedings{wu2024mccs,
  title={Mccs: A service-based approach to collective communication for multi-tenant cloud},
  author={Wu, Yongji and Xu, Yechen and Chen, Jingrong and Wang, Zhaodong and Zhang, Ying and Lentz, Matthew and Zhuo, Danyang},
  booktitle={Proceedings of the ACM SIGCOMM 2024 Conference},
  pages={679--690},
  year={2024}
}

@inproceedings{gangidi2024rdma,
author = {Gangidi, Adithya and Miao, Rui and Zheng, Shengbao and Bondu, Sai Jayesh and Goes, Guilherme and Morsy, Hany and Puri, Rohit and Riftadi, Mohammad and Shetty, Ashmitha Jeevaraj and Yang, Jingyi and Zhang, Shuqiang and Fernandez, Mikel Jimenez and Gandham, Shashidhar and Zeng, Hongyi},
title = {RDMA over Ethernet for Distributed Training at Meta Scale},
year = {2024},
isbn = {9798400706141},
publisher = {Association for Computing Machinery},
address = {New York, NY, USA},
url = {https://doi.org/10.1145/3651890.3672233},
doi = {10.1145/3651890.3672233},
booktitle = {Proceedings of the ACM SIGCOMM 2024 Conference},
pages = {57–70},
numpages = {14},
location = {Sydney, NSW, Australia},
series = {ACM SIGCOMM '24}
}

@inproceedings {bai2023empowering,
author = {Wei Bai and Shanim Sainul Abdeen and Ankit Agrawal and Krishan Kumar Attre and Paramvir Bahl and Ameya Bhagat and Gowri Bhaskara and Tanya Brokhman and Lei Cao and Ahmad Cheema and Rebecca Chow and Jeff Cohen and Mahmoud Elhaddad and Vivek Ette and Igal Figlin and Daniel Firestone and Mathew George and Ilya German and Lakhmeet Ghai and Eric Green and Albert Greenberg and Manish Gupta and Randy Haagens and Matthew Hendel and Ridwan Howlader and Neetha John and Julia Johnstone and Tom Jolly and Greg Kramer and David Kruse and Ankit Kumar and Erica Lan and Ivan Lee and Avi Levy and Marina Lipshteyn and Xin Liu and Chen Liu and Guohan Lu and Yuemin Lu and Xiakun Lu and Vadim Makhervaks and Ulad Malashanka and David A. Maltz and Ilias Marinos and Rohan Mehta and Sharda Murthi and Anup Namdhari and Aaron Ogus and Jitendra Padhye and Madhav Pandya and Douglas Phillips and Adrian Power and Suraj Puri and Shachar Raindel and Jordan Rhee and Anthony Russo and Maneesh Sah and Ali Sheriff and Chris Sparacino and Ashutosh Srivastava and Weixiang Sun and Nick Swanson and Fuhou Tian and Lukasz Tomczyk and Vamsi Vadlamuri and Alec Wolman and Ying Xie and Joyce Yom and Lihua Yuan and Yanzhao Zhang and Brian Zill},
title = {Empowering Azure Storage with {RDMA}},
booktitle = {20th USENIX Symposium on Networked Systems Design and Implementation (NSDI 23)},
year = {2023},
isbn = {978-1-939133-33-5},
address = {Boston, MA},
pages = {49--67},
url = {https://www.usenix.org/conference/nsdi23/presentation/bai},
publisher = {USENIX Association},
month = apr
}

@inproceedings {gao2021cloud,
author = {Yixiao Gao and Qiang Li and Lingbo Tang and Yongqing Xi and Pengcheng Zhang and Wenwen Peng and Bo Li and Yaohui Wu and Shaozong Liu and Lei Yan and Fei Feng and Yan Zhuang and Fan Liu and Pan Liu and Xingkui Liu and Zhongjie Wu and Junping Wu and Zheng Cao and Chen Tian and Jinbo Wu and Jiaji Zhu and Haiyong Wang and Dennis Cai and Jiesheng Wu},
title = {When Cloud Storage Meets {RDMA}},
booktitle = {18th USENIX Symposium on Networked Systems Design and Implementation (NSDI 21)},
year = {2021},
isbn = {978-1-939133-21-2},
pages = {519--533},
url = {https://www.usenix.org/conference/nsdi21/presentation/gao},
publisher = {USENIX Association},
month = apr
}

@inproceedings{cao2024crux,
  title={Crux: Gpu-efficient communication scheduling for deep learning training},
  author={Cao, Jiamin and Guan, Yu and Qian, Kun and Gao, Jiaqi and Xiao, Wencong and Dong, Jianbo and Fu, Binzhang and Cai, Dennis and Zhai, Ennan},
  booktitle={Proceedings of the ACM SIGCOMM 2024 Conference},
  pages={1--15},
  year={2024}
}

@article{meng2019micronf,
  title={MicroNF: An efficient framework for enabling modularized service chains in NFV},
  author={Meng, Zili and Bi, Jun and Wang, Haiping and Sun, Chen and Hu, Hongxin},
  journal={IEEE Journal on Selected Areas in Communications},
  volume={37},
  number={8},
  pages={1851--1865},
  year={2019},
  publisher={IEEE}
}

@inproceedings{gallo2018clicknf,
  title={{ClickNF}: a Modular Stack for Custom Network Functions},
  author={Gallo, Massimo and Laufer, Rafael},
  booktitle={2018 USENIX Annual Technical Conference (USENIX ATC 18)},
  pages={745--757},
  year={2018}
}

@inproceedings{ee2006modular,
  title={A Modular Network Layer for Sensornets.},
  author={Ee, Cheng Tien and Fonseca, Rodrigo and Kim, Sukun and Moon, Daekyeong and Tavakoli, Arsalan and Culler, David E and Shenker, Scott and Stoica, Ion},
  booktitle={OSDI},
  volume={6},
  pages={249--262},
  year={2006}
}

@misc{tofino,
author = {Intel},
  url = {https://www.intel.com/content/www/us/en/products/details/network-io/intelligent-fabric-processors/tofino-2.html},
title = {Intel Tofino 2},
    year = {2024},
}

@ARTICLE{hoefler2025network,
  author={Hoefler, Torsten and Khalilov, Mikhail and Clark, Josiah and Anubolu, Surendra and Kalkunte, Mohan and Schramm, Karen and Spada, Eric and Roweth, Duncan and Underwood, Keith and Caulfield, Adrian and Kabbani, Abdul and Rastegari, Amirreza},
  journal={Computer},
  title={In-Network Collective Operations: Game Changer or Challenge for AI Workloads?},
  year={2026},
  volume={59},
  number={1},
  pages={24-33},
  doi={10.1109/MC.2025.3616048}}

@techreport{nvidia_nvlink_2014,
	title = {{NVIDIA NVLink High-Speed Interconnect: Application Performance}},
	shorttitle = {{NVLink} Application Performance},
	institution = {NVIDIA Corporation},
	type = {Whitepaper},
	date = {2014-11},
	author = {{NVIDIA Corporation}},

}

@misc{ualink_consortium,
    title = {{UAlink Consortium}},
    author = {{UAlink Consortium}},
    year = {2024},
    url = {https://ualinkconsortium.org/},
    howpublished = {Online Consortium Website},
}

@inproceedings{li2025revisiting,
author = {Li, Wenxue and Liu, Xiangzhou and Zhang, Yunxuan and Wang, Zihao and Gu, Wei and Qian, Tao and Zeng, Gaoxiong and Ren, Shoushou and Huang, Xinyang and Ren, Zhenghang and Liu, Bowen and Zhang, Junxue and Chen, Kai and Liu, Bingyang},
title = {Revisiting RDMA Reliability for Lossy Fabrics},
year = {2025},
isbn = {9798400715242},
publisher = {Association for Computing Machinery},
address = {New York, NY, USA},
url = {https://doi.org/10.1145/3718958.3750480},
doi = {10.1145/3718958.3750480},
booktitle = {Proceedings of the ACM SIGCOMM 2025 Conference},
pages = {85–98},
numpages = {14},
location = {S\~{a}o Francisco Convent, Coimbra, Portugal},
series = {SIGCOMM '25}
}

@techreport{ibta_rocev2_2014,
    title = {Supplement to InfiniBand™ Architecture Specification Volume 1 Release 1.2.1: Annex A17: RoCEv2},
    author = {{InfiniBand Trade Association}},
    institution = {InfiniBand Trade Association},
    type = {Technical Specification Supplement},
    number = {Release 1.2.1, Annex A17},
    date = {2014-09-02},
    copyright = {2010, InfiniBand Trade Association},
    note = {Proprietary document; available via InfiniBand Trade Association membership},
}

@misc{dong2024boosting,
      title={Enhancing Large-Scale AI Training Efficiency: The C4 Solution for Real-Time Anomaly Detection and Communication Optimization},
      author={Jianbo Dong and Bin Luo and Jun Zhang and Pengcheng Zhang and Fei Feng and Yikai Zhu and Ang Liu and Zian Chen and Yi Shi and Hairong Jiao and Gang Lu and Yu Guan and Ennan Zhai and Wencong Xiao and Hanyu Zhao and Man Yuan and Siran Yang and Xiang Li and Jiamang Wang and Rui Men and Jianwei Zhang and Chang Zhou and Dennis Cai and Yuan Xie and Binzhang Fu},
      year={2025},
      eprint={2406.04594},
      archivePrefix={arXiv},
      primaryClass={cs.DC},
      url={https://arxiv.org/abs/2406.04594},
}

@article{hoefler2024hammingmesh,
  title={Hammingmesh: A network topology for large-scale deep learning},
  author={Hoefler, Torsten and Bonato, Tommaso and De Sensi, Daniele and Di Girolamo, Salvatore and Li, Shigang and Heddes, Marco and Goel, Deepak and Castro, Miguel and Scott, Steve},
  journal={Communications of the ACM},
  volume={67},
  number={12},
  pages={97--105},
  year={2024},
  publisher={ACM New York, NY, USA}
}

@inproceedings{li2024understanding,
  title={Understanding communication characteristics of distributed training},
  author={Li, Wenxue and Liu, Xiangzhou and Li, Yuxuan and Jin, Yilun and Tian, Han and Zhong, Zhizhen and Liu, Guyue and Zhang, Ying and Chen, Kai},
  booktitle={Proceedings of the 8th Asia-Pacific Workshop on Networking},
  pages={1--8},
  year={2024}
}

@inproceedings{wang2023topoopt,
  title={{TopoOpt}: Co-optimizing network topology and parallelization strategy for distributed training jobs},
  author={Wang, Weiyang and Khazraee, Moein and Zhong, Zhizhen and Ghobadi, Manya and Jia, Zhihao and Mudigere, Dheevatsa and Zhang, Ying and Kewitsch, Anthony},
  booktitle={20th USENIX Symposium on Networked Systems Design and Implementation (NSDI 23)},
  pages={739--767},
  year={2023}
}

@inproceedings{qi2025sglb,
author = {Qi, Chenchen and Wu, Wenfei and Wang, Yongcan and He, Keqiang and Kao, Yu-Hsiang (Sean) and He, Zongying and Yen, Chen-Yu and Jiang, Zhuo and Luo, Feng and Anubolu, Surendra and Gao, Yanjin and Lin, Bingfeng and Ni, Wenda and Yang, Yiming and Wei, Donglin and Zhou, Boyang and Wang, Jian and Ding, Shan},
title = {SGLB: Scalable and Robust Global Load Balancing in Commodity AI Clusters},
year = {2025},
isbn = {9798400715242},
publisher = {Association for Computing Machinery},
address = {New York, NY, USA},
url = {https://doi.org/10.1145/3718958.3750527},
doi = {10.1145/3718958.3750527},
booktitle = {Proceedings of the ACM SIGCOMM 2025 Conference},
pages = {626–644},
numpages = {19},
location = {S\~{a}o Francisco Convent, Coimbra, Portugal},
series = {SIGCOMM '25}
}

@inproceedings{qian2024alibaba,
author = {Qian, Kun and Xi, Yongqing and Cao, Jiamin and Gao, Jiaqi and Xu, Yichi and Guan, Yu and Fu, Binzhang and Shi, Xuemei and Zhu, Fangbo and Miao, Rui and Wang, Chao and Wang, Peng and Zhang, Pengcheng and Zeng, Xianlong and Ruan, Eddie and Yao, Zhiping and Zhai, Ennan and Cai, Dennis},
title = {Alibaba HPN: A Data Center Network for Large Language Model Training},
year = {2024},
isbn = {9798400706141},
publisher = {Association for Computing Machinery},
address = {New York, NY, USA},
url = {https://doi.org/10.1145/3651890.3672265},
doi = {10.1145/3651890.3672265},
booktitle = {Proceedings of the ACM SIGCOMM 2024 Conference},
pages = {691–706},
numpages = {16},
location = {Sydney, NSW, Australia},
series = {ACM SIGCOMM '24}
}

@inproceedings{meng2025astral,
author = {Meng, Qingkai and Zheng, Hao and Zhang, Zhenhui and Lao, ChonLam and Huang, Chengyuan and Li, Baojia and Zhu, Ziyuan and Lu, Hao and Dang, Weizhen and Lin, Zitong and Zhang, Weifeng and Liu, Lingfeng and Gong, Yuanyuan and He, Chunzhi and Hu, Xiaoyuan and Xia, Yinben and Li, Xiang and He, Zekun and Wang, Yachen and Zou, Xianneng and Yang, Kun and Antichi, Gianni and Chen, Guihai and Tian, Chen},
title = {Astral: A Datacenter Infrastructure for Large Language Model Training at Scale},
year = {2025},
isbn = {9798400715242},
publisher = {Association for Computing Machinery},
address = {New York, NY, USA},
url = {https://doi.org/10.1145/3718958.3750521},
doi = {10.1145/3718958.3750521},
booktitle = {Proceedings of the ACM SIGCOMM 2025 Conference},
pages = {609–625},
numpages = {17},
location = {S\~{a}o Francisco Convent, Coimbra, Portugal},
series = {SIGCOMM '25}
}

@inproceedings {wang2025simai,
author = {Xizheng Wang and Qingxu Li and Yichi Xu and Gang Lu and Dan Li and Li Chen and Heyang Zhou and Linkang Zheng and Sen Zhang and Yikai Zhu and Yang Liu and Pengcheng Zhang and Kun Qian and Kunling He and Jiaqi Gao and Ennan Zhai and Dennis Cai and Binzhang Fu},
title = {{SimAI}: Unifying Architecture Design and Performance Tuning for {Large-Scale} Large Language Model Training with Scalability and Precision},
booktitle = {22nd USENIX Symposium on Networked Systems Design and Implementation (NSDI 25)},
year = {2025},
isbn = {978-1-939133-46-5},
address = {Philadelphia, PA},
pages = {541--558},
url = {https://www.usenix.org/conference/nsdi25/presentation/wang-xizheng-simai},
publisher = {USENIX Association},
month = apr
}

@inproceedings{lao2021atp,
  title={{ATP}: In-network aggregation for multi-tenant learning},
  author={Lao, ChonLam and Le, Yanfang and Mahajan, Kshiteej and Chen, Yixi and Wu, Wenfei and Akella, Aditya and Swift, Michael},
  booktitle={18th USENIX Symposium on Networked Systems Design and Implementation (NSDI 21)},
  pages={741--761},
  year={2021}
}

@inproceedings{sapio2021scaling,
  title={Scaling distributed machine learning with {In-Network} aggregation},
  author={Sapio, Amedeo and Canini, Marco and Ho, Chen-Yu and Nelson, Jacob and Kalnis, Panos and Kim, Changhoon and Krishnamurthy, Arvind and Moshref, Masoud and Ports, Dan and Richt{\'a}rik, Peter},
  booktitle={18th USENIX Symposium on Networked Systems Design and Implementation (NSDI 21)},
  pages={785--808},
  year={2021}
}

@INPROCEEDINGS{graham2016scalable,
  author={Graham, Richard L. and Bureddy, Devendar and Lui, Pak and Rosenstock, Hal and Shainer, Gilad and Bloch, Gil and Goldenerg, Dror and Dubman, Mike and Kotchubievsky, Sasha and Koushnir, Vladimir and Levi, Lion and Margolin, Alex and Ronen, Tamir and Shpiner, Alexander and Wertheim, Oded and Zahavi, Eitan},
  booktitle={2016 First International Workshop on Communication Optimizations in HPC (COMHPC)},
  title={Scalable Hierarchical Aggregation Protocol (SHArP): A Hardware Architecture for Efficient Data Reduction},
  year={2016},
  volume={},
  number={},
  pages={1-10},
  doi={10.1109/COMHPC.2016.006}}

@inproceedings{li2019accelerating,
  title={Accelerating distributed reinforcement learning with in-switch computing},
  author={Li, Youjie and Liu, Iou-Jen and Yuan, Yifan and Chen, Deming and Schwing, Alexander and Huang, Jian},
  booktitle={Proceedings of the 46th International Symposium on Computer Architecture},
  pages={279--291},
  year={2019}
}

@article{gebara2021network,
  title={In-network aggregation for shared machine learning clusters},
  author={Gebara, Nadeen and Ghobadi, Manya and Costa, Paolo},
  journal={Proceedings of Machine Learning and Systems},
  volume={3},
  pages={829--844},
  year={2021}
}

@inproceedings{segal2021soar,
  title={SOAR: Minimizing network utilization with bounded in-network computing},
  author={Segal, Raz and Avin, Chen and Scalosub, Gabriel},
  booktitle={Proceedings of the 17th International Conference on emerging Networking EXperiments and Technologies},
  pages={16--29},
  year={2021}
}

@inproceedings{li2024cepheus,
  title={Cepheus: accelerating datacenter applications with high-performance roce-capable multicast},
  author={Li, Wenxue and Zhang, Junyi and Liu, Yufei and Zeng, Gaoxiong and Wang, Zilong and Zeng, Chaoliang and Zhou, Pengpeng and Wang, Qiaoling and Chen, Kai},
  booktitle={2024 IEEE International Symposium on High-Performance Computer Architecture (HPCA)},
  pages={908--921},
  year={2024},
  organization={IEEE}
}

@INPROCEEDINGS{huang2023mc,
  author={Huang, Chengyuan and Gao, Yixiao and Chen, Wei and Li, Duoxing and Xiao, Yibo and Zhang, Ruyi and Tian, Chen and Wang, Xiaoliang and Dou, Wanchun and Chen, Guihai and Wang, Yi and Xiao, Fu},
  booktitle={2023 IEEE 31st International Conference on Network Protocols (ICNP)},
  title={MC-RDMA: Improving Replication Performance of RDMA-based Distributed Systems with Reliable Multicast Support},
  year={2023},
  volume={},
  number={},
  pages={1-11},
  doi={10.1109/ICNP59255.2023.10355619}}

@inproceedings{klenk2020network,
  title={An in-network architecture for accelerating shared-memory multiprocessor collectives},
  author={Klenk, Benjamin and Jiang, Nan and Thorson, Greg and Dennison, Larry},
  booktitle={2020 ACM/IEEE 47th Annual International Symposium on Computer Architecture (ISCA)},
  pages={996--1009},
  year={2020},
  organization={IEEE}
}

@inproceedings{liu2023network,
  title={In-network aggregation with transport transparency for distributed training},
  author={Liu, Shuo and Wang, Qiaoling and Zhang, Junyi and Wu, Wenfei and Lin, Qinliang and Liu, Yao and Xu, Meng and Canini, Marco and Cheung, Ray CC and He, Jianfei},
  booktitle={Proceedings of the 28th ACM International Conference on Architectural Support for Programming Languages and Operating Systems, Volume 3},
  pages={376--391},
  year={2023}
}

@inproceedings{zhao2023enabling,
  title={Enabling switch memory management for distributed training with in-network aggregation},
  author={Zhao, Bohan and Liu, Chang and Dong, Jianbo and Cao, Zheng and Nie, Wei and Wu, Wenfei},
  booktitle={IEEE INFOCOM 2023-IEEE conference on computer communications},
  pages={1--10},
  year={2023},
  organization={IEEE}
}

@inproceedings{zhao2024training,
  title={Training job placement in clusters with statistical in-network aggregation},
  author={Zhao, Bohan and Xu, Wei and Liu, Shuo and Tian, Yang and Wang, Qiaoling and Wu, Wenfei},
  booktitle={Proceedings of the 29th ACM International Conference on Architectural Support for Programming Languages and Operating Systems, Volume 1},
  pages={420--434},
  year={2024}
}

@inproceedings{di2021risc,
  title={A RISC-V in-network accelerator for flexible high-performance low-power packet processing},
  author={Di Girolamo, Salvatore and Kurth, Andreas and Calotoiu, Alexandru and Benz, Thomas and Schneider, Timo and Ber{\'a}nek, Jakub and Benini, Luca and Hoefler, Torsten},
  booktitle={2021 ACM/IEEE 48th Annual International Symposium on Computer Architecture (ISCA)},
  pages={958--971},
  year={2021},
  organization={IEEE}
}

@inproceedings{graham2020scalable,
author = {Graham, Richard L. and Levi, Lion and Burredy, Devendar and Bloch, Gil and Shainer, Gilad and Cho, David and Elias, George and Klein, Daniel and Ladd, Joshua and Maor, Ophir and Marelli, Ami and Petrov, Valentin and Romlet, Evyatar and Qin, Yong and Zemah, Ido},
title = {Scalable Hierarchical Aggregation and Reduction Protocol (SHARP)TM Streaming-Aggregation Hardware Design and Evaluation},
year = {2020},
isbn = {978-3-030-50742-8},
publisher = {Springer-Verlag},
address = {Berlin, Heidelberg},
url = {https://doi.org/10.1007/978-3-030-50743-5_3},
doi = {10.1007/978-3-030-50743-5_3},
booktitle = {High Performance Computing: 35th International Conference, ISC High Performance 2020, Frankfurt/Main, Germany, June 22–25, 2020, Proceedings},
pages = {41–59},
numpages = {19},
location = {Frankfurt am Main, Germany}
}

@inproceedings{li2023a2tp,
  title={A2TP: Aggregator-aware in-network aggregation for multi-tenant learning},
  author={Li, Zhaoyi and Huang, Jiawei and Li, Yijun and Xu, Aikun and Zhou, Shengwen and Liu, Jingling and Wang, Jianxin},
  booktitle={Proceedings of the Eighteenth European Conference on Computer Systems},
  pages={639--653},
  year={2023}
}

@inproceedings{li2023pa,
  title={PA-ATP: Progress-Aware Transmission Protocol for In-Network Aggregation},
  author={Li, Zhaoyi and Huang, Jiawei and Zhang, Tao and Zhou, Shengwen and Wang, Qile and Li, Yijun and Liu, Jingling and Jiang, Wanchun and Wang, Jianxin},
  booktitle={2023 IEEE 31st International Conference on Network Protocols (ICNP)},
  pages={1--11},
  year={2023},
  organization={IEEE}
}

@article{rottenstreich2024edge,
  title={Edge-disjoint tree allocation for multi-tenant cloud security in datacenter topologies},
  author={Rottenstreich, Ori and Yallouz, Jose},
  journal={IEEE/ACM Transactions on Networking},
  volume={32},
  number={4},
  pages={2858--2874},
  year={2024},
  publisher={IEEE}
}

@INPROCEEDINGS{dong2025mina,
  author={Dong, Shichen and Niu, Zhixiong and Zhang, Mingchao and Xu, Zhiying and Hu, Chuntao and Zhu, Pengzhi and Song, Qingchun and Qu, Lei and Cheng, Peng and Nguyen, Cam-Tu and Sun, Shaoling and Xu, Xiaohu and Xiong, Yongqiang and Wang, Wei and Wang, Xiaoliang},
  booktitle={IEEE INFOCOM 2025 - IEEE Conference on Computer Communications},
  title={Mina: Fine-Grained In-network Aggregation Resource Scheduling for Machine Learning Service},
  year={2025},
  volume={},
  number={},
  pages={1-10},
  doi={10.1109/INFOCOM55648.2025.11044657}}

@inproceedings{segal2022constrained,
  title={Constrained in-network computing with low congestion in datacenter networks},
  author={Segal, Raz and Avin, Chen and Scalosub, Gabriel},
  booktitle={IEEE INFOCOM 2022-IEEE Conference on Computer Communications},
  pages={1639--1648},
  year={2022},
  organization={IEEE}
}

@book{tlaplus,
author = {Lamport, Leslie},
title = {Specifying systems: The {TLA+} language and tools for hardware and software engineers},
year = {2002},
month = jun,
publisher = {Addison-Wesley}
}

@inproceedings{wang2023roar,
  title={Roar: A router microarchitecture for in-network allreduce},
  author={Wang, Ruiqi and Dong, Dezun and Lei, Fei and Ma, Junchao and Wu, Ke and Lu, Kai},
  booktitle={Proceedings of the 37th International Conference on Supercomputing},
  pages={423--436},
  year={2023}
}

@inproceedings{he2023generic,
  title={A generic service to provide in-network aggregation for key-value streams},
  author={He, Yongchao and Wu, Wenfei and Le, Yanfang and Liu, Ming and Lao, ChonLam},
  booktitle={Proceedings of the 28th ACM International Conference on Architectural Support for Programming Languages and Operating Systems, Volume 2},
  pages={33--47},
  year={2023}
}

@inproceedings{huang2025traci,
  title={TRACI: Network Acceleration of Input-Dynamic Communication for Large-Scale Deep Learning Recommendation Model},
  author={Huang, Guyue and Li, Hao and Qin, Le and Huang, Jiayi and Kang, Yangwook and Ding, Yufei and Xie, Yuan},
  booktitle={Proceedings of the 52nd Annual International Symposium on Computer Architecture},
  pages={1880--1893},
  year={2025}
}

@misc{liu2024deepseek,
      title={DeepSeek-V3 Technical Report},
      author={DeepSeek-AI and Aixin Liu and Bei Feng and Bing Xue and Bingxuan Wang and Bochao Wu and Chengda Lu and Chenggang Zhao and Chengqi Deng and Chenyu Zhang and Chong Ruan and Damai Dai and Daya Guo and Dejian Yang and Deli Chen and Dongjie Ji and Erhang Li and Fangyun Lin and Fucong Dai and Fuli Luo and Guangbo Hao and Guanting Chen and Guowei Li and H. Zhang and Han Bao and Hanwei Xu and Haocheng Wang and Haowei Zhang and Honghui Ding and Huajian Xin and Huazuo Gao and Hui Li and Hui Qu and J. L. Cai and Jian Liang and Jianzhong Guo and Jiaqi Ni and Jiashi Li and Jiawei Wang and Jin Chen and Jingchang Chen and Jingyang Yuan and Junjie Qiu and Junlong Li and Junxiao Song and Kai Dong and Kai Hu and Kaige Gao and Kang Guan and Kexin Huang and Kuai Yu and Lean Wang and Lecong Zhang and Lei Xu and Leyi Xia and Liang Zhao and Litong Wang and Liyue Zhang and Meng Li and Miaojun Wang and Mingchuan Zhang and Minghua Zhang and Minghui Tang and Mingming Li and Ning Tian and Panpan Huang and Peiyi Wang and Peng Zhang and Qiancheng Wang and Qihao Zhu and Qinyu Chen and Qiushi Du and R. J. Chen and R. L. Jin and Ruiqi Ge and Ruisong Zhang and Ruizhe Pan and Runji Wang and Runxin Xu and Ruoyu Zhang and Ruyi Chen and S. S. Li and Shanghao Lu and Shangyan Zhou and Shanhuang Chen and Shaoqing Wu and Shengfeng Ye and Shengfeng Ye and Shirong Ma and Shiyu Wang and Shuang Zhou and Shuiping Yu and Shunfeng Zhou and Shuting Pan and T. Wang and Tao Yun and Tian Pei and Tianyu Sun and W. L. Xiao and Wangding Zeng and Wanjia Zhao and Wei An and Wen Liu and Wenfeng Liang and Wenjun Gao and Wenqin Yu and Wentao Zhang and X. Q. Li and Xiangyue Jin and Xianzu Wang and Xiao Bi and Xiaodong Liu and Xiaohan Wang and Xiaojin Shen and Xiaokang Chen and Xiaokang Zhang and Xiaosha Chen and Xiaotao Nie and Xiaowen Sun and Xiaoxiang Wang and Xin Cheng and Xin Liu and Xin Xie and Xingchao Liu and Xingkai Yu and Xinnan Song and Xinxia Shan and Xinyi Zhou and Xinyu Yang and Xinyuan Li and Xuecheng Su and Xuheng Lin and Y. K. Li and Y. Q. Wang and Y. X. Wei and Y. X. Zhu and Yang Zhang and Yanhong Xu and Yanhong Xu and Yanping Huang and Yao Li and Yao Zhao and Yaofeng Sun and Yaohui Li and Yaohui Wang and Yi Yu and Yi Zheng and Yichao Zhang and Yifan Shi and Yiliang Xiong and Ying He and Ying Tang and Yishi Piao and Yisong Wang and Yixuan Tan and Yiyang Ma and Yiyuan Liu and Yongqiang Guo and Yu Wu and Yuan Ou and Yuchen Zhu and Yuduan Wang and Yue Gong and Yuheng Zou and Yujia He and Yukun Zha and Yunfan Xiong and Yunxian Ma and Yuting Yan and Yuxiang Luo and Yuxiang You and Yuxuan Liu and Yuyang Zhou and Z. F. Wu and Z. Z. Ren and Zehui Ren and Zhangli Sha and Zhe Fu and Zhean Xu and Zhen Huang and Zhen Zhang and Zhenda Xie and Zhengyan Zhang and Zhewen Hao and Zhibin Gou and Zhicheng Ma and Zhigang Yan and Zhihong Shao and Zhipeng Xu and Zhiyu Wu and Zhongyu Zhang and Zhuoshu Li and Zihui Gu and Zijia Zhu and Zijun Liu and Zilin Li and Ziwei Xie and Ziyang Song and Ziyi Gao and Zizheng Pan},
      year={2025},
      eprint={2412.19437},
      archivePrefix={arXiv},
      primaryClass={cs.CL},
      url={https://arxiv.org/abs/2412.19437},
}

@inproceedings{rajbhandari2020zero,
  title={Zero: Memory optimizations toward training trillion parameter models},
  author={Rajbhandari, Samyam and Rasley, Jeff and Ruwase, Olatunji and He, Yuxiong},
  booktitle={SC20: International Conference for High Performance Computing, Networking, Storage and Analysis},
  pages={1--16},
  year={2020},
  organization={IEEE}
}

@misc{zhao2023pytorch,
      title={PyTorch FSDP: Experiences on Scaling Fully Sharded Data Parallel},
      author={Yanli Zhao and Andrew Gu and Rohan Varma and Liang Luo and Chien-Chin Huang and Min Xu and Less Wright and Hamid Shojanazeri and Myle Ott and Sam Shleifer and Alban Desmaison and Can Balioglu and Pritam Damania and Bernard Nguyen and Geeta Chauhan and Yuchen Hao and Ajit Mathews and Shen Li},
      year={2023},
      eprint={2304.11277},
      archivePrefix={arXiv},
      primaryClass={cs.DC},
      url={https://arxiv.org/abs/2304.11277},
}

@inproceedings{khalilov2024network,
  title={Network-offloaded bandwidth-optimal broadcast and Allgather for distributed AI},
  author={Khalilov, Mikhail and Di Girolamo, Salvatore and Chrapek, Marcin and Nudelman, Rami and Bloch, Gil and Hoefler, Torsten},
  booktitle={SC24: International Conference for High Performance Computing, Networking, Storage and Analysis},
  pages={1--17},
  year={2024},
  organization={IEEE}
}

@article{hoefler2014energy,
  title={Energy, memory, and runtime tradeoffs for implementing collective communication operations},
  author={Hoefler, Torsten and Moor, Dmitry},
  journal={Supercomputing frontiers and innovations},
  volume={1},
  number={2},
  pages={58--75},
  year={2014}
}

@inproceedings{hoefler2007implementation,
  title={Implementation and performance analysis of non-blocking collective operations for MPI},
  author={Hoefler, Torsten and Lumsdaine, Andrew and Rehm, Wolfgang},
  booktitle={Proceedings of the 2007 ACM/IEEE conference on Supercomputing},
  pages={1--10},
  year={2007}
}

@article{kohler2000click,
  title={The Click modular router},
  author={Kohler, Eddie and Morris, Robert and Chen, Benjie and Jannotti, John and Kaashoek, M Frans},
  journal={ACM Transactions on Computer Systems (TOCS)},
  volume={18},
  number={3},
  pages={263--297},
  year={2000},
  publisher={ACM New York, NY, USA}
}

@inproceedings{dobrescu2009routebricks,
  title={RouteBricks: Exploiting parallelism to scale software routers},
  author={Dobrescu, Mihai and Egi, Norbert and Argyraki, Katerina and Chun, Byung-Gon and Fall, Kevin and Iannaccone, Gianluca and Knies, Allan and Manesh, Maziar and Ratnasamy, Sylvia},
  booktitle={Proceedings of the ACM SIGOPS 22nd symposium on Operating systems principles},
  pages={15--28},
  year={2009}
}

@inproceedings{panda2016netbricks,
  title={{NetBricks}: Taking the V out of {NFV}},
  author={Panda, Aurojit and Han, Sangjin and Jang, Keon and Walls, Melvin and Ratnasamy, Sylvia and Shenker, Scott},
  booktitle={12th USENIX Symposium on Operating Systems Design and Implementation (OSDI 16)},
  pages={203--216},
  year={2016}
}

@INPROCEEDINGS{wan2025generic,
  author={Wan, Xinchen and Li, Luyang and Tian, Han and Liao, Xudong and Huang, Xinyang and Zeng, Chaoliang and Wang, Zilong and Yang, Xinyu and Cheng, Ke and Ning, Qingsong and Liu, Guyue and Luo, Layong and Chen, Kai},
  booktitle={IEEE INFOCOM 2025 - IEEE Conference on Computer Communications},
  title={A Generic and Efficient Communication Framework for Message-Level In-Network Computing},
  year={2025},
  volume={},
  number={},
  pages={1-10},
  doi={10.1109/INFOCOM55648.2025.11044670}}

@inproceedings{xia2024accelerating,
  title={Accelerating and securing federated learning with stateless in-network aggregation at the edge},
  author={Xia, Junxu and Wu, Wenfei and Luo, Lailong and Cheng, Geyao and Guo, Deke and Nian, Qifeng},
  booktitle={2024 IEEE 44th International Conference on Distributed Computing Systems (ICDCS)},
  pages={692--702},
  year={2024},
  organization={IEEE}
}

@inproceedings{yang2022using,
  title={Using trio: juniper networks' programmable chipset-for emerging in-network applications},
  author={Yang, Mingran and Baban, Alex and Kugel, Valery and Libby, Jeff and Mackie, Scott and Kananda, Swamy Sadashivaiah Renu and Wu, Chang-Hong and Ghobadi, Manya},
  booktitle={Proceedings of the ACM SIGCOMM 2022 Conference},
  pages={633--648},
  year={2022}
}

@inproceedings{blocher2021switches,
  title={Switches for HIRE: Resource scheduling for data center in-network computing},
  author={Bl{\"o}cher, Marcel and Wang, Lin and Eugster, Patrick and Schmidt, Max},
  booktitle={Proceedings of the 26th ACM International Conference on Architectural Support for Programming Languages and Operating Systems},
  pages={268--285},
  year={2021}
}

@inproceedings{de2021flare,
  title={Flare: Flexible in-network allreduce},
  author={De Sensi, Daniele and Di Girolamo, Salvatore and Ashkboos, Saleh and Li, Shigang and Hoefler, Torsten},
  booktitle={Proceedings of the International Conference for High Performance Computing, Networking, Storage and Analysis},
  pages={1--16},
  year={2021}
}

@article{zhu2025dina,
  title={DINA: Toward Determined In-Network Aggregation for Distributed Machine Learning},
  author={Zhu, Haowen and Guo, Zehua and Ye, Minghao},
  journal={IEEE Transactions on Networking},
  year={2025},
  publisher={IEEE}
}

@article{fang2023grid,
  title={GRID: Gradient routing with in-network aggregation for distributed training},
  author={Fang, Jin and Zhao, Gongming and Xu, Hongli and Wu, Changbo and Yu, Zhuolong},
  journal={IEEE/ACM Transactions on Networking},
  volume={31},
  number={5},
  pages={2267--2280},
  year={2023},
  publisher={IEEE}
}

@inproceedings{zhao2023netrpc,
  title={{NetRPC}: Enabling {In-Network} computation in remote procedure calls},
  author={Zhao, Bohan and Wu, Wenfei and Xu, Wei},
  booktitle={20th USENIX symposium on networked systems design and implementation (NSDI 23)},
  pages={199--217},
  year={2023}
}

@inproceedings{xu2023clickinc,
author = {Xu, Wenquan and Zhang, Zijian and Feng, Yong and Song, Haoyu and Chen, Zhikang and Wu, Wenfei and Liu, Guyue and Zhang, Yinchao and Liu, Shuxin and Tian, Zerui and Liu, Bin},
title = {ClickINC: In-network Computing as a Service in Heterogeneous Programmable Data-center Networks},
year = {2023},
isbn = {9798400702365},
publisher = {Association for Computing Machinery},
address = {New York, NY, USA},
url = {https://doi.org/10.1145/3603269.3604835},
doi = {10.1145/3603269.3604835},
booktitle = {Proceedings of the ACM SIGCOMM 2023 Conference},
pages = {798–815},
numpages = {18},
location = {New York, NY, USA},
series = {ACM SIGCOMM '23}
}

@inproceedings{kim2024tccl,
  title={{TCCL}: Discovering Better Communication Paths for {PCIe GPU} Clusters},
  author={Kim, Heehoon and Ryu, Junyeol and Lee, Jaejin},
  booktitle={Proceedings of the 29th ACM International Conference on Architectural Support for Programming Languages and Operating Systems, Volume 3},
  pages={999--1015},
  year={2024}
}

@misc{nccl,
  title = {{NCCL}},
  author = {NVIDIA},
  note = {\url{https://github.com/NVIDIA/nccl}},
  year  = {2024}
}

@misc{rccl,
  title = {{RCCL}},
  author = {AMD},
  note = {\url{https://github.com/ROCm/rccl}},
  year  = {2024}
}

@misc{oneccl,
  title = {{oneAPI} Collective Communications Library ({oneCCL})},
  author = {Intel},
  note = {\url{https://github.com/oneapi-src/oneCCL}},
  year  = {2024}
}

@article{dong2021accl,
author = {Dong, Jianbo and Wang, Shaochuang and Feng, Fei and Cao, Zheng and Pan, Heng and Tang, Lingbo and Li, Pengcheng and Li, Hao and Ran, Qianyuan and Guo, Yiqun and Gao, Shanyuan and Long, Xin and Zhang, Jie and Li, Yong and Xia, Zhisheng and Song, Liuyihan and Zhang, Yingya and Pan, Pan and Wang, Guohui and Jiang, Xiaowei},
title = {ACCL: Architecting Highly Scalable Distributed Training Systems With Highly Efficient Collective Communication Library},
year = {2021},
issue_date = {Sept.-Oct. 2021},
publisher = {IEEE Computer Society Press},
address = {Washington, DC, USA},
volume = {41},
number = {5},
issn = {0272-1732},
url = {https://doi.org/10.1109/MM.2021.3091475},
doi = {10.1109/MM.2021.3091475},
journal = {IEEE Micro},
month = sep,
pages = {85–92},
numpages = {8}
}

@misc{msccl,
  title = {{MSCCL}},
  author = {Microsoft},
  note = {\url{https://github.com/microsoft/msccl}},
  year  = {2023}
}

@inproceedings{shah2023taccl,
  title={{TACCL}: Guiding Collective Algorithm Synthesis using Communication Sketches},
  author={Shah, Aashaka and Chidambaram, Vijay and Cowan, Meghan and Maleki, Saeed and Musuvathi, Madan and Mytkowicz, Todd and Nelson, Jacob and Saarikivi, Olli and Singh, Rachee},
  booktitle={20th USENIX Symposium on Networked Systems Design and Implementation (NSDI 23)},
  pages={593--612},
  year={2023}
}

@article{hidayetoglu2024hiccl,
  title={{HiCCL}: A Hierarchical Collective Communication Library},
  author={Hidayetoglu, Mert and de Gonzalo, Simon Garcia and Slaughter, Elliott and Surana, Pinku and Hwu, Wen-mei and Gropp, William and Aiken, Alex},
  journal={arXiv preprint arXiv:2408.05962},
  year={2024}
}

@misc{uec2024specupdate,
  author = {{Ultra Ethernet Consortium}},
  title  = {Ultra Ethernet Specification Update},
  year   = {2024},
  month  = {August},
  day    = {29},
  url    = {https://ultraethernet.org/ultra-ethernet-specification-update/},
  note   = {Accessed: 2026-02-06},
  howpublished = {Ultra Ethernet Consortium Blog}
}

@misc{broadcom_tomahawk5,
  author       = {{Broadcom Inc.}},
  title        = {{StrataXGS Tomahawk 5 Series: 51.2 Tb/s Ethernet Switch ASIC Family}},
  howpublished = {\url{https://www.broadcom.com/products/ethernet-connectivity/switching/strataxgs/bcm78920-series}},
  year         = {2026},
  note         = {Accessed: 2026-02-06}
}

@misc{omnetpp,
  title        = {OMNeT++ Discrete Event Simulator},
  author       = {{OMNeT++ Community}},
  howpublished = {\url{https://omnetpp.org}},
  year         = {2024},
  note         = {Version 6.2}
}

@misc{ns3,
  title        = {ns-3 Network Simulator},
  author       = {{nsnam}},
  howpublished = {\url{https://www.nsnam.org}},
  year         = {2024},
  note         = {Version 3.42}
}

@online{libpcap,
  author  = {{The Tcpdump Group}},
  title   = {{libpcap}: Portable packet capture library},
  url     = {https://www.tcpdump.org/},
  urldate = {2026-02-07}
}

@online{soft-roce,
  author  = {{Linux man-pages project}},
  title   = {\texttt{rxe}(7): Software RDMA over Ethernet (RoCE) driver},
  url     = {https://man7.org/linux/man-pages/man7/rxe.7.html},
  urldate = {2026-02-07},
  note    = {Documents Linux kernel module \texttt{rdma\_rxe} (Soft-RoCE/RXE)}
}

@inproceedings{openmpi,
  author    = {Edgar Gabriel and Graham E. Fagg and George Bosilca and Thara Angskun and Jack J. Dongarra and Jeffrey M. Squyres and Vishal Sahay and Prabhanjan Kambadur and Brian Barrett and Andrew Lumsdaine and Ralph H. Castain and David J. Daniel and Richard L. Graham and Timothy S. Woodall},
  title     = {Open {MPI}: Goals, Concept, and Design of a Next Generation {MPI} Implementation},
  booktitle = {Proceedings of the 11th European {PVM/MPI} Users' Group Meeting},
  address   = {Budapest, Hungary},
  month     = sep,
  year      = {2004}
}

@online{openstack,
  author  = {{OpenInfra Foundation}},
  title   = {{OpenStack}: Open source cloud computing infrastructure},
  url     = {https://www.openstack.org/},
  urldate = {2026-02-07}
}

@misc{libibverbs,
  author       = {{Linux RDMA Community}},
  title        = {{libibverbs}: Userspace {InfiniBand} Verbs Library},
  howpublished = {\url{https://github.com/linux-rdma/rdma-core/tree/master/libibverbs}},
  year         = {2024},
  note         = {Part of \texttt{rdma-core}; provides the \texttt{ibv\_*} API for RDMA device management, QP/CQ/MR operations, and data transfer}
}

@inproceedings{ovs,
  author    = {Ben Pfaff and Justin Pettit and Teemu Koponen and Ethan J. Jackson and Andy Zhou and Jarno Rajahalme and Jesse Gross and Alex Wang and Jonathan Stringer and Pravin Shelar and Keith Amidon and Martin Casado},
  title     = {The Design and Implementation of {Open vSwitch}},
  booktitle = {12th USENIX Symposium on Networked Systems Design and Implementation (NSDI '15)},
  year      = {2015},
  pages     = {117--130},
  publisher = {USENIX Association},
  url       = {https://www.usenix.org/system/files/conference/nsdi15/nsdi15-paper-pfaff.pdf}
}

\clearpage\newpage 
\appendix

\noindent Appendices are supporting material that has not been peer-reviewed.

\section{Semantics of Collectives}
\label{sec:app:collectives}

\noindent
\textbf{AllReduce.} Its semantics are to sum up (or perform other reduction operations on) data vectors from all ranks and distribute the result to all ranks. In \sysname IncTree, data streams from leaves to the root, being aggregated at each intermediate node and sent to the parent until reaching the root. The root streams the final result back, and intermediate nodes multicast the result to their children until it reaches the leaves (\cref{fig:aggtree-routing}a).

\noindent
\textbf{Reduce.} Its semantics are to sum up all ranks' data and deliver the result to one specific rank (the root). \sysname configures all non-receivers to stream data toward the receiver along the aggregation tree. Intermediate nodes aggregate incoming data streams into one and forward it toward the receiver. The receiver obtains the intermediate results and adds its own data to produce the final result (\cref{fig:aggtree-routing}b).

\noindent
\textbf{Broadcast.} Its semantics are to send one rank's data to all other ranks. In \sysname, the sender streams data into the tree, and intermediate nodes replicate the data, forwarding it to all neighbors except the incoming interface, ensuring all ranks eventually receive the data copy (\cref{fig:aggtree-routing}c).

\noindent
\textbf{Barrier.} Its semantics are to synchronize the state among members. It is performed as an AllReduce operation with an empty data payload, leveraging the synchronization property of the IncTree.

\noindent
\textbf{ReduceScatter.} Its semantics are to reduce data from all members, split the result, and scatter fragments to members. \sysname performs it as multiple Reduce operations.

\noindent
\textbf{AllGather.} Its semantics are to gather data from all members, concatenate it, and return the concatenated data to all members. \sysname performs this as multiple Broadcast operations.

\section{IncEngine Algorithms}
\label{sec:app:algorithms}

\begin{algorithm}[t]
\caption{Algorithm and control flow of Mode-II}
\label{alg:mode-II}

\small

\SetKwFunction{FLookup}{LookupTable}
\SetKwFunction{FCheck}{CheckDuplicate}
\SetKwFunction{FAgg}{AggregateData}
\SetKwFunction{FClear}{RecycleBuffer}
\SetKwFunction{FDuplicate}{ReplicateData}
\SetKwFunction{FTranslate}{TranslateHeader}
\SetKwFunction{FForward}{Forward}
\SetKwProg{Fn}{Function}{:}{}

\SetKwProg{Struct}{struct}{}{end}

\Struct{Pipe}{ 
    $payload$: Address of MTU array; \tcc{in SRAM, \cref{sec:resource-management}} 
    $degree$: Address of Int array; \tcc{in SRAM, \cref{sec:resource-management}} 
}


\Struct{EndPoint}{
    \tcp{configuration}
    $remote$: Endpoint\;
    $ip$: IP address;\ \ \  $qp$: QP Number;\ \ \  $port$: switch port\;
    \tcp{receive states}
    $arrived$: Array of Bit;  
}


\Struct{Context}{ 
    \tcp{persistent states}
    $N$: Int of array size;\  
    $eps$: Array of EndPoint;\  
    $pipe$: Pipe\;
    
    \tcp{invocation states}
    $collective$: Enum;\  
    $root$: Int;\ 
    $firstPsn$: Int;\  
    $lastPsn$: Int;
}

\While{receive a packet pkt}{
    (ctx, ep, class, fanin, outs) = \FLookup{pkt}\;
    pkt.idx = pkt.psn\%{ctx.N}\;
    pkt.idx2 = (pkt.psn+ctx.W)\%{ctx.N}\;
    \If{class = UP\_DATA}{
        isDup = \FCheck{pkt, ep}\;
        \lIf{not isDup}{
            \FAgg{ctx.pipe, pkt}
        }
        \lIf{ctx.pipe.degree[pkt.idx] < fanin}{
            \textbf{continue}
        }
        pkt.payload = ctx.pipe.payload[pkt.idx]\;
        \FClear{ctx.pipe, pkt.idx2, pkt.idx2 + 1}\;
    }
    
    \For{(pkt', out) $\in$ \FDuplicate{pkt, outs}}{
        pkt' = \FTranslate{pkt', out}\;
        \FForward{pkt', out.port}\;
    }
}


\Fn{\FCheck{pkt, ep}}{
    v=ep.arrived[pkt.idx]; \ \ ep.arrived[pkt.idx]=1\; 
    \Return{v==1}\;
}

\Fn{\FAgg{pipe, pkt}}{
    pipe.payload[pkt.idx] += pkt.payload\;
    pipe.degree[pkt.idx] += 1\;
}

\Fn{\FClear{pipe, start, end}}{
    \For{i $\in$ [start, end)}{ 
        pipe.payload[i]=pipe.degree[i]=0\;
    }
}
\end{algorithm}

\begin{algorithm}[t]
\caption{Algorithm and control flow of Mode-III (changes highlighted)}
\label{alg:mode-III}

\footnotesize
\SetKwFunction{FLookup}{LookupTable}
\SetKwFunction{FCheck}{CheckDuplicate}
\SetKwFunction{FAgg}{AggregateData}
\SetKwFunction{FClear}{\hl{RecycleBuffer}}
\SetKwFunction{FDuplicate}{ReplicateData}
\SetKwFunction{FTranslate}{TranslateHeader}
\SetKwFunction{FForward}{Forward}

\SetKwFunction{FReceiveAck}{\hl{ReceiveAck}}
\SetKwFunction{FSendAck}{\hl{SendAck}}
\SetKwFunction{FRetrans}{\hl{Retransmission}}
\SetKwFunction{FMakePkt}{MakePacket}
\SetKwFunction{FMin}{Min}
\SetKwFunction{FMax}{Max}

\SetKwProg{Fn}{Function}{:}{}

\SetKwProg{Struct}{struct}{}{end}
\SetKw{Continue}{continue}
\SetKw{Goto}{goto}
\SetKw{Forward}{FORWARD:}

\Struct{Pipe}{
    $payload$: Address of MTU array\; 
    $degree$: Address of Int array\; 
    \hl{$fromEps$: Array of EndPoint}\; 
    \hl{$toEps$: Array of EndPoint}\;
    \hl{$psnStart$: Int}\;
}
\Struct{EndPoint}{
    \tcp{configuration}
    $remote$: Endpoint\;
    $ip$: IP address;\  $qp$: QP Number;\  $port$: switch port\;
    \tcp{receive states}
    $arrived$: Array of Bit; \ \ \ \ 
    \hl{$epsn$: Int}; \ \ \ \ \hl{$fromPipe$: Pipe}\; 
    \tcp{send states}
    \hl{$resend$: Timer}; \ \ \hl{$lastAcked$: Int}\;
    \hl{$maxPsnSent$: Int}; \ \ \hl{$toPipe$: Pipe}\;
}
\Struct{Context}{
    $N$: Int of array size; \ \ \ \ $eps$: Array of EndPoint\;
    $collective$: Enum;\ $root$: Int;\ $first\_psn$: Int;\ $last\_psn$: Int\;
    \hl{$aggPipe$: Pipe}; \ \ \ \ \hl{$bcastPipe$: Pipe}\;
}


\While{receive a packet pkt}{
    (ctx, ep, class, fanin, outs) = LookupTable(pkt, table)\;
    pkt.idx = pkt.PSN \% ctx.N\;
    
    \If{pkt is ACK}{
        \FReceiveAck{pkt, ep}\;
        pipe = ep.fromPipe;\ \ eps = pipe.toEps\; 
        psnStart0 = pipe.psnStart\; 
        pipe.psnStart = \FMin{\{e.lastAcked $\mid$ e $\in$ eps\}} + 1\;
        \FClear{pipe, psnStart0, pipe.psnStart}\;
        \Continue\;
    }
    
    pipe = ep.toPipe;\ 
    toSend = $\emptyset$\;
    \If{pkt.psn $\ge$ pipe.psnStart + N}{
        \Continue;
    }
    
    isDup $\gets$ CheckDuplicate(pkt, ep)\;
    \lWhile{ep.arrived[ep.epsn] == 1}{
        ep.epsn++
    }
    {
    toSend.add( \FSendAck{pkt, ep} )\;
    }
    \lIf{isDup}{ \Goto \textbf{FORWARD}  }
    
    \FAgg{pkt, pipe}\;
    \lIf{pipe.degree[pkt.idx] < fanin}{ \Goto \textbf{FORWARD}  } 
    
    pkt.payload = pipe.payload[pkt.idx]\;
    \For{(pkt', out) $\in$ \FDuplicate{pkt, outs}}{
        pkt' = \FTranslate{pkt', out}\;
        out.maxPsnSent = \FMax{out.maxPsnSent, pkt'.psn}\;
        toSend.add(pkt', out.port)\; 
        \hl{out.resend.set(TIMEOUT)\;}
    }

    \textbf{FORWARD:} \\   
    \lFor{(pkt', port) $\in$ toSend}{
        \FForward{pkt', port}   
    }
}

\Fn{\FReceiveAck{pkt, ep}}{
    ep.lastAcked = \FMax{ep.lastAcked, pkt.psn}\;
    ep.resend.set( ep.maxPsnSent > ep.lastAcked ? TIMEOUT : 0 )\;
}
\Fn{\FSendAck{pkt, ep}}{
    pkt'=\FMakePkt{ep.ip, ep.remote.ip, ep.remote.qp, ep.epsn-1, ep.epsn-1==pkt.psn?ACK:NAK}\;
	\Return (pkt', ep.port)
}

\Fn{\FRetrans{timer, ep, ctx}}{
    \For{idx $\in$ [ep.lastAcked + 1, ep.maxPsnSent]}{
        \If{ep.fromPipe.degree[idx\%ctx.N] != fanin}{\Continue;}
        pkt = \FMakePkt{ep.ip, ep.remote.ip, ep.remote.qp, idx, ep.fromPipe.payload[idx \% ctx.N]}\;
        \FForward{pkt};\ 
        timer.set(TIMEOUT)\; 
    }
}
\end{algorithm}











Algorithm~\ref{alg:mode-II} shows the IncEngine algorithm for Mode-II, and Algorithm~\ref{alg:mode-III} shows that for Mode-III. Algorithm~\ref{alg:mode-III} can evolve from Algorithm~\ref{alg:mode-II} with the changes highlighted.

\section{Impact from RoCE's Reliability}
\label{sec:app:reliability}


Mode-II is compatible with both versions of RoCE reliability mechanisms. \textbf{(1)} Go-Back-N (GBN): Timeouts and NAKs trigger the sender to retransmit from the first missing PSN. \sysname remains idempotent when processing retransmissions. An optimization is to maintain an $epsn$ (expected PSN) in the switch endpoint state, dropping out-of-order packets immediately. This matches the receiver's logic (which would NAK out-of-order packets anyway) and reduces switch buffer access costs.
\textbf{(2)} Selective Retransmission (SR): SACKs describe ranges of out-of-order packets. \sysname switches and receiver buffers can handle out-of-order arrival. The complexity arises in Broadcast ACK aggregation, where merging SACK ranges requires computing the intersection of multiple intervals; if the switch does not support interval intersection, it can discard the SACK ranges, and the system falls back to GBN. Given the low packet loss rate in data centers and effective congestion control and flow control (e.g., credit-based or priority-based), these lossy corner cases are infrequent.


Mode-III does not confine the selection of the reliability mechanism. Both GBN and SR can work correctly in switch nodes. The module {\tt Retransmission} needs to adapt the corresponding packet sending to the selection.


\section{Other Possible IncEngine Modes} 
\label{sec:app:other-modes}

There could be other modes, representing a tradeoff between complexity, cost, and performance. For example, in Mode-II AllReduce, the switch node can buffer broadcasted result packets and return the result directly when data packet retransmission happens; however, this change consequently requires the upward data packet workflow to check whether the broadcast buffer has the result ready to send, which is workable but entails a bit more complexity. 

For another example, in Mode-II, if RoCE applies Go-back-N (GBN) in retransmission, the switch can also maintain an {\tt epsn} state (expected PSN) to optimize traffic; it filters all out-of-order packets to reduce network traffic (because out-of-order packets are dropped and retransmitted by the NIC). 
The design space is large, and all modes should be verified before being developed and deployed (\cref{subsec:correctness}).

\section{Switch Microarchitecture}
\label{sec:app:microarchitecture}

\begin{figure}[h]
    \centering
    \includegraphics[width=0.9\linewidth]{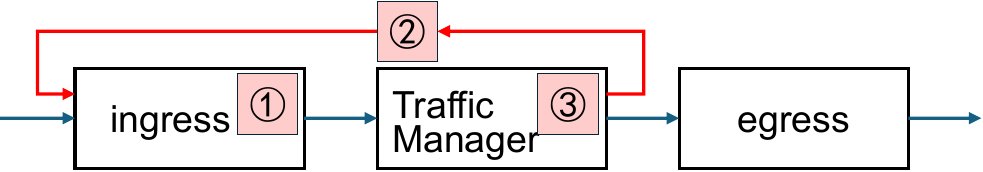}
    \caption{\sysname in switch micro architecture}
    \label{fig:microarchitecture}
\end{figure}


As shown in \cref{fig:microarchitecture}, modern high-performance switches typically utilize a pipelined architecture comprising an Ingress Pipeline, a Traffic Manager (TM), and an Egress Pipeline. The Ingress Pipeline performs packet parsing and forwarding decisions, while the TM serves as the data plane core, managing on-chip shared memory for buffering and scheduling. The Egress Pipeline completes the process with packet rewriting and encapsulation. Within this framework, the IncEngine can be integrated through three distinct architectural patterns: Ingress, Loopback, or Bypass.

Ingress Integration (\ding{172}) embeds the IncEngine logic directly into the ingress stages, allowing computation and modification to occur before packets reach the TM. This approach, exemplified by programmable targets like Tofino, enables line-rate processing with minimal latency. However, it is fundamentally constrained by pipeline resources, such as limited ALU stages, and requires packet recirculation to handle large payloads, which can significantly consume internal switching bandwidth.

Alternatively, Loopback Integration (\ding{173}) treats the IncEngine as a "look-aside" accelerator attached to a dedicated loopback port. Packets identified by the Ingress Pipeline are routed to this port for processing before being re-submitted to the ingress stage~\cite{wang2023roar}. While this mode offers high compatibility and supports complex operations without modifying the core TM logic, it inevitably increases latency due to multiple pipeline passes and places additional strain on internal switching capacity.

Finally, Bypass Integration (\ding{174}) positions the IncEngine as an auxiliary component connected directly to the on-chip shared memory via a dedicated interface. In this configuration, the Ingress Pipeline tags INC packets, triggering the TM to notify the engine to aggregate data directly from memory and write results back. This design balances the latency of moving data in and out of TM and the complexity of redesigning TM hardware.

\section{Model Checking}
\label{sec:app:model-checking}









\subsection{Implementation}

\begin{table}[h]
\centering
\caption{[Model Checking] LOC and Verification Time}
\label{tab:mc-loc}


\begin{tabular}{c c c c}

\hline
\textbf{Mode/Primitive} & \textbf{TLA+ LOC} & \textbf{Time} \\
\hline
Mode-II/AllReduce &  1222 & 20s\\
Mode-II/Reduce &  1139 & 3s\\
Mode-II/Broadcast  & 1108 & 10s \\
Mode-III/AllReduce & 1741 & 146min \\
Mode-III/Reduce  & 1394 & 9s \\
Mode-III/Broadcast  & 1386 & 113s \\


\hline

\end{tabular}
\end{table}


We build a model checker~\cite{bai2026verifying} based on TLA+ specification~\cite{tlaplus}. The model checker has its own specification language to simplify describing networked systems and overcomes the challenges of state explosion with state pruning.

We model the IncEngine logic of Mode-II\&III and verify them with the model checker. Our implementation covers three primitives—AllReduce, Reduce, and Broadcast. Table~\ref{tab:mc-loc} summarizes the lines of code (LOC) and the verification time.

\subsection{Experiment Settings}



\begin{table}[h]
\centering
\caption{[Model Checking] State Space Size}
\label{tab:mc-state}


\resizebox{\linewidth}{!}{
\begin{tabular}{c c c c}

\hline
\textbf{Mode/Primitive} & \textbf{Diameter} & \textbf{Total} & \textbf{Distinct} \\
\hline
Mode-II/AllReduce & 77 & 3981257 & 901703 \\
Mode-II/Reduce & 57 & 52105 & 14136 \\
Mode-II/Broadcast & 60 & 1763880 & 320128 \\
Mode-III/AllReduce & 115 & 2242781644 & 439476508 \\
Mode-III/Reduce & 65 & 2291727 & 447972 \\
Mode-III/Broadcast & 67 & 67902618 & 9911424 \\

\hline

\end{tabular}
}
\end{table}

We evaluate model checking efficiency on a workstation with 75 cores of \SI{2.3}{GHz} Intel(R) Xeon(R) Silver 4316 CPU and \SI{128}{GB} DRAM.
The network topology is set to be a Tree-3-2.
Each client sends three request packets. At most one packet loss may occur at any arbitrary moment.

\subsection{Experiment Results}

\parab{Efficiency.} Table~\ref{tab:mc-loc} and Table~\ref{tab:mc-state} show the verification time and state space size of model checking. The diameter refers to the maximum distance from the initial state to all reachable states. ``Total'' means the total count of explored states, while ``Distinct'' indicates the de-duplicated count.

The most time-consuming setting completes in \SI{146}{min}, and all other settings complete in \SI{3}{s}-\SI{113}{s}. We regard these execution times as acceptable, because they are a one-time cost during design, instead of runtime overhead.

\parab{Correctness.} We verified our final design under different system configurations. Results show that the design guarantees correctness under all possible concurrent executions, even in the presence of packet loss, out-of-order delivery, and duplication.

Since ReduceScatter and AllGather can be regarded as multiple independent Reduce and Broadcast operations, their correctness is guaranteed by this independence, together with the verified correctness of Reduce and Broadcast.

The topology is small but suffices to capture bugs. Due to node symmetry and state locality, a larger topology does not provide new protocol behaviors.

\subsection{Design Pitfalls}
\label{subsec:app:pitfalls}
We present some representative pitfalls of \sysname's historical design, discovered through model checking.

\parab{Iterate Designing Mode-III.} With the intuition that a per-hop reliable mode is possible, we conducted several trials. First, we add hop-by-hop acknowledgment and retransmission directly on Mode-II. The model checker reports that the switch clearing buffer (one-window away) could falsely erase data for next-hop transmission.

We then consider adding a broadcast buffer on each switch and using broadcast packets to clear the aggregation buffer and aggregated packets to clear the broadcast buffer. We later found that this solution consumes path BDP resources, which does not show an advantage over Mode-II. More critically, the model checker reports an error regarding the root design. If the root uses a single buffer, that buffer receives  UP\_DATA packets and sends DOWN\_DATA, causing the same problem as in the first trial; if the root uses separate buffers for aggregation and broadcast, the model checker reports an error in the design where the switch copies data from the aggregation buffer to the broadcast buffer upon aggregation success, because this copy could overwrite an unfinished broadcast transmission. The solution is that the root should generate a local \texttt{DOWN\_DATA} packet as if receiving from its parent, attempt to write to the broadcast buffer, and regenerate the packet on failure timeout.

Eventually, we recognize that ``a buffer slot can be cleared only if its outgoing endpoints receive ACKs'', confirming the next hop owning the data, and ``an incoming endpoint can only write the emptied buffer slot''. With these two intuitions, we propose the refined pipe abstraction, which maintains a valid PSN range to coordinate the incoming and outgoing endpoints. It can represent aggregation's many-to-one pattern, broadcast's one-to-many pattern, and the AllReduce root's many-to-many pattern.

\parab{NAK Necessity.}
In Mode‑III, the NAK mechanism is strongly suggested to be enabled; otherwise, a single packet loss cannot be handled quickly and can stall the entire aggregation tree and trigger timeout‑based retransmission in downstream nodes, degrading system performance.

The NAK mechanism also needs to be refined, as one packet loss could cause all subsequent packets to trigger NAKs, leading to redundant retransmissions. Given the limited processing capability of switches, we adopt the simplest implementation: maintaining a \texttt{nak\_sent} flag per flow. When an in‑order packet is received, the flag is set to \texttt{false}. When an out‑of‑order packet is received, the switch checks this flag—if it is \texttt{false}, the switch replies with a NAK and sets the flag to \texttt{true}; otherwise, the packet is silently dropped.

\begin{table*}[t]
\centering
\small
\caption{[Testbed, Tofino] Algorithm Throughput (Gbps) of Collectives (Barrier in queries per second (QPS)).}
\label{tab:app:testbed:six_collectives}
\setlength{\tabcolsep}{3pt}
\begin{tabular}{@{}c|c *{10}{|c} @{}}
\hline
\multirow{2}{*}{\textbf{Collective}} & \multirow{2}{*}{\textbf{Solution}} & \multicolumn{10}{c}{\textbf{Message Size (Byte)}} \\
\cline{3-12}
 & & \textbf{4K} & \textbf{16K} & \textbf{64K} & \textbf{256K} & \textbf{1M} & \textbf{4M} & \textbf{16M} & \textbf{64M} & \textbf{256M} & \textbf{1G} \\
\hline
\multirow{8}{*}{AllReduce}
 & \sysname-II    & 5.42 & 16.8 & 41.7 & 59.5 & 60.8 & 62.7 & 64.4 & 61.7 & 66.6 & 65.3 \\
 & SwitchML       & 3.85 & 13.1 & 32.7 & 52.3 & 61.6 & 64.4 & 65.1 & 65.4 & 65.4 & 65.4 \\
 & ATP 7-to-1     & 1.98 & 7.21 & 21.2 & 41.3 & 54.2 & 58.7 & 60.0 & 60.3 & 60.4 & 60.4 \\
 & ATP 8-to-1     & 1.92 & 6.47 & 15.9 & 25.1 & 29.3 & 30.5 & 30.9 & 31.0 & 31.0 & 30.9 \\
 & NCCL-Ring-1KB  & 0.015 & 0.058 & 0.243 & 0.910 & 3.47 & 13.3 & 29.1 & 41.0 & 46.3 & 48.0 \\
 & NCCL-Ring-256B & 0.015 & 0.060 & 0.251 & 0.820 & 3.71 & 12.1 & 26.6 & 36.0 & 39.7 & 41.1 \\
 & NCCL-Tree-1KB  & 0.015 & 0.060 & 0.244 & 0.890 & 3.17 & 11.1 & 24.9 & 31.9 & 33.5 & 34.1 \\
 & NCCL-Tree-256B & 0.015 & 0.060 & 0.223 & 0.860 & 3.12 & 10.1 & 21.4 & 27.4 & 29.0 & 29.3 \\
\hline
\multirow{3}{*}{Reduce}
 & \sysname-II & 4.19 & 16.6 & 37.2 & 55.6 & 71.6 & 74.4 & 67.7 & 66.3 & 68.4 & 66.7 \\
 & NCCL-1KB    & 0.016 & 0.065 & 0.263 & 1.02 & 3.87 & 14.3 & 40.7 & 65.0 & 67.1 & 67.3 \\
 & NCCL-256B   & 0.015 & 0.060 & 0.240 & 0.997 & 3.59 & 12.7 & 36.3 & 57.7 & 59.8 & 59.6 \\
\hline
\multirow{3}{*}{ReduceScatter}
 & \sysname-II & 3.97 & 15.6 & 39.6 & 62.5 & 61.4 & 74.3 & 66.6 & 71.6 & 67.2 & 66.3 \\
 & NCCL-1KB    & 0.014 & 0.056 & 0.218 & 0.926 & 3.34 & 12.9 & 41.1 & 72.3 & 89.6 & 95.6 \\
 & NCCL-256B   & 0.014 & 0.056 & 0.240 & 0.927 & 3.65 & 12.7 & 36.0 & 62.2 & 77.7 & 81.3 \\
\hline
\multirow{3}{*}{Broadcast}
 & \sysname-II & 4.45 & 16.1 & 41.2 & 72.9 & 86.2 & 89.8 & 90.9 & 92.1 & 88.7 & 89.0 \\
 & NCCL-1KB    & 0.015 & 0.061 & 0.240 & 0.913 & 3.52 & 12.9 & 40.0 & 65.6 & 66.6 & 66.6 \\
 & NCCL-256B   & 0.015 & 0.063 & 0.240 & 0.990 & 3.61 & 12.9 & 35.6 & 55.5 & 60.0 & 62.9 \\
\hline
\multirow{3}{*}{AllGather}
 & \sysname-II & 4.37 & 16.7 & 39.1 & 72.6 & 86.3 & 90.0 & 91.7 & 92.1 & 90.7 & 90.6 \\
 & NCCL-1KB    & 0.015 & 0.059 & 0.232 & 0.883 & 3.76 & 13.6 & 39.9 & 70.3 & 89.4 & 95.6 \\
 & NCCL-256B   & 0.014 & 0.058 & 0.235 & 0.885 & 3.70 & 12.3 & 37.3 & 62.0 & 74.2 & 80.7 \\
\hline
\multirow{2}{*}{Barrier}
 & \sysname-II & \multicolumn{10}{c}{165K QPS} \\
 & NCCL        & \multicolumn{10}{c}{0.458K QPS} \\
\hline
\end{tabular}
\end{table*}

\section{Testbed Evaluation}
\label{sec:app:testbed}

\subsection{Implementation}

\parab{Tofino.}
We implemented the \sysname server-side library in C++ and its switch logic in P4, managed by a lightweight Python-based control plane. The P4 program was compiled and deployed on a programmable Intel Tofino switch for hardware-level execution. 

When integrating \sysname CCL with PyTorch, the message buffer in \sysname is a registered RDMA buffer, and the application data is in the application buffer, not in the registered buffers. Copying data between buffers is costly. We also pipeline the memory copy with the RDMA message send/receive.


As the Tofino switch only handles integer numbers, when performing collectives on floating-point numbers, we (de)quantize the number with a fixed scaling factor; if value summation overflows, the switch rounds the value to the maximum integer value; by end-to-end tests, such (de)quantization does not affect training accuracy.
We use the same method to handle data (de)quantization as ATP~\cite{lao2021atp}.


\parab{NP (Network Processor).} We also develop IncEngine on a Huawei NetEngine 8000~\cite{sun2026rearchitecting,zheng2025p4}, which is an NP switch like Juniper Trio~\cite{yang2022using}. The switch has numerous network processor (NP) cores with shared SRAM. At runtime, the switch runs numerous parallel threads on the cores; it distributes packets to threads for processing and forwarding. The NP switch uses its own NPC++ language to write user programs. The language and the architecture allow for more flexible programmability than Tofino; for example, it supports loops and also enables timeout logic by allowing a thread to loop indefinitely and check the clock to trigger a timeout.

\subsection{Experiment Settings}

\parab{Workloads.} 
We evaluate the collective and application performance of \sysname. The environment settings are as \cref{subsec:settings}. For collective evaluation, we run all six collectives with message sizes ranging from \SI{4}{KB} to \SI{1}{GB} with a scaling factor of 4. For \sysname, collectives requiring payload aggregation (AllReduce, Reduce, and ReduceScatter) use a \SI{256}{B} packet payload due to limited switch aggregation hardware resources on Tofino, while Broadcast and AllGather use a \SI{1}{KB} payload that matches NCCL's default. For application evaluation, we run an LLM training job with tensor and data parallelism, as shown in \cref{tab:app:testbed:training}.


\parab{Baselines.} In the collective evaluation, we compare \sysname against SwitchML, ATP, and NCCL. We evaluate two ATP configurations: in ATP 7-to-1, one GPU serves as the Parameter Server (PS) to receive in-network aggregated data and perform local aggregation before broadcasting the results; ATP 8-to-1 follows a similar logic, but the PS also transmits data to the switch, emulating scenarios where the PS is placed on GPU workers. For NCCL, we vary the MTU (\SI{1}{KB} and \SI{256}{B}) and routing algorithms (Ring and Tree) to assess their impact. Note that the Tree algorithm is only supported by AllReduce; furthermore, we omit MTU and algorithm tuning for the Barrier collective as they do not affect its performance. Importantly, SwitchML and ATP only support the AllReduce collective.

For application evaluation, we utilize the same baselines, specifically using ATP 7-to-1 to represent ATP and NCCL-Ring-1KB to represent NCCL.




\begin{table}[t]
\caption{[Testbed] Training job iteration time (s), context\_length=2k, batch\_size=256, DP=8 / TP=8.}
\label{tab:app:testbed:training}

\setlength{\tabcolsep}{2pt}
\centering 
\small
\begin{tabular}{@{}c|c *{3}{|c} @{}}
\hline
\textbf{Settings} & \textbf{Solution} & \textbf{GPT-3 Large} & \textbf{OPT-350m} & \textbf{Llama-3.2-1B} \\
\hline
\multirow{4}{*}{DP=8}
 & \textbf{\sysname-II} & 16.7 & 12.2 & 24.5 \\
 & \textbf{SwitchML}    & 16.7 & 12.3 & 24.6 \\
 & \textbf{ATP}         & 16.8 & 12.3 & 24.7 \\
 & \textbf{NCCL}        & 16.9 & 12.4 & 24.8 \\
\hline
\multirow{4}{*}{TP=8}
 & \textbf{\sysname-II} & 28.3 & 20.0 & 35.1 \\
 & \textbf{SwitchML}    & 31.6 & 22.5 & 38.8 \\
 & \textbf{ATP}         & 34.5 & 24.8 & 42.1 \\
 & \textbf{NCCL}        & 38.1 & 27.5 & 46.1 \\
\hline
\end{tabular}
\end{table}
\begin{table}[t]
    \centering
    \caption{[Testbed] Tofino Resource Usage}
    \label{tab:testbed:resource}
  \small
  
\begin{tabular}{@{}c *{4}{|c} @{}}
\hline
    \textbf{SRAM}  & \textbf{128KB} & \textbf{512KB} & \textbf{2MB} & \textbf{8MB} \\
\hline
    Hash Bit     &  5.65$\%$  &  6.05$\%$   &  6.45$\%$ &  6.85$\%$ \\
    Gateway      & 22.92$\%$  & 22.92$\%$  & 22.92$\%$  & 22.92$\%$ \\
    SRAM         &  7.92$\%$  &  7.92$\%$   & 11.25$\%$ & 31.67$\%$ \\
    TCAM         &  1.39$\%$  &  1.39$\%$   &  1.39$\%$ &  1.39$\%$ \\  
    VLIW Instr   &  8.59$\%$  &  8.59$\%$   &  8.59$\%$ &  8.59$\%$ \\
    Map RAM      & 12.33$\%$  & 12.33$\%$   & 17.88$\%$ & 51.91$\%$ \\
    Meter ALU    & 72.92$\%$  & 72.92$\%$   & 72.92$\%$ & 72.92$\%$ \\
    PHV          & 34.80$\%$  & 34.80$\%$   & 34.80$\%$ & 34.80$\%$ \\
\hline
    \end{tabular}
\end{table}

\begin{table*}[t]
\centering
\small
\caption{[Testbed, NP] Throughput of full-function collectives (Gbps; Barrier in QPS).}
\label{tab:app:testbed:six_collectives_np_full_function}
\setlength{\tabcolsep}{3pt}
\begin{tabular}{@{}c|c *{10}{|c} @{}}
\hline
\multirow{2}{*}{\textbf{Collective}} &
\multirow{2}{*}{\textbf{Solution}} &
\multicolumn{10}{c}{\textbf{Message Size (Byte)}} \\
\cline{3-12}
 & & \textbf{4K} & \textbf{16K} & \textbf{64K} & \textbf{256K} &
\textbf{1M} & \textbf{4M} & \textbf{16M} & \textbf{64M} & \textbf{256M} & \textbf{1G} \\
\hline
\multirow{2}{*}{AllReduce}
 & \sysname-II  & 1.12 & 2.63 & 9.40 & 10.5 & 11.2 & 12.3 & 12.8 & 12.4 & 12.1 & 11.7 \\
 & \sysname-III & 2.29 & 3.95 & 10.4 & 11.9 & 12.4 & 12.8 & 12.7 & 12.0 & 13.8 & 12.6 \\
\hline
\multirow{2}{*}{Reduce}
 & \sysname-II  & 1.56 & 3.06 & 9.89 & 11.3 & 11.0 & 11.9 & 12.8 & 12.7 & 12.4 & 12.5 \\
 & \sysname-III & 3.30 & 5.09 & 10.9 & 11.3 & 11.1 & 12.1 & 12.1 & 12.9 & 13.0 & 12.5 \\
\hline
\multirow{2}{*}{ReduceScatter}
 & \sysname-II  & 2.44 & 4.60 & 10.9 & 11.2 & 12.6 & 11.3 & 11.6 & 12.1 & 12.2 & 11.8 \\
 & \sysname-III & 3.11 & 5.42 & 10.4 & 11.6 & 11.9 & 12.4 & 11.6 & 11.4 & 12.1 & 12.5 \\
\hline
\multirow{2}{*}{Broadcast}
 & \sysname-II  & 2.39 & 8.62 & 26.5 & 47.4 & 63.6 & 69.0 & 71.4 & 72.9 & 73.1 & {72.3} \\
 & \sysname-III & 4.70 & 9.70 & 28.7 & 52.6 & 65.9 & 69.1 & 71.7 & 73.3 & 73.2 & 72.8 \\
\hline
\multirow{2}{*}{AllGather}
 & \sysname-II  & 3.24 & 12.1 & 34.3 & 62.4 & 81.7 & 90.1 & 92.2 & 94.8 & 94.2 & 94.5 \\
 & \sysname-III & 6.53 & 11.8 & 37.5 & 67.5 & 86.4 & 89.6 & 94.1 & 94.8 & 94.2 & 93.9 \\
\hline
\multirow{2}{*}{Barrier}
 & \sysname-II  & \multicolumn{10}{c}{72992 QPS} \\
 & \sysname-III & \multicolumn{10}{c}{86956 QPS} \\
\hline
\end{tabular}
\end{table*}

\begin{table*}[t]
\centering
\small
\caption{[Testbed, NP] Throughput of partial-function collectives (Gbps; Barrier in QPS).}
\label{tab:app:testbed:six_collectives_np_no_sram_access}
\setlength{\tabcolsep}{3pt}
\begin{tabular}{@{}c|c *{10}{|c} @{}}
\hline
\multirow{2}{*}{\textbf{Collective}} &
\multirow{2}{*}{\textbf{Solution}} &
\multicolumn{10}{c}{\textbf{Message Size (Byte)}} \\
\cline{3-12}
 & & \textbf{4K} & \textbf{16K} & \textbf{64K} & \textbf{256K} &
\textbf{1M} & \textbf{4M} & \textbf{16M} & \textbf{64M} & \textbf{256M} & \textbf{1G} \\
\hline
\multirow{2}{*}{AllReduce}
 & \sysname-II  & 1.13 & 6.87 & 25.0 & 49.9 & 66.0 & 71.9 & 72.6 & 72.7 & 71.5 & 72.4 \\
 & \sysname-III & 2.19 & 9.11 & 27.3 & 52.9 & 66.7 & 71.8 & 71.0 & 72.7 & 72.3 & 72.5 \\
\hline
\multirow{2}{*}{Reduce}
 & \sysname-II  & 1.56 & 6.06 & 25.9 & 46.9 & 64.0 & 70.9 & 72.8 & 72.7 & 71.4 & 73.6 \\
 & \sysname-III & 4.30 & 10.1 & 28.9 & 51.3 & 66.1 & 72.1 & 73.1 & 72.9 & 72.3 & 72.5 \\
\hline
\multirow{2}{*}{ReduceScatter}
 & \sysname-II  & 3.57 & 8.25 & 36.6 & 61.6 & 84.3 & 90.2 & 95.3 & 94.2 & 94.1 & 92.1 \\
 & \sysname-III & 6.69 & 12.2 & 39.1 & 65.3 & 86.7 & 92.1 & 96.2 & 94.3 & 95.6 & 92.8 \\
\hline
\multirow{2}{*}{Broadcast}
 & \sysname-II  & 2.39 & 8.62 & 26.5 & 47.4 & 63.6 & 69.0 & 71.4 & 72.9 & 73.1 & \textbf{72.3} \\
 & \sysname-III & 4.70 & 9.70 & 28.7 & 52.6 & 65.9 & 69.1 & 71.7 & 73.3 & 73.2 & 72.8 \\
\hline
\multirow{2}{*}{AllGather}
 & \sysname-II  & 3.24 & 12.1 & 34.3 & 62.4 & 81.7 & 90.1 & 92.2 & 94.8 & 94.2 & 94.5 \\
 & \sysname-III & 6.53 & 11.8 & 37.5 & 67.5 & 86.4 & 89.6 & 94.1 & 94.8 & 94.2 & 93.9 \\
\hline
\multirow{2}{*}{Barrier}
 & \sysname-II  & \multicolumn{10}{c}{72992 QPS} \\
 & \sysname-III & \multicolumn{10}{c}{86956 QPS} \\
\hline
\end{tabular}
\end{table*}

\subsection{Performance on Tofino}

\parab{Collective Communication.}
\cref{tab:app:testbed:six_collectives} illustrates the performance of the collective operation. For small messages, \sysname outperforms the others because its RDMA latency is lower than the DPDK latency used by SwitchML and ATP. ATP's Parameter Server (PS) introduces additional network hops, resulting in slightly higher latency compared to SwitchML. NCCL exhibits the highest latency and lowest bandwidth due to significant server-side overhead.

For large messages, all three in-network computing schemes demonstrate higher bandwidth than NCCL by compressing network traffic. Furthermore, \sysname's RDMA achieves slightly higher network bandwidth utilization than DPDK. In the ATP 8-to-1 configuration, a bottleneck occurs because the PS must simultaneously transmit data and results to the switch; consequently, the bandwidth is halved compared to ATP 7-to-1, falling even below that of NCCL.

For other collectives delivering small messages, NCCL continues to underperform compared to \sysname due to its control latency.
For large messages, \sysname runs Reduce and Broadcast faster than NCCL because INC saves on traffic volume as well as worker I/O overhead; \sysname runs Broadcast faster than Reduce because the former does not cause computation on the switch. When running ReduceScatter and AllGather, \sysname runs them as sequential Reduces and Broadcasts and thus achieves similar performance; however, NCCL runs Reduces and Broadcasts in parallel and achieves higher throughput. Even so, \sysname uses less CPU for I/O and generates less traffic volume, and it can also be implemented in parallel mode (with switch memory split to Reduces/Broadcast).




\parab{Application Performance.}
The application-level evaluation in \cref{tab:app:testbed:training} yields results consistent with those observed in the collective communication analysis, with the performance gain effectively reducing the communication time within the end-to-end application execution time. INC schemes outperform NCCL, with \sysname specifically demonstrating superiority over SwitchML and ATP in leveraging RoCE hardware to offload transmission.

For Data Parallelism (DP), the performance gap is relatively small because communication accounts for a smaller fraction of the total execution time in 3D parallel training. Conversely, for Tensor Parallelism (TP), the performance differences are more pronounced due to the higher proportion of communication overhead.





\parab{Overhead.}
\cref{tab:testbed:resource} illustrates \sysname's switch resource utilization. As the total size of the INC buffer increases, the SRAM consumption grows significantly.

\subsection{Performance on NP}
\label{ssec:np-performance}


\cref{tab:app:testbed:six_collectives_np_full_function} shows the throughput of collectives on NP. Take AllReduce as an example, the per-rank throughput is not high (\SI{11.7}{Gbps}/ \SI{12.6}{Gbps} for \SI{1}{GB} message in Mode-II/III). The NP router's architecture contains many cores with a shared SRAM; when multiple cores concurrently process packets, INC's workflow requires them to access SRAM for payload aggregation, where the concurrent accesses cause contention and the system is bottlenecked at the SRAM access bandwidth.

We comment out the payload operations (aggregation and/or write back) in the program, and show the system performance in \cref{tab:app:testbed:six_collectives_np_no_sram_access}. The AllReduce throughput reaches \SI{72.4}{Gbps} and \SI{72.5}{Gbps}, respectively for Mode-II and Mode-III.
The result indicates that the control logic of \sysname-II/III is correct, but the system bottleneck is the SRAM access bandwidth. It also reveals that future INC switches should overcome the challenge of designing high-throughput SRAM access for vector-summation in INC.




\section{Emulation}
\label{sec:app:emulation}

\subsection{Implementation}

We implement six collective communication primitives 
across three \sysname modes on a virtual-machine testbed using \texttt{libibverbs}~\cite{libibverbs} and \texttt{libpcap}~\cite{libpcap}, demonstrating the interoperability between \sysname and RDMA verbs.

\parab{\sysname-I.}
In Mode-I, the switch establishes native RDMA Queue Pair (QP) connections via the \texttt{libibverbs} API and performs aggregation at the message level upon receiving complete RDMA messages. The RDMA transport layer handles all reliability semantics natively, eliminating the need for custom PSN tracking or ACK generation. The switch application layer maintains a set of aggregation slots, which are reused through modular indexing. Control information is conveyed via the 32-bit Immediate Data field encoded as \texttt{[slot\_id:20][primitive:2][op:2][sender:4][root:4]}. The 
switch routes messages according to its role (\textsc{Leaf} or \textsc{Root}) and the primitive type: a \textsc{Leaf} switch aggregates data from its local hosts before forwarding it to the parent switch, while the \textsc{Root} switch aggregates contributions from all child switches and broadcasts the result back down the tree.

\parab{\sysname-II.}
In Mode-II, the switch operates as a \emph{transparent middlebox}, intercepting RoCEv2 packets via \texttt{libpcap}, performing payload aggregation and header rewriting before forwarding without terminating the RDMA connection. The host RDMA stack retains full responsibility for reliability. The switch maintains the context. The core state array is indexed by $\texttt{PSN} \bmod \texttt{SWITCH\_ARRAY\_LENGTH}$, and a 32-bit \texttt{arrival\_state} bitmap tracks the arrival status of each connection. The I/O framework in Mode-II is built upon a single-threaded event loop using Linux \texttt{epoll}. Each physical network interface is assigned a \texttt{pcap} handle (configured in promiscuous mode, non-blocking, with a \SI{256}{MB} kernel buffer) and registered with \texttt{epoll}. The main loop processes up to 256 packets per device per iteration, each packet dispatched to a P4-style three-stage \texttt{pipeline}: \emph{parser} (extracting ETH/IP/UDP/BTH headers) $\rightarrow$ \emph{match} (determining direction and PSN window membership) $\rightarrow$ \emph{action} (primitive-specific logic).

\parab{\sysname-III.}
In Mode-III, the switch \textbf{terminates RDMA connections}---from the host's perspective, the switch acts as an endpoint with hop-by-hop reliability. It intercepts raw Ethernet frames via \texttt{libpcap}, parses the complete RoCEv2 protocol stack (ETH $\rightarrow$ IPv4 $\rightarrow$ UDP $\rightarrow$ BTH $\rightarrow$ Payload $\rightarrow$ ICRC), performs aggregation at per-PSN granularity, and constructs new packets for downstream broadcast. The switch manages the PSN space, generates ACK/NAK responses, and implements timeout-based retransmission.

\parab{Baseline.} In addition to the three modes described above, we deploy OpenMPI~4.1.5~\cite{openmpi} in the same environment as a baseline.
\emph{MPI} routes traffic through a simple L3 forwarding switch (\texttt{soft\_switch}) that relays IP packets between subnets via \texttt{libpcap} without any payload inspection or modification.


\subsection{Experiment Settings}

All three modes are evaluated on the same Tree-3-2 topology, deployed across eight OpenStack~\cite{openstack} virtual machines interconnected by an OVS virtual network. RDMA transport uses Soft-RoCE~\cite{soft-roce} (the \texttt{rdma\_rxe} kernel module). Message sizes increase by a factor of four, covering both small and large transfers; the payload per packet is 4KB. The sliding window, aggregation slot size, and CQ size are all configured large enough to avoid performance bottlenecks. Each experiment is run 10 times, and we report the average throughput. The startup order is: Spine (ROOT) $\rightarrow$ Leaf 1 $\rightarrow$ Leaf 2 $\rightarrow$ four hosts in parallel.

\subsection{Development Workload}


\begin{table}[t]
\centering
\small
\caption{Development Workload}
\label{tab:dev-workload}
\begin{tabular}{@{}c|c|c|c|c|c@{}}
\hline
\textbf{Mode} &
\textbf{Switch} &
\textbf{Host} &
\multicolumn{2}{c|}{\textbf{Misc.}} &
\textbf{Total} \\
\hline
\sysname-I & 801 & 499 & 466 & \multirow{3}{*}{587} & 2353 \\
\cline{1-4}\cline{6-6}
\sysname-II & 1246 & \multirow{2}{*}{970} & / & & 2803 \\
\cline{1-2}\cline{4-4}\cline{6-6}
\sysname-III & 1184 & & / & & 2741 \\
\hline
\end{tabular}
\end{table}


Table~\ref{tab:dev-workload} counts only the clean VM emulation implementation, excluding tests, scripts, topology setup files, and baseline implementations. \sysname-I places the host and switch roles in one RDMA endpoint program; we separate the role-specific code and count role-independent helpers as miscellaneous code. \sysname-II and \sysname-III share the 970-LOC host runtime and the 587-LOC packet/protocol library, but use different switch-side ACK paths. \sysname-II implements non-termination by reflecting or aggregating downstream ACKs before releasing upstream senders, while \sysname-III terminates ACKs at each hop. Thus, \sysname-II and \sysname-III reuse 1,557 LOC, and all three modes share 587 LOC of packet parsing and protocol definitions.

\subsection{Performance}

\begin{table*}[t]
\centering
\small
\caption{[Emulation] Throughput of collectives (Mbps; Barrier in QPS).}
\label{tab:app:emulation:six_collectives}
\setlength{\tabcolsep}{3pt}
\begin{tabular}{@{}c|c *{9}{|c} @{}}
\hline
\multirow{2}{*}{\textbf{Collective}} &
\multirow{2}{*}{\textbf{Solution}} &
\multicolumn{9}{c}{\textbf{Message Size (Byte)}} \\
\cline{3-11}
 & & \textbf{4K} & \textbf{16K} & \textbf{64K} & \textbf{256K} &
\textbf{1M} & \textbf{4M} & \textbf{16M} & \textbf{64M} & \textbf{256M} \\
\hline
\multirow{4}{*}{AllReduce}
 & \sysname-I   & 112 & 317 & 596 & 744 & 787 & 876 & 915 & 1015 & 1177 \\
 & \sysname-II  & 110 & 202 & 361 & 355 & 447 & 485 & 483 & 529  & 559  \\
 & \sysname-III & 159 & 256 & 367 & 420 & 464 & 495 & 512 & 616  & 661  \\
 & MPI          & 74.9 & 186 & 331 & 346 & 496 & 519 & 515 & 508  & 517  \\
\hline
\multirow{4}{*}{Reduce}
 & \sysname-I   & 130 & 437 & 893 & 1327 & 1469 & 1570 & 1621 & 1710 & 1724 \\
 & \sysname-II  & 125 & 322 & 565 & 656  & 729  & 787  & 749  & 744  & 900  \\
 & \sysname-III & 230 & 567 & 925 & 840  & 862  & 1037 & 847  & 1042 & 1239 \\
 & MPI          & 64.5 & 234 & 523 & 732  & 796  & 852  & 846  & 853  & 834  \\
\hline
\multirow{4}{*}{ReduceScatter}
 & \sysname-I   & 104 & 355 & 870 & 1240 & 1586 & 1549 & 1571 & 1581 & 1590 \\
 & \sysname-II  & 86.2 & 302 & 566 & 631  & 695  & 773  & 729  & 782  & 851  \\
 & \sysname-III & 115 & 395 & 563 & 694  & 745  & 786  & 834  & 971  & 1226 \\
 & MPI          & 78.8 & 275 & 561 & 635  & 884  & 986  & 907  & 902  & 920  \\
\hline
\multirow{4}{*}{Broadcast}
 & \sysname-I   & 145 & 383  & 867 & 1044 & 1307 & 1798 & 2047 & 2033 & 1949 \\
 & \sysname-II  & 119 & 297  & 485 & 572  & 614  & 637  & 832  & 843  & 894  \\
 & \sysname-III & 183 & 309  & 690 & 705  & 725  & 840  & 842  & 894  & 957  \\
 & MPI          & 266 & 1102 & 829 & 732  & 988  & 756  & 744  & 831  & 841  \\
\hline
\multirow{4}{*}{AllGather}
 & \sysname-I   & 109 & 346 & 847 & 1363 & 1368 & 1470 & 1547 & 1811 & 2267 \\
 & \sysname-II  & 104 & 391 & 614 & 775  & 866  & 888  & 908  & 1249 & 1351 \\
 & \sysname-III & 130 & 481 & 592 & 761  & 882  & 935  & 918  & 1102 & 1381 \\
 & MPI          & 89.8 & 247 & 579 & 674  & 833  & 953  & 958  & 955  & 955  \\
\hline
\multirow{4}{*}{Barrier}
 & \sysname-I   & \multicolumn{9}{c}{4033 QPS} \\
 & \sysname-II  & \multicolumn{9}{c}{4499 QPS} \\
 & \sysname-III & \multicolumn{9}{c}{5679 QPS} \\
 & MPI          & \multicolumn{9}{c}{4170 QPS} \\
\hline
\end{tabular}
\end{table*}

Table~\ref{tab:app:emulation:six_collectives} shows the collective performance of \sysname and the baseline. Note that the emulation is designed to demonstrate the interoperability between \sysname switches and RoCE NICs. All system components run on the same server, sharing CPU cores; the performance numbers do not represent the system performance running on multiple hardware components.



\parab{AllReduce.}
Mode-I achieves the highest throughput among the three modes, reaching \SI{1177}{Mbps} at \SI{256}{MB}---roughly $2\times$ that of Mode-II and Mode-III.
This advantage stems from message-level aggregation (\SI{64}{KB} per RDMA message), which reduces the number of switch-side I/O operations by up to $64\times$ compared with the per-packet modes, along with the associated CPU overhead.
Mode-I is faster than Mode-II and Mode-III because it uses RDMA directly for message-level transmission on the switch side, whereas the latter two rely on libpcap for per-packet processing and thus incur higher CPU overhead.
Mode-II and Mode-III achieve \SI{559}{Mbps} and \SI{661}{Mbps} respectively, reflecting the per-packet processing overhead of \texttt{libpcap}-based interception.
Mode-III slightly outperforms Mode-II because its connection-termination design enables independent PSN management and avoids head-of-line blocking from the host retransmission timer.
All three modes outperform MPI because in-network aggregation reduces data transmission. 

\parab{Reduce.}
Mode-I again leads, peaking at \SI{1724}{Mbps} at \SI{256}{MB}.
The unidirectional nature of Reduce (no broadcast phase) allows Mode-I to approach the raw RDMA link rate.
Mode-II, Mode-III, and MPI plateau at \SI{900}{Mbps}, \SI{1239}{Mbps}, and \SI{834}{Mbps} respectively, consistent with the switch-side per-packet processing ceiling imposed by libpcap, as observed in AllReduce.

\parab{ReduceScatter.}
With a computation load similar to Reduce, ReduceScatter shows comparable performance.
Mode-I, Mode-II, Mode-III, and MPI achieve bandwidths of \SI{1590}{Mbps}, \SI{851}{Mbps}, \SI{1226}{Mbps}, and \SI{920}{Mbps}, respectively.

\parab{Broadcast.}
Broadcast has the same communication volume as Reduce but transfers data in the opposite direction.
Without aggregation computation, it achieves higher performance than Reduce while preserving the same ranking across modes.
Mode-I, Mode-II, Mode-III, and MPI achieve bandwidths of \SI{1949}{Mbps}, \SI{894}{Mbps}, \SI{957}{Mbps}, and \SI{841}{Mbps}, respectively.

\parab{AllGather.}
AllGather has the same communication volume as ReduceScatter but transfers data in the opposite direction. Without aggregation computation, it achieves higher performance than ReduceScatter while preserving the same ranking across modes. Mode-I, Mode-II, Mode-III, and MPI achieve bandwidths of \SI{2267}{Mbps}, \SI{1351}{Mbps}, \SI{1381}{Mbps}, and \SI{955}{Mbps}, respectively.

\parab{Barrier.}
The Barrier QPS of the three EPIC modes is close to their small-message AllReduce performance, reaching 4,033, 4,499, and 5,679 QPS for Mode-I, Mode-II, and Mode-III, respectively.
MPI uses a dedicated Barrier topology instead of reusing AllReduce, achieving 4,170 QPS, which is comparable to Mode-I and Mode-II.

\parab{Mode-I outperforms Mode-II/III at large message sizes.}
The throughput ceiling of Mode-II and Mode-III is fundamentally limited by the \texttt{libpcap} user-space packet processing path.
Every RoCEv2 packet traverses the kernel capture buffer, crosses the kernel--user boundary, and is parsed, aggregated, header-rewritten, ICRC-recomputed, and re-injected---all in software.
With a per-packet payload of only \SI{1024}{B} (Soft-RoCE active MTU), a \SI{256}{MB} transfer generates over 256K packets per connection, and the switch must process each one individually.
This per-packet overhead establishes a hard throughput ceiling that cannot be overcome by increasing the message size.

Mode-I bypasses this bottleneck entirely by operating at the \emph{message} level.
The switch establishes native RDMA QP connections and receives complete 64\,KB messages via the kernel RDMA stack, reducing the number of switch-side I/O operations by $64\times$ compared with Mode-II/III.
Aggregation is performed on whole messages rather than individual packets, and the RDMA transport layer handles segmentation, reassembly, and reliability transparently.
As a result, Mode-I scales to \SI{1177}{Mbps} (AllReduce) and \SI{1724}{Mbps} (Reduce), approaching the effective Soft-RoCE link capacity.

\begin{table*}[t]
\centering
\small
\caption{[Packet Simulation] Algorithm Throughput (Gbps) of Collectives (Barrier in QPS).}
\label{tab:app:pkt-sim:six_collectives}
\setlength{\tabcolsep}{3pt}
\begin{tabular}{@{}c|c *{10}{|c} @{}}
\hline
\multirow{2}{*}{\textbf{Collective}} & \multirow{2}{*}{\textbf{Solution}} & \multicolumn{10}{c}{\textbf{Message Size (Byte)}} \\
\cline{3-12}
 & & \textbf{4K} & \textbf{16K} & \textbf{64K} & \textbf{256K} & \textbf{1M} & \textbf{4M} & \textbf{16M} & \textbf{64M} & \textbf{256M} & \textbf{1G} \\
\hline
\multirow{3}{*}{AllReduce}
 & \sysname-II  & 6.10 & 20.4 & 48.8 & 75.1 & 84.9 & 87.5 & 88.09 & 88.3 & 88.3 & 88.3 \\
 & \sysname-III & 10.2 & 30.5 & 61.0 & 81.4 & 88.8 & 89.0 & 89.73 & 89.9 & 89.9 & 89.9 \\
 & Ring         & 1.1  & 4.1  & 13.2 & 30.0 & 44.1 & 50.0 & 51.7  & 52.1 & 52.2 & 52.3
 \\
\hline
\multirow{3}{*}{Reduce}
 & \sysname-II  & 10.2 & 30.5 & 61.0 & 81.4 & 88.8 & 90.8 & 91.3 & 91.5 & 91.5 & 91.5 \\
 & \sysname-III & 10.2 & 30.5 & 61.0 & 81.4 & 88.8 & 90.8 & 91.3 & 91.5 & 91.5 & 91.5 \\
 & Chain        & 1.8  & 5.1  & 9.4  & 19.5 & 47.6 & 74.4 & 86.5 & 90.2 & 91.2 & 91.4
 \\
\hline
\multirow{3}{*}{ReduceScatter}
 & \sysname-II  & 10.2 & 30.5 & 61.0 & 81.4 & 88.8 & 90.8 & 91.3 & 91.5 & 91.5 & 91.5 \\
 & \sysname-III & 10.2 & 30.5 & 61.0 & 81.4 & 88.8 & 90.8 & 91.3 & 91.5 & 91.5 & 91.5 \\
 & Ring         & 2.2  & 8.1  & 27.1 & 61.0 & 88.8 & 100.2 & 103.4 & 104.3 & 104.5 & 104.6
 \\
\hline
\multirow{3}{*}{Broadcast}
  & \sysname-II  & 10.2 & 30.5 & 61.0 & 81.4 & 88.8 & 90.8 & 91.3 & 91.5 & 91.5 & 91.5 \\
 & \sysname-III & 10.2 & 30.5 & 61.0 & 81.4 & 88.8 & 90.8 & 91.3 & 91.5 & 91.5 & 91.5 \\
 & Chain        & 1.8  & 5.1  & 9.4  & 19.5 & 47.6 & 74.4 & 86.5 & 90.2 & 91.2 & 91.4
 \\
\hline
\multirow{3}{*}{AllGather}
 & \sysname-II  & 10.2 & 30.5 & 61.0 & 81.4 & 88.8 & 90.8 & 91.3 & 91.5 & 91.5 & 91.5 \\
 & \sysname-III & 10.2 & 30.5 & 61.0 & 81.4 & 88.8 & 90.8 & 91.3 & 91.5 & 91.5 & 91.5 \\
 & Ring         & 2.2 & 8.1 & 27.1 & 61.0 & 88.8 & 100.2 & 103.4 & 104.3 & 104.5 & 104.6
 \\
\hline
\multirow{3}{*}{Barrier}
 & \sysname-II  & \multicolumn{9}{c}{200K QPS} \\
 & \sysname-III & \multicolumn{9}{c}{333K QPS} \\
 & ToRank0      & \multicolumn{9}{c}{142K QPS} \\
\hline
\end{tabular}
\end{table*}



\begin{table*}[h]
\centering
\caption{[Packet Simulation] Throughput vs Loss Rate}
\label{tab:app:pkt-sim:tput-loss-rate}
\small
\begin{tabular}{@{}c *{8}{|c} @{}}
\hline
\textbf{Loss Rate @ 1 link (\%)} & \textbf{0\%} & \textbf{0.001\%} & \textbf{0.003\%} & \textbf{0.01\%} & \textbf{0.03\%} & \textbf{0.1\%} & \textbf{0.3\%} & \textbf{1\%}  \\
\hline
\sysname-II  & 88.3 & 88.3 & 88.2 & 88.1 & 87.6 & 86.7 & 68.1 & 52.7 \\
\sysname-III & 89.9 & 89.9 & 89.9 & 89.7 & 89.4 & 88.3 & 87.9 & 87.6 \\
\hline
\end{tabular}
\end{table*}

\begin{table*}[h]
\centering
\caption{[Packet Simulation] Throughput (Gbps) vs Lossy Links (0.100\% loss)}
\label{tab:app:pkt-sim:tput-loss-links}
\small
\begin{tabular}{@{}c *{9}{|c} @{}}
\hline
\textbf{Links} & \textbf{0} & \textbf{1} & \textbf{2} & \textbf{3} & \textbf{4} & \textbf{5} & \textbf{6} & \textbf{7} & \textbf{8} \\
\hline
\sysname-II  & 88.3 & 86.7 & 82.3 & 68.1 & 65.5 & 62.9 & 60.6 & 58.4 & 56.4 \\
\sysname-III & 89.9 & 88.3 & 88.4 & 87.6 & 88.0 & 87.4 & 87.8 & 87.3 & 87.6 \\
\hline
\end{tabular}
\end{table*}


\begin{table*}[t]
\centering
\small
\caption{[Packet Simulation, Mode-III] AllReduce Algorithm Throughput (Gbps), with/without switch replying CNP}
\label{tab:app:simulation:switch-cnp}
\begin{tabular}{@{}c *{8}{|c} @{}}
\hline
\textbf{Msg. Size (B)} & \textbf{1K} & \textbf{4K} &  \textbf{16K} &  \textbf{256K} & \textbf{1M} &  \textbf{4M} & \textbf{16M}  & \textbf{64M} \\
\hline
Switch Replying CNP & 14.0 & 38.7 & 70.4 & 88.1 & 94.4 & 96.0 & 95.9 & 90.5  \\
Switch Not Replying CNP & 14.0 & 38.7 & 70.4 & 88.1 & 94.4 & 44.6 & 43.3 & 20.3 \\
\hline
\end{tabular}
\end{table*}

\begin{table*}[h]
\centering
\small
\caption{[SimAI/NS3 Simulation] Large Model Settings}
\label{tab:large-model-settings}
\begin{tabular}{@{}c *{8}{|c} @{}}
\hline
\textbf{Model} & \textbf{\makecell{GPU \\Flops}} & \textbf{\makecell{num \\ layers}} & \textbf{\makecell{hidden \\ size}} & \textbf{\makecell{num \\ parameters}} & \textbf{\makecell{num \\ tokens \\ per\_seq}} & \textbf{\makecell{batch \\ size}} & \textbf{\makecell{dtype \\ size}} & \textbf{\makecell{TP, \\DP, PP}} \\
\hline
GPT-3-175B & 312T & 96 & 12288 & 175B & 2048 & 128 & 2bytes & 4,32,8 \\
GPT-3-13B & 312T & 40 & 5120 & 13B & 2048 & 128 & 2bytes & 8,16,1 \\
Llama-65B & 312T & 80 & 8192 & 65B & 4096 & 128 & 2bytes & 8,16,1 \\
Llama-7B & 312T & 32 & 4096 & 6.7B & 4096 & 128 & 2bytes & 8,16,1 \\
\hline
\end{tabular}
\end{table*}

\begin{table*}[h]
\centering
\caption{[SimAI/NS3 Simulation] Model Training Iteration Time}
\label{tab:app:pkt-sim:model-training-time}
\small
\begin{tabular}{@{}c *{4}{|c} @{}}
\hline
\textbf{Model} & \textbf{GPT-3-175B} & \textbf{GPT-3-13B} & \textbf{Llama-65B} & \textbf{Llama-7B}\\
\hline
\sysname-II & 3.75s & 1.15s & 8.83s & 1.13s \\
\sysname-III & 3.75s & 1.16s & 8.86s & 1.13s \\
Ring & 4.57s & 1.40s & 10.1s & 1.26s \\
\hline
\end{tabular}
\end{table*}

\section{Packet-Level Simulation}
\label{sec:app:packet-simulation}

\subsection{Implementation}
We implement a packet-level simulator based on ns-3 (version 3.43) to evaluate the performance of \sysname-II, \sysname-III, and Ring across six collective communication primitives: AllReduce, Reduce, ReduceScatter, Broadcast, AllGather, and Barrier.

\subsection{Experiment Settings}
For \sysname-II and \sysname-III, we use a star topology with 8 ranks connected to a single switch. For Ring, 8 ranks form a ring topology.
Each link has \SI{100}{Gbps} bandwidth and \SI{1}{\micro\second} latency. We evaluate message sizes ranging from \SI{4}{KB} to \SI{1}{GB}. We vary the packet loss rate from 0\% to 1\% on a single link, and the number of lossy links from 0 to 8 with a fixed 0.1\% loss rate per link.

\subsection{Collective Performance}



\cref{tab:app:pkt-sim:six_collectives} reports the throughput of \sysname-II, \sysname-III, and Ring across all six collectives.
The performance difference between \sysname and Ring stems from two factors: \textit{theoretical bandwidth efficiency} at large messages and \textit{hop count} at small messages. Let $N$ denote message size, $B$ link bandwidth, and $K$ the number of ranks.
Since throughput is inversely proportional to completion time, we analyze performance through theoretical completion time.

\parab{Large messages: theoretical completion time matters.}
At large message sizes where bandwidth dominates, performance is determined by the theoretical completion time.
For AllReduce, \sysname completes in $N/B$ while Ring requires $2(K-1)N/(KB)$, resulting in \sysname achieving 1.7$\times$ throughput (\SI{89.9}{Gbps} vs \SI{52.3}{Gbps} at \SI{1}{GB}).
For other collectives (Reduce, Broadcast, ReduceScatter, AllGather), the theoretical gap is smaller or absent—Ring's $(K-1)N/(KB)$ versus \sysname's $N/B$ for scatter/gather operations, and identical $N/B$ for Reduce and Broadcast. The algorithm bandwidth of AllGather and ReduceScatter is higher than the \SI{100}{Gbps} link bandwidth because their traffic volume ($(K-1)N/K$) is smaller than the message size ($N$).
Consequently, all approaches converge to near line-rate ($\sim$\SI{91}{Gbps}) at \SI{1}{GB} for these collectives.

\parab{Small messages: hop count matters.}
At small message sizes where latency dominates, \sysname's single-hop communication through the switch consistently outperforms Ring's $O(K)$ hops.
This advantage is most pronounced at 4KB messages, where \sysname achieves 6--10$\times$ higher throughput than Ring across all collectives.
The gap narrows at medium sizes as pipelining effects begin to benefit Ring.

\parab{\sysname-III vs \sysname-II.}
\sysname-III achieves marginally higher throughput than \sysname-II in AllReduce due to its decoupled aggregation-broadcast design, which enables better pipelining.
For Barrier, \sysname-III reaches 333K QPS versus \sysname-II's 200K QPS.


\subsection{Loss Tolerance}

We evaluate \sysname-II and \sysname-III under packet loss by measuring AllReduce throughput.
To isolate the retransmission mechanism's impact, we disable congestion control (i.e., the sending window does not shrink upon packet loss), focusing purely on how each design handles retransmissions.

Table~\ref{tab:app:pkt-sim:tput-loss-rate} varies the packet loss rate on rank~0's uplink, where \sysname-III consistently outperforms \sysname-II across all loss rates.
At a 0.1\% loss rate, \sysname-III achieves \SI{88.3}{Gbps} versus \sysname-II's \SI{86.7}{Gbps}.
The gap widens significantly at higher loss rates: at a 1\% loss, \sysname-III maintains \SI{87.6}{Gbps}, while \sysname-II drops to only \SI{52.7}{Gbps}.

Table~\ref{tab:app:pkt-sim:tput-loss-links} extends this setting to multiple ranks by fixing the loss rate at 0.1\% and varying the number of lossy uplinks: with all 8 links experiencing loss, \sysname-III achieves \SI{87.6}{Gbps} versus \sysname-II's \SI{56.4}{Gbps}.
The performance gap stems from different retransmission mechanisms.
In \sysname-II, a packet loss on any uplink triggers fast retransmissions from all participating ranks. In contrast, \sysname-III enables the switch to actively detect packet loss and issue NAKs to notify only the corresponding rank for retransmission. This significantly reduces retransmission overhead compared to \sysname-II.

\subsection{Rate Synchronization}
We further evaluate rate synchronization for \sysname-III by running AllReduce on Tree-2-8 with \SI{200}{Gbps} link capacity, where one link is congested by background traffic to \SI{100}{Gbps} and DCQCN is enabled on the ranks. \cref{tab:app:simulation:switch-cnp} reports the AllReduce goodput with and without switch-generated CNPs. The results show that congestion-control-based rate synchronization slows down faster ranks before they overrun the pipe window, avoiding excessive packet drops and retransmissions and improving the overall collective throughput.

\subsection{Large Model Training Performance}


We run SimAI to simulate large model training, with \sysname-ns3 as the backend. The topology is fat-tree. Model settings are shown in \cref{tab:large-model-settings}. \cref{tab:app:pkt-sim:model-training-time} shows the model training iteration time.


\sysname-II accelerates GPT-3-175B, GPT-3-13B, Llama-65B, and Llama-7B by 22\%, 22\%, 14\%, and 12\%, respectively. In most cases, the performance gain is significant (up to 22\%); even in some instances where the relative gain is small, the absolute reduction of JCT is still considerable, considering the scale.


\section{Flow-Level Simulation}
\label{sec:app:flow-simulation}


\subsection{Implementation}

\parab{Simulator.}
We implement a flow-level simulator based on OMNeT++~\cite{omnetpp} with approximately 3.9K lines of C++ code, designed to model single-tenant and multi-tenant large-scale distributed training workloads over a fat-tree interconnect, with explicit support for different INC resource allocation policies. The simulator operates at the granularity of communication groups (i.e., collective operations), rather than packets, which enables scalable evaluation of job-level performance metrics such as job completion time (JCT) under realistic cluster scales.

The simulator explicitly models both rank behaviors and network behaviors. In each simulation step, ranks alternate between computation and communication phases, while the network allocates bandwidth and INC resources subject to link capacity and switch-level constraints. During communication, the network applies a waterfilling-based algorithm to estimate per-group throughput and advances data transmission accordingly.

Each job consists of multiple iterations, and each iteration triggers a fixed set of communication groups. INC resource allocation decisions are made independently for each communication group based on the selected execution mode and current resource availability.

\begin{table*}[t]
\centering
\small
\caption{[Flow Simulation] JCT of model training in a 128-GPU Fat-tree; one unit of switch memory can saturate one group.}
\label{tab:app:flow-sim:fattree-scaleup}
\begin{tabular}{@{}c|c|*{4}{c}|*{4}{c}@{}}
\hline
\multirow{3}{*}{\textbf{Model}} & \multirow{3}{*}{\textbf{Policy}}
 & \multicolumn{8}{c}{\textbf{Switch SRAM (Unit)}} \\ \cline{3-10}
 & & \multicolumn{4}{c|}{\textbf{without scale-up}} & \multicolumn{4}{c}{\textbf{with scale-up}} \\ \cline{3-10}
 & & \textbf{4} & \textbf{8} & \textbf{16} & \textbf{32} & \textbf{4} & \textbf{8} & \textbf{16} & \textbf{32} \\
\hline
\multirow{4}{*}{GPT-3-175B}
 & Ring         & 253 & 253 & 253 & 253 & 106  & 106  & 106  & 106  \\
 & EDT          & 179 & 179 & 179 & 179 & 84.6 & 84.6 & 84.6 & 84.6 \\
 & Spatial Mux  & 179 & 158 & 137 & 137 & 84.6 & 63.6 & 63.6 & 63.6 \\
 & Temporal Mux & 158 & 137 & 137 & 137 & 84.6 & 63.6 & 63.6 & 63.6 \\
\hline
\multirow{4}{*}{GPT-3-13B}
 & Ring         & 33.6 & 33.6 & 33.6 & 33.6 & 7.94 & 7.94 & 7.94 & 7.94 \\
 & EDT          & 20.7 & 20.7 & 20.7 & 20.7 & 6.38 & 6.38 & 6.38 & 6.38 \\
 & Spatial Mux  & 20.7 & 19.1 & 17.5 & 17.5 & 6.38 & 4.82 & 4.82 & 4.82 \\
 & Temporal Mux & 19.1 & 17.5 & 17.5 & 17.5 & 6.38 & 4.82 & 4.82 & 4.82 \\
\hline
\multirow{4}{*}{Llama-65B}
 & Ring         & 211 & 211 & 211 & 211 & 47.6 & 47.6 & 47.6 & 47.6 \\
 & EDT          & 129 & 129 & 129 & 129 & 39.8 & 39.8 & 39.8 & 39.8 \\
 & Spatial Mux  & 129 & 121 & 113 & 113 & 110  & 39.8 & 32.0 & 32.0 \\
 & Temporal Mux & 121 & 113 & 113 & 113 & 110  & 39.8 & 32.0 & 32.0 \\
\hline
\multirow{4}{*}{Llama-7B}
 & Ring         & 37.8 & 37.8 & 37.8 & 37.8 & 5.01 & 5.01 & 5.01 & 5.01 \\
 & EDT          & 21.3 & 21.3 & 21.3 & 21.3 & 4.20 & 4.20 & 4.20 & 4.20 \\
 & Spatial Mux  & 21.3 & 20.5 & 19.7 & 19.7 & 4.20 & 3.40 & 3.40 & 3.40 \\
 & Temporal Mux & 20.5 & 19.7 & 19.7 & 19.7 & 4.20 & 3.40 & 3.40 & 3.40 \\
\hline
\end{tabular}
\end{table*}

\subsection{Experiment Settings}

\parab{Policies and Baseline.} We implement four execution policies and settings corresponding to different INC strategies.

\textbf{(1) Ring without INC.} All communication groups execute using conventional collective communication without in-network aggregation. Communication volumes are not reduced, and all traffic traverses the network without consuming INC resources.

\textbf{(2) EDT.} We implement an edge-disjoint tree based INC policy. A communication group is granted INC if and only if none of the links on its aggregation tree are currently used by another INC-enabled group. Link usage is tracked explicitly. If any link conflict exists, the group falls back to non-INC execution. INC allocation is therefore enforced at link granularity, ensuring strict edge disjointness among concurrent INC trees.

\textbf{(3) Spatial Mux.} Each switch maintains a fixed number of INC resource slots. A communication group is granted INC if all switches along its aggregation tree have at least one available slot. Once allocated, INC resources are reserved for the entire lifetime of the job and are not released until the job completes. This models the static partitioning of switch INC resources across concurrent jobs.

\textbf{(4) Temporal Mux.} INC allocation follows the same admission conditions as spatial multiplexing, but resources are allocated and released dynamically at the granularity of individual communication groups. After a communication group completes, its occupied INC resources are immediately released and may be reused by other groups. This enables time-sharing of INC resources across jobs and improves overall utilization under multi-tenant contention.

For each communication group, the simulator determines whether INC is enabled according to the selected mode and current resource availability. If INC is granted, the effective communication volume is reduced accordingly; otherwise, the group executes in non-INC mode. The simulator advances until all communication groups and computation phases of a job complete, at which point the job completion time is recorded.

\parab{Topology and Server.}
We model a three-tier fat-tree topology consisting of leaf, spine, and core switches. Each switch has 32 ports, and each link has a bandwidth capacity of \SI{100}{Gbps}. Communication groups are mapped onto aggregation trees spanning the fat-tree topology. Switches are associated with a limited amount of INC-related resources, abstracted differently by each INC policy.

A server is set up with multiple GPUs and NICs, each GPU bonded with a NIC. In the runtime, a rank owns one GPU and its associated NIC. A server's NICs connect to fat-tree leaf switches sequentially. We provide options to enable scale-up networks between GPUs within one server (e.g., NVLink); intra-server communication among these GPUs is handled by the scale-up fabric and does not consume fat-tree network bandwidth (inter-server communication is still carried over the fat-tree network).


In the single-tenant setting, the topology is configured to support a single 128-GPU training job. Each leaf switch connects to 8 servers. A pod contains 4 leaf switches and 4 spine switches with full leaf–spine connectivity, and each spine switch connects to 4 core switches. Under this configuration, each pod interconnects $4 \times 8 = 32$ GPUs, and the system consists of 4 pods in total, providing exactly 128 GPUs. This setting exposes both intra-pod and inter-pod communication paths for a large single-tenant workload.

In the multi-tenant setting, we scale up the topology to support concurrent jobs. Each leaf switch connects to 16 servers, and each pod consists of 16 leaf switches and 16 spine switches with full leaf–spine connectivity, while each spine switch connects to 8 core switches. As a result, each pod interconnects $16 \times 16 = 256$ GPUs, and the overall system comprises 8 pods, forming a shared network fabric for multi-tenant training with increased contention at the spine and core layers.


\parab{Workloads.}
We evaluate our design with 3D parallel training in single-tenant settings, where the configurations are the same as \cref{tab:large-model-settings}.
We evaluate \sysname in multi-tenant settings using three workload traces that differ in how job compositions and network pressures are constructed.

Trace 1 is a manually configured synthetic workload that serves as a controlled baseline. In this trace, we explicitly define the job size distribution. We consider five representative job sizes, namely 8, 16, 32, 64, and 128 GPUs, and assign them fixed proportions of 30\%, 30\%, 25\%, 10\%, and 5\% of the total jobs, respectively. By construction, Trace 1 provides a stable and interpretable workload composition, allowing us to isolate the impact of network configurations and INC strategies without interference from trace-specific artifacts.

Trace 2 is derived from the Alibaba Lingjun dataset~\cite{cao2024crux}, which provides CSV summaries of distributed training jobs collected from a large-scale production GPU cluster. We extract job sizes and their relative proportions from these summaries, and use the resulting distribution to generate synthetic workloads for simulation. Trace 2, therefore, reflects the job mix observed under the default production network configuration.

Trace 3 is constructed by keeping the same workload composition as Trace 2, while modifying the network topology to increase communication pressure. Specifically, we reduce the number of core-layer switches by half, which increases the likelihood of cross-pod communication and contention in the upper layers of the fat-tree. This trace enables us to study the behavior of different INC strategies under heightened network bottlenecks while holding the workload mix constant.

\subsection{Performance}
\parab{Single-Tenant.}
We first compare the three INC-based schemes with the conventional ring-based collective communication. As shown in \cref{tab:app:flow-sim:fattree-scaleup}, when switch SRAM is at least 8 units, all INC schemes
achieve lower job completion time (JCT) than the ring baseline; even with only 4 SRAM units, most configurations still benefit from INC. This demonstrates the benefit of in-network aggregation.

We then analyze the impact of limited switch memory on JCT. When switch memory is scarce, the JCT performance ordering follows EDT, spatial multiplexing, and temporal multiplexing, with temporal multiplexing achieving the lowest JCT due to its more flexible allocation of switch resources. As switch memory becomes sufficient, the JCT gap between spatial and temporal multiplexing diminishes, and both schemes achieve comparable performance, while EDT remains inferior due to its stricter allocation constraints.

\begin{figure*}[th]
\centering

\pgfplotsset{every axis/.style={
    xlabel={Job Completion Time (s)},
    ylabel={CDF},
    xtick align=inside,
    ymajorgrids=true,
    xmajorgrids=true,
    font=\footnotesize,
    grid style=dashed,
}}

\begin{subfigure}{0.33\textwidth}
\centering
\begin{tikzpicture}
\begin{axis}[
    name=axis0,
    title={\textbf{All Jobs}}, title style={yshift=-7pt,},
    ymin=0, ymax=1,
    ytick={0,0.2,0.4,0.6,0.8,1.0},
    legend style={
        at={(0.43,1.23)},
        anchor=south,
        legend columns=-1,
        column sep=0.4em,
        row sep=0.1em,
        font=\scriptsize
    },
    width=1\linewidth,
    height=0.64\linewidth,
]
    \addplot[solid, color=blue,  mark=o,        mark repeat=6, mark phase=0]
    table[x expr={\thisrowno{2}==1 ? \thisrowno{0} : nan}, y expr={\thisrowno{2}==1 ? \thisrowno{1} : nan}]{plots/flow_simulation/data_fattree_tenant_trace1_all.dat};
    \addplot[solid, color=red,   mark=+,        mark repeat=6, mark phase=2]
    table[x expr={\thisrowno{2}==2 ? \thisrowno{0} : nan}, y expr={\thisrowno{2}==2 ? \thisrowno{1} : nan}]{plots/flow_simulation/data_fattree_tenant_trace1_all.dat};
    \addplot[solid, color=black, mark=square,   mark repeat=6, mark phase=4]
    table[x expr={\thisrowno{2}==3 ? \thisrowno{0} : nan}, y expr={\thisrowno{2}==3 ? \thisrowno{1} : nan}]{plots/flow_simulation/data_fattree_tenant_trace1_all.dat};
    \addplot[solid, color=orange,mark=triangle, mark repeat=6, mark phase=1]
    table[x expr={\thisrowno{2}==4 ? \thisrowno{0} : nan}, y expr={\thisrowno{2}==4 ? \thisrowno{1} : nan}]{plots/flow_simulation/data_fattree_tenant_trace1_all.dat};
\end{axis}
\begin{axis}[
    name=axis1,
    yshift=-0.6\linewidth,
    title={\textbf{Tail 60\% Jobs}}, title style={yshift=-7pt,},
    ymin=0.4, ymax=1,
    width=1\linewidth,
    height=0.64\linewidth,
]
    \addplot[solid, color=blue,  mark=o,        mark repeat=6, mark phase=0]
    table[x expr={\thisrowno{2}==1 ? \thisrowno{0} : nan}, y expr={\thisrowno{2}==1 ? \thisrowno{1} : nan}]{plots/flow_simulation/data_fattree_tenant_trace1_tail.dat};
    \addplot[solid, color=red,   mark=+,        mark repeat=6, mark phase=2]
    table[x expr={\thisrowno{2}==2 ? \thisrowno{0} : nan}, y expr={\thisrowno{2}==2 ? \thisrowno{1} : nan}]{plots/flow_simulation/data_fattree_tenant_trace1_tail.dat};
    \addplot[solid, color=black, mark=square,   mark repeat=6, mark phase=4]
    table[x expr={\thisrowno{2}==3 ? \thisrowno{0} : nan}, y expr={\thisrowno{2}==3 ? \thisrowno{1} : nan}]{plots/flow_simulation/data_fattree_tenant_trace1_tail.dat};
    \addplot[solid, color=orange,mark=triangle, mark repeat=6, mark phase=1]
    table[x expr={\thisrowno{2}==4 ? \thisrowno{0} : nan}, y expr={\thisrowno{2}==4 ? \thisrowno{1} : nan}]{plots/flow_simulation/data_fattree_tenant_trace1_tail.dat};
\end{axis}
\end{tikzpicture}
\caption{Trace 1}
\label{fig:app:jct_cdf_trace1}
\end{subfigure}
\hfill
\begin{subfigure}{0.33\textwidth}
\centering
\begin{tikzpicture}
\begin{axis}[
    name=axis0,
    title={\textbf{All Jobs}}, title style={yshift=-7pt,},
    ymin=0, ymax=1,
    ytick={0,0.2,0.4,0.6,0.8,1.0},
    legend style={
        at={(0.55,1.16)},
        anchor=south,
        legend columns=-1,
        column sep=0.4em,
        row sep=0.1em,
        font=\scriptsize
    },
    width=1\linewidth,
    height=0.64\linewidth,
]
    \addplot[solid, color=blue,  mark=o,        mark repeat=6, mark phase=0]
    table[x expr={\thisrowno{2}==1 ? \thisrowno{0} : nan}, y expr={\thisrowno{2}==1 ? \thisrowno{1} : nan}]{plots/flow_simulation/data_fattree_tenant_trace2_all.dat};
    \addplot[solid, color=red,   mark=+,        mark repeat=6, mark phase=2]
    table[x expr={\thisrowno{2}==2 ? \thisrowno{0} : nan}, y expr={\thisrowno{2}==2 ? \thisrowno{1} : nan}]{plots/flow_simulation/data_fattree_tenant_trace2_all.dat};
    \addplot[solid, color=black, mark=square,   mark repeat=6, mark phase=4]
    table[x expr={\thisrowno{2}==3 ? \thisrowno{0} : nan}, y expr={\thisrowno{2}==3 ? \thisrowno{1} : nan}]{plots/flow_simulation/data_fattree_tenant_trace2_all.dat};
    \addplot[solid, color=orange,mark=triangle, mark repeat=6, mark phase=1]
    table[x expr={\thisrowno{2}==4 ? \thisrowno{0} : nan}, y expr={\thisrowno{2}==4 ? \thisrowno{1} : nan}]{plots/flow_simulation/data_fattree_tenant_trace2_all.dat};
\legend{Ring, EDT, Spatial Mux, Temporal Mux}
\end{axis}
\begin{axis}[
    name=axis1,
    yshift=-0.6\linewidth,
    title={\textbf{Tail 15\% Jobs}}, title style={yshift=-7pt,},
    ymin=0.85, ymax=1,
    width=1\linewidth,
    height=0.64\linewidth,
]
    \addplot[solid, color=blue,  mark=o,        mark repeat=6, mark phase=0]
    table[x expr={\thisrowno{2}==1 ? \thisrowno{0} : nan}, y expr={\thisrowno{2}==1 ? \thisrowno{1} : nan}]{plots/flow_simulation/data_fattree_tenant_trace2_tail.dat};
    \addplot[solid, color=red,   mark=+,        mark repeat=6, mark phase=2]
    table[x expr={\thisrowno{2}==2 ? \thisrowno{0} : nan}, y expr={\thisrowno{2}==2 ? \thisrowno{1} : nan}]{plots/flow_simulation/data_fattree_tenant_trace2_tail.dat};
    \addplot[solid, color=black, mark=square,   mark repeat=6, mark phase=4]
    table[x expr={\thisrowno{2}==3 ? \thisrowno{0} : nan}, y expr={\thisrowno{2}==3 ? \thisrowno{1} : nan}]{plots/flow_simulation/data_fattree_tenant_trace2_tail.dat};
    \addplot[solid, color=orange,mark=triangle, mark repeat=6, mark phase=1]
    table[x expr={\thisrowno{2}==4 ? \thisrowno{0} : nan}, y expr={\thisrowno{2}==4 ? \thisrowno{1} : nan}]{plots/flow_simulation/data_fattree_tenant_trace2_tail.dat};
\end{axis}
\end{tikzpicture}
\caption{Trace 2}
\label{fig:app:jct_cdf_trace2}
\end{subfigure}
\hfill
\begin{subfigure}{0.33\textwidth}
\centering
\begin{tikzpicture}
\begin{axis}[
    name=axis0,
    title={\textbf{All Jobs}}, title style={yshift=-7pt,},
    ymin=0, ymax=1,
    ytick={0,0.2,0.4,0.6,0.8,1.0},
    legend style={
        at={(0.43,1.23)},
        anchor=south,
        legend columns=-1,
        column sep=0.4em,
        row sep=0.1em,
        font=\scriptsize
    },
    width=1\linewidth,
    height=0.64\linewidth,
]
    \addplot[solid, color=blue,  mark=o,        mark repeat=6, mark phase=0]
    table[x expr={\thisrowno{2}==1 ? \thisrowno{0} : nan}, y expr={\thisrowno{2}==1 ? \thisrowno{1} : nan}]{plots/flow_simulation/data_fattree_tenant_trace3_all.dat};
    \addplot[solid, color=red,   mark=+,        mark repeat=6, mark phase=2]
    table[x expr={\thisrowno{2}==2 ? \thisrowno{0} : nan}, y expr={\thisrowno{2}==2 ? \thisrowno{1} : nan}]{plots/flow_simulation/data_fattree_tenant_trace3_all.dat};
    \addplot[solid, color=black, mark=square,   mark repeat=6, mark phase=4]
    table[x expr={\thisrowno{2}==3 ? \thisrowno{0} : nan}, y expr={\thisrowno{2}==3 ? \thisrowno{1} : nan}]{plots/flow_simulation/data_fattree_tenant_trace3_all.dat};
    \addplot[solid, color=orange,mark=triangle, mark repeat=6, mark phase=1]
    table[x expr={\thisrowno{2}==4 ? \thisrowno{0} : nan}, y expr={\thisrowno{2}==4 ? \thisrowno{1} : nan}]{plots/flow_simulation/data_fattree_tenant_trace3_all.dat};
\end{axis}
\begin{axis}[
    name=axis1,
    yshift=-0.6\linewidth,
    title={\textbf{Tail 20\% Jobs}}, title style={yshift=-7pt},
    ymin=0.8, ymax=1,
    width=1\linewidth,
    height=0.64\linewidth,
]
    \addplot[solid, color=blue,  mark=o,        mark repeat=6, mark phase=0]
    table[x expr={\thisrowno{2}==1 ? \thisrowno{0} : nan}, y expr={\thisrowno{2}==1 ? \thisrowno{1} : nan}]{plots/flow_simulation/data_fattree_tenant_trace3_tail.dat};
    \addplot[solid, color=red,   mark=+,        mark repeat=6, mark phase=2]
    table[x expr={\thisrowno{2}==2 ? \thisrowno{0} : nan}, y expr={\thisrowno{2}==2 ? \thisrowno{1} : nan}]{plots/flow_simulation/data_fattree_tenant_trace3_tail.dat};
    \addplot[solid, color=black, mark=square,   mark repeat=6, mark phase=4]
    table[x expr={\thisrowno{2}==3 ? \thisrowno{0} : nan}, y expr={\thisrowno{2}==3 ? \thisrowno{1} : nan}]{plots/flow_simulation/data_fattree_tenant_trace3_tail.dat};
    \addplot[solid, color=orange,mark=triangle, mark repeat=6, mark phase=1]
    table[x expr={\thisrowno{2}==4 ? \thisrowno{0} : nan}, y expr={\thisrowno{2}==4 ? \thisrowno{1} : nan}]{plots/flow_simulation/data_fattree_tenant_trace3_tail.dat};
\end{axis}
\end{tikzpicture}
\caption{Trace 3}
\label{fig:app:jct_cdf_trace3}
\end{subfigure}

\caption{[Flow Simulation] CDF of JCT in Three Traces on a 2048-GPU Fat-tree}
\label{fig:app:jct_cdf_traces}
\end{figure*}
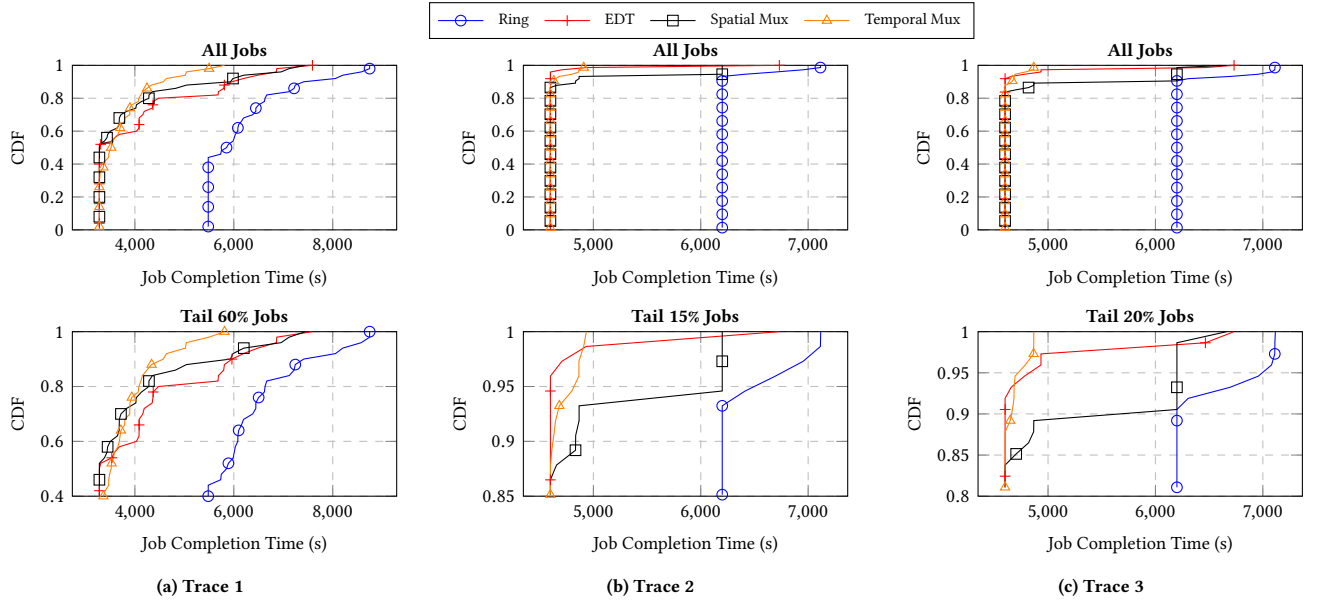

\begin{table}[t]
\centering
\caption{[Flow Simulation] Average JCT of Multi-tenant Jobs on 2048-GPU Fat-tree}
\label{tab:app:flow-sim:fattree:tenant}
\begin{tabular}{@{}l *{3}{c} @{}}
\hline
\textbf{Workloads} & \textbf{Trace1} & \textbf{Trace2} & \textbf{Trace3} \\
\hline
Ring          & 6190 & 6245 & 6267 \\
EDT             & 4108 & 4635 & 4666 \\
Spatial Mux & 3920 & 4722 & 4791 \\
Temporal Mux     & 3750 & 4621 & 4618 \\
\hline
\end{tabular}
\end{table}

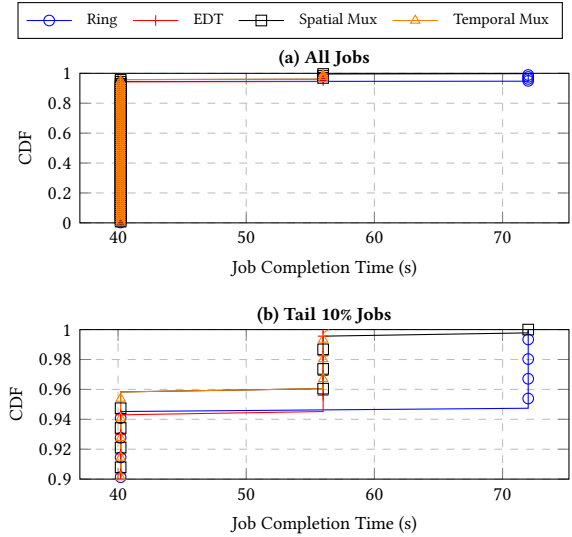
\begin{figure}[th]
\centering

\pgfplotsset{every axis/.style={
    height=0.42\linewidth,
    width=0.95\linewidth,
    xlabel={Job Completion Time (s)},
    ylabel={CDF},
    xtick align=inside,
    ymajorgrids=true,
    xmajorgrids=true,
    font=\footnotesize,
    grid style=dashed,
}}

\begin{tikzpicture}
\begin{axis}[
    name=axis0,
    title={\textbf{(a) All Jobs}}, title style={yshift=-7pt},
    ymin=0, ymax=1,
    ytick={0,0.2,0.4,0.6,0.8,1.0},
    legend style={
        at={(0.43,1.23)},
        anchor=south,
        legend columns=-1,        
        column sep=0.4em,        
        row sep=0.1em,           
        font=\scriptsize         
    },
]

    \addplot[solid, color=blue,  mark=o,        mark repeat=6, mark phase=0]
    table[
      x expr={\thisrowno{2}==1 ? \thisrowno{0} : nan},
      y expr={\thisrowno{2}==1 ? \thisrowno{1} : nan}
    ]{plots/flow_simulation/data_scaleup_tenant_all.dat};
    
    \addplot[solid, color=red,   mark=+,        mark repeat=6, mark phase=2]
    table[
      x expr={\thisrowno{2}==2 ? \thisrowno{0} : nan},
      y expr={\thisrowno{2}==2 ? \thisrowno{1} : nan}
    ]{plots/flow_simulation/data_scaleup_tenant_all.dat};
    
    \addplot[solid, color=black, mark=square,   mark repeat=6, mark phase=4]
    table[
      x expr={\thisrowno{2}==3 ? \thisrowno{0} : nan},
      y expr={\thisrowno{2}==3 ? \thisrowno{1} : nan}
    ]{plots/flow_simulation/data_scaleup_tenant_all.dat};
    
    \addplot[solid, color=orange,mark=triangle, mark repeat=6, mark phase=1]
    table[
      x expr={\thisrowno{2}==4 ? \thisrowno{0} : nan},
      y expr={\thisrowno{2}==4 ? \thisrowno{1} : nan}
    ]{plots/flow_simulation/data_scaleup_tenant_all.dat};

\legend{Ring, EDT, Spatial Mux, Temporal Mux}
\end{axis}
\begin{axis}[
    name=axis1,
    yshift=-0.4\linewidth,
    title={\textbf{(b) Tail 10\% Jobs}}, title style={yshift=-7pt},
    ymin=0.9, ymax=1,
    ytick={0.90,0.92,0.94,0.96,0.98,1.00},
    legend style={
        at={(0.5,1.08)},
        anchor=south,
        legend columns=2,        
        column sep=0.4em,        
        row sep=0.1em,           
        font=\scriptsize         
    },
]

    \addplot[solid, color=blue,  mark=o,        mark repeat=6, mark phase=0]
    table[
      x expr={\thisrowno{2}==1 ? \thisrowno{0} : nan},
      y expr={\thisrowno{2}==1 ? \thisrowno{1} : nan}
    ]{plots/flow_simulation/data_scaleup_tenant_tail.dat};
    
    \addplot[solid, color=red,   mark=+,        mark repeat=6, mark phase=2]
    table[
      x expr={\thisrowno{2}==2 ? \thisrowno{0} : nan},
      y expr={\thisrowno{2}==2 ? \thisrowno{1} : nan}
    ]{plots/flow_simulation/data_scaleup_tenant_tail.dat};
    
    \addplot[solid, color=black, mark=square,   mark repeat=6, mark phase=4]
    table[
      x expr={\thisrowno{2}==3 ? \thisrowno{0} : nan},
      y expr={\thisrowno{2}==3 ? \thisrowno{1} : nan}
    ]{plots/flow_simulation/data_scaleup_tenant_tail.dat};
    
    \addplot[solid, color=orange,mark=triangle, mark repeat=6, mark phase=1]
    table[
      x expr={\thisrowno{2}==4 ? \thisrowno{0} : nan},
      y expr={\thisrowno{2}==4 ? \thisrowno{1} : nan}
    ]{plots/flow_simulation/data_scaleup_tenant_tail.dat};

\end{axis}
\end{tikzpicture}

\caption{[Flow Simulation] 8-GPU's JCT CDF of Trace 2 on 2048-GPU with scale-up.}
\label{fig:app:jct_cdf_scaleup}
\end{figure}

\begin{table}[th]
\centering
\caption{[Flow Simulation] 8-GPU Jobs' JCT in Trace 2 on 2048-GPU Fat-tree with and without scale-up}
\label{tab:app:flow-sim:scaleup:tenant}
\begin{tabular}{@{}l *{2}{c} @{}}
\hline
\textbf{Workload} & \textbf{without scale-up} & \textbf{with scale-up}  \\
\hline
Ring          & 72.0 & 41.9 \\
EDT           & 56.0 & 41.1  \\
Spatial Mux   & 56.2 & 40.9 \\
Temporal Mux  & 56.0 & 40.9  \\
\hline
\end{tabular}
\end{table}

\parab{Multi-Tenant.} \cref{fig:app:jct_cdf_traces} and \cref{tab:app:flow-sim:fattree:tenant} show the JCT of multi-tenant jobs with different policies. 
The three policies with INC outperform the Ring algorithm in all cases because INC generally reduces communication time. \cref{tab:app:flow-sim:scaleup:tenant} and \cref{fig:app:jct_cdf_scaleup} exemplify the 8-GPU jobs in Trace 2, showing that some jobs spanning servers can benefit from INC; when scale-up networks exist, all jobs benefit from high bandwidth, and INC still shows acceleration.

Temporal Mux can significantly improve tail latency compared to Spatial Mux because it utilizes switch SRAM more efficiently; however, Temporal Mux allows jobs to reserve resources and exclude others. Some jobs (with resources in Spatial Mux) experience a higher JCT in Temporal Mux due to resources being reserved by others.

There are cases where EDT outperforms Spatial Mux (\cref{fig:app:jct_cdf_trace2} and \cref{fig:app:jct_cdf_trace3}). EDT constrains the number of supported jobs by disjoint trees, while Spatial Mux is constrained by the switch memory quota. If EDT allocates more trees for jobs, it would outperform Spatial Mux; in this case, the EDT switch actually consumes more SRAM than Spatial Mux.

\section{FPGA Implementation}
\label{sec:app:fpga}

\begin{table}[th]
\centering
\caption{[FPGA, Cluster Vendor] Resource Consumption with FP32}
\label{tab:fpga:lenovo:resource}
\begin{tabular}{@{}c *{3}{|c} @{}}
\hline
\textbf{Buffer Size (MB)}  & \textbf{ALM} & \textbf{Memory Bits} & \textbf{DSP} \\
\hline
2 & 265969 & 54484032 & 256   \\
4 & 271160 & 92753792 & 256  \\
8 & 287376 & 169571264 & 256 \\
16 & 306642 & 322347776 & 256  \\
\hline
\end{tabular}
\end{table}

\begin{table*}[th]
\centering
\caption{[FPGA, Institute A] Overall Resource Consumption}
\label{tab:fpga:nudt:resource_overall}
\begin{tabular}{@{}c *{5}{|c} @{}}
\hline
\textbf{\makecell{Data \\Type}} & \textbf{\makecell{Buffer\\ Size}} & \textbf{\makecell{LUTs\\ (1,728,000)}} & \textbf{\makecell{LUTRAM \\(791,040)}} & \textbf{\makecell{FF \\(3,456,000)}} & \textbf{\makecell{BRAM\\ (2,688)}} \\ 
\hline
Int32 & 1MB & 867479(50.20\%) & 637900(80.64\%) & 169382(4.90\%) & 235(8.74\%) \\
Int32 & 512KB & 863426(49.97\%) & 637900(80.64\%) & 165024(4.78\%) & 122(4.54\%) \\
Int32 & 256KB & 863433(49.97\%) & 637900(80.64\%) & 165026(4.78\%) & 122(4.54\%) \\
Int32 & 128KB & 468170(27.09\%) & 319116(40.34\%) & 158769(4.69\%) & 122(4.54\%) \\
FP32 & 1MB & 987308(57.14\%) & 638924(80.77\%) & 343724(9.95\%) & 235(8.74\%) \\
FP32 & 512KB & 983245(56.90\%) & 638924(80.77\%) & 339366(9.82\%) & 122(4.54\%) \\
FP32 & 256KB & 983250(56.90\%) & 638924(80.77\%) & 339369(9.82\%) & 122(4.54\%) \\
FP32 & 128KB & 588010(34.03\%) & 320140(40.47\%) & 333111(9.64\%) & 122(4.54\%) \\
\hline
\end{tabular}
\end{table*}
\begin{table*}[th]
\centering
\caption{[FPGA, Institute A] IncEngine Resource Consumption}
\label{tab:fpga:nudt:resource_agg}
\begin{tabular}{@{}c *{5}{|c} @{}}
\hline
\textbf{\makecell{Data \\Type}} & \textbf{\makecell{Buffer\\ Size}} & \textbf{\makecell{LUTs\\ (1,728,000)}} & \textbf{\makecell{LUTRAM \\(791,040)}} & \textbf{\makecell{FF \\(3,456,000)}} & \textbf{\makecell{BRAM\\ (2,688)}} \\ 
\hline
Int32 & 1MB & 866023(50.12\%) & 637568(80.60\%) & 144989(4.20\%) & 235(8.74\%) \\
Int32 & 512KB & 850970(49.25\%) & 637568(80.60\%) & 140631(4.07\%) & 122(4.54\%) \\
Int32 & 256KB & 850973(49.25\%) & 637568(80.60\%) & 140629(4.07\%) & 122(4.54\%) \\
Int32 & 128KB & 462529(26.77\%) & 318784(40.30\%) & 140583(4.07\%) & 122(4.54\%) \\
FP32 & 1MB & 974852(56.42\%) & 638592(80.73\%) & 319331(9.24\%) & 235(8.74\%) \\
FP32 & 512KB & 970789(56.18\%) & 638592(80.73\%) & 314973(9.11\%) & 122(4.54\%) \\
FP32 & 256KB & 970804(56.18\%) & 638592(80.73\%) & 314971(9.11\%) & 122(4.54\%) \\
FP32 & 128KB & 582370(33.70\%) & 319808(40.43\%) & 314971(9.11\%) & 122(4.54\%) \\
\hline
\end{tabular}
\end{table*}

\subsection{Implementation from a Cluster Vendor}

As \sysname Mode-II presents the lowest barrier to implementation, we collaborated with a cluster vendor to facilitate its FPGA-based realization and evaluation. The implementation was developed using Verilog on the Intel Agilex™ 7 FPGA M-Series 039 (R47A) platform. 

To assess the performance of the proposed IncEngine, we conducted cycle-accurate RTL simulations focused on throughput, per-packet latency, and resource utilization. Experimental results indicate that for an 8-rank configuration, the total latency from the arrival of the final packet to the commencement of aggregated result output is approximately \SI{230}{ns}. This total duration encompasses the entire processing pipeline of the parser and deparser, with the core functional latency of the aggregation engine specifically accounting for \SI{53}{ns}.

The hardware resource overhead is primarily influenced by the buffer capacity and supported data types. The current design supports the FP32 data type with a default buffer specification of \SI{8}{MB}, which can be configured for either 16 ranks × 2 groups or 32 ranks × 1 group. Under the FP32 configuration, as the buffer capacity scales from \SI{2}{MB} to \SI{16}{MB}, the Adaptive Logic Module (ALM) consumption ranges from 265,969 to 306,642, while memory bit utilization increases from approximately \SI{54.5}{Mb} to \SI{322.3}{Mb}. Across all buffer sizes, the Digital Signal Processor (DSP) utilization remains constant at 256 units (\cref{tab:fpga:lenovo:resource}).

\subsection{Implementation from an Institute}

The \sysname specification was delivered to a research institute for comprehensive FPGA implementation and performance evaluation. The system is implemented on the Xilinx Virtex UltraScale+ VU13P platform, utilizing approximately 6,500 lines of Verilog code for the entire logic, with the IncEngine itself comprising 5,500 lines.

To ensure high-throughput processing, two critical FPGA-specific optimizations are integrated: first, a block-based SIMD parallel computing method is employed to divide large bit-width payloads into independent sub-vectors, thereby overcoming the bottlenecks of element-wise serial processing; second, a speculation-based consistency control mechanism is implemented to resolve read-after-write (RAW) hazards arising from varying data path lengths between initial and retransmitted packets without stalling the pipeline.

Experimental evaluations conducted at a \SI{250}{MHz} clock frequency demonstrate that the system achieves line-rate \SI{100}{Gbps} throughput across various packet types, with end-to-end latencies for \SI{1}{KB} payloads ranging from \SI{96}{ns} for direct forwarding to \SI{244}{ns} for FP32 upstream aggregation. \cref{tab:fpga:nudt:resource_overall} and \cref{tab:fpga:nudt:resource_agg} show the overall and IncEngine's resource consumption.

Detailed resource utilization analysis shows that for a \SI{1}{MB} buffer configuration, the Int32 implementation consumes 867,479 LUTs (50.20\%) and 235 BRAMs (8.74\%), while the FP32 configuration increases the demand to 987,308 LUTs (57.14\%) and 343,724 Flip-Flops (9.95\%), reflecting the increased complexity of floating-point operations.

\section{Evaluation by Chip Vendor}
\label{sec:app:chip}

The \sysname specification was delivered to a chip vendor for a comprehensive hardware implementation assessment. The core logic of the engine was developed using Verilog, comprising 66,377 lines of code to define the Register-Transfer Level (RTL) hardware description, which was subsequently verified through cycle-accurate SystemVerilog simulations. 

Architecturally, the design prioritizes high-performance collective communications, supporting critical operations such as Broadcast, AllReduce, and ReduceScatter with internal FP32 precision to ensure numerical consistency. The design also supports other data types such as FP16, BF16, and INT32, but internally, all data types except INT32 are converted to FP32 for computation and then converted back after computation.
The design chooses reproducible computation (buffering all data and then computing).

Experimental results from the RTL simulations indicate that each packet incurs a processing latency of \SI{50}{ns}, with a single engine providing a throughput of \SI{3.2}{Tbps}. When integrated into the switching chip, a cluster of eight IncEngine units achieves an aggregate processing capacity of \SI{25.6}{Tbps}. Each IncEngine supports 64 communication groups, each group with 16 members.

Regarding hardware overhead, the design was synthesized using a \SI{28}{nm} process technology. A single IncEngine instance, configured with 512 FP32 ALUs, 512 UINT ALUs, and a \SI{1}{MB} payload buffer, occupies an area of \SI{4.89}{\milli\meter\squared}. Consequently, the total area overhead for the eight integrated engines amounts to \SI{39.12}{\milli\meter\squared}, demonstrating a scalable and efficient footprint for high-bandwidth in-network computing.
The chip specification includes:
\begin{itemize}
    \item \textbf{Collective Types:} Broadcast, AllGather, Reduce, ReduceScatter, AllReduce;
    \item \textbf{Operation Types:} SUM, MIN, MAX;
    \item \textbf{Data Types:} FP32, FP16, BF16, INT32;
    \item \textbf{Capacity:} \SI{25.6}{Tbps}.
\end{itemize}

\end{document}